\documentclass{aa}  

\usepackage{graphicx}
\usepackage{txfonts}
\usepackage{multirow}
\usepackage{colortbl}

\usepackage{hyperref}  
\hypersetup{colorlinks=true,linkcolor=[rgb]{1.,0.2,0.2},citecolor=[rgb]{0.1,0.4,1.},filecolor=[rgb]{0.7,0.2,0.2},urlcolor=[rgb]{0.7,0.2,0.2}}

\usepackage{color}
\definecolor{blue}{rgb}{0., 0., 1}

\newcommand{\nvalone}{\textrm{N}\textsc{v}}
\newcommand{\civalone}{\textrm{C}\textsc{iv}}

\newcommand{\heiialone}{\textrm{He}\textsc{ii}}

\newcommand{\ciiialone}{\textrm{C}\textsc{iii}}

\newcommand{\oiidoub}{[\textrm{O}\textsc{ii}]\ensuremath{\lambda3727,3729}}

\newcommand{\mgii}{\textrm{Mg}\textsc{ii}~\ensuremath{\lambda2800}}

\newcommand{\oiiialone}{[\textrm{O}\textsc{iii}]}

\newcommand{\oiiidoub}{[\textrm{O}~\textsc{iii}]\ensuremath{\lambda\lambda4959,5007}}
 
\newcommand{\ha}{\ifmmode {\rm H}\alpha \else H$\alpha$\fi}
\newcommand{\hb}{\ifmmode {\rm H}\beta \else H$\beta$\fi}
\newcommand{\lya}{\ifmmode {\rm Ly}\alpha \else Ly$\alpha$\fi}
\newcommand{\pg}{\ifmmode {\rm P}\gamma \else Pa$\gamma$\fi}
\newcommand{\lyb}{\ifmmode {\rm Ly}\beta \else Ly$\beta$\fi}
\newcommand{\lyg}{\ifmmode {\rm Ly}\gamma \else Ly$\gamma$\fi}
\newcommand{\ciii}{\textrm{C}\textsc{iii}]\ensuremath{\lambda1908}}
\newcommand{\ciiiblue}{\textrm{C}\textsc{iii}]\ensuremath{\lambda1907}}

\newcommand{\ciiidoub}{\textrm{C}\textsc{iii}]\ensuremath{\lambda\lambda1907,1909}}

\newcommand{\siiiuvem}{[\textrm{Si}\textsc{iii}]\ensuremath{\lambda1883,1892}}

\newcommand{\nv}{\textrm{N}\textsc{v}\ensuremath{\lambda1240}}
\newcommand{\nvdoub}{[\textrm{N}\textsc{V}]\ensuremath{\lambda\lambda1239,1243}}
\newcommand{\civ}{\textrm{C}\textsc{iv}\ensuremath{\lambda1548,1550}}

\newcommand{\heii}{\textrm{He}\textsc{ii}\ensuremath{\lambda1640}}
\newcommand{\oiiiuv}{\textrm{O}\textsc{iii}]\ensuremath{\lambda1661,1666}}

\newcommand{\flyc}{\ifmmode  \mathrm{f}_\mathrm{esc}\mathrm{(LyC)} \else $\mathrm{f}_\mathrm{esc}\mathrm{(LyC)}$\fi}

\def\kms{km s$^{-1}$}

\def\ergs{\ifmmode \mathrm{erg\hspace{1mm}s}^{-1} \else erg s$^{-1}$\fi}
\def\ergscm{erg s$^{-1}$ cm$^{-2}$}
\def\micron{\ifmmode \mu\mathrm{m} \else $\mu$m\fi}
\def\msun{\ifmmode \mathrm{M}_{\odot} \else M$_{\odot}$\fi}
\def\msunyr{\ifmmode \mathrm{M}_{\odot} \hspace{1mm}{\rm yr}^{-1} \else $\mathrm{M}_{\odot}$ yr$^{-1}$\fi}
\def\zsun{\ifmmode Z_{\odot} \else Z$_{\odot}$\fi}
\def\lsun{\ifmmode L_{\odot} \else L$_{\odot}$\fi}
\def\mstar{\ifmmode \mathrm{M}_{\star} \else M$_{\star}$\fi}

\begin{document}

\titlerunning{Star-forming complexes at cosmological distance}
\title{The MUSE Deep Lensed Field on the Hubble Frontier Field MACS~J0416}
\subtitle{Star-forming complexes at cosmological distances}

\authorrunning{E. Vanzella et al.}
\author{
E.~Vanzella \inst{\ref{inafbo}} \fnmsep\thanks{E-mail: \href{mailto:eros.vanzella@inaf.it}{eros.vanzella@inaf.it}}\fnmsep\thanks{Based on observations collected at the European Southern Observatory for Astronomical research in the Southern Hemisphere under ESO programmes ID 0100.A-0763(A) (PI E. Vanzella), 094.A-0115B (PI J. Richard), 094.A-0525(A) (PI F.E. Bauer).},
G. B. Caminha \inst{\ref{kapteyn}},
P. Rosati \inst{\ref{unife},\ref{inafbo}},
A. Mercurio \inst{\ref{inafna}},
M. Castellano \inst{\ref{inafrome}},
M. Meneghetti \inst{\ref{inafbo}},
C. Grillo \inst{\ref{unimi}},
E. Sani \inst{\ref{ESO}}, \\
P. Bergamini \inst{\ref{inafbo}},
F. Calura \inst{\ref{inafbo}},
K. Caputi \inst{\ref{kapteyn}},
S. Cristiani \inst{\ref{inafts}},
G. Cupani \inst{\ref{inafts}},
A. Fontana \inst{\ref{inafrome}},
R. Gilli \inst{\ref{inafbo}},
A. Grazian \inst{\ref{inafpd}},
M. Gronke \inst{\ref{JHU}} \fnmsep\thanks{Hubble Fellow}, \\
M. Mignoli \inst{\ref{inafbo}},
M. Nonino \inst{\ref{inafts}},
L. Pentericci \inst{\ref{inafrome}},
P. Tozzi \inst{\ref{arcetri}},
T. Treu \inst{\ref{UCLA}},
I. Balestra \inst{\ref{OMEGA}}, \and
M. Dijkstra
}

\institute{
INAF -- OAS, Osservatorio di Astrofisica e Scienza dello Spazio di Bologna, via Gobetti 93/3, I-40129 Bologna, Italy \label{inafbo} 
\and
Kapteyn Astronomical Institute, University of Groningen, Postbus 800, 9700 AV Groningen, The Netherlands \label{kapteyn}
\and
Dipartimento di Fisica e Scienze della Terra, Universit\`a degli Studi di Ferrara, via Saragat 1, I-44122 Ferrara, Italy \label{unife}
\and
INAF -- Osservatorio Astronomico di Capodimonte, Via Moiariello 16, I-80131 Napoli, Italy \label{inafna}
\and
INAF -- Osservatorio Astronomico di Roma, Via Frascati 33, I-00078 Monte Porzio Catone (RM), Italy \label{inafrome}
\and
Dipartimento di Fisica, Universit\`a  degli Studi di Milano, via Celoria 16, I-20133 Milano, Italy \label{unimi}
\and
European Southern Observatory, Alonso de Cordova 3107, Casilla 19, Santiago 19001, Chile \label{ESO}
\and
INAF -- Osservatorio Astronomico di Trieste, via G. B. Tiepolo 11, I-34143, Trieste, Italy \label{inafts}
\and
INAF -- Osservatorio Astronomico di Padova, Vicolo Osservatorio 5, 35122, Padova, Italy \label{inafpd}
\and
Department of Physics $\&$ Astronomy, Johns Hopkins University, Baltimore, MD 21218, USA \label{JHU}
\and
INAF -- Osservatorio Astrofisico di Arcetri, Largo E. Fermi, I-50125, Firenze, Italy \label{arcetri}
\and
Department of Physics and Astronomy, University of California, Los Angeles, CA 90095, USA \label{UCLA}
\and
OmegaLambdaTec GmbH, Lichtenbergstrasse 8, 85748 Garching bei Munchen, Germany \label{OMEGA}
}

\date{} 

 
  \abstract
   {A census of faint and tiny star forming complexes at high redshift is key to improving our understanding of  reionizing sources, galaxy growth, and the formation of globular clusters.}
   {We present the MUSE Deep Lensed Field (MDLF) program, which is aimed at unveiling the very faint population of high redshift sources that are magnified by strong gravitational lensing and to significantly increase the number of constraints for the lens model.}
   {We describe Deep MUSE observations of 17.1 hours of integration on a single pointing over the Hubble Frontier Field galaxy cluster MACS~J0416, providing line flux limits down to 
   $2\times 10^{-19}$ \ergscm\ within 300 \kms\ and continuum detection down to magnitude 26, both at the three sigma level at $\lambda = 7000$~\AA. 
   For point sources with a magnification ($\mu$) greater than 2.5 (7.7), the MLDF depth is equivalent to integrating more than 100 (1000) hours in blank fields, as well as  complementing non-lensed studies of very faint high-z sources.  
   The source-plane effective area of the MDLF with $\mu>6.3$ is $<50$\% of the image-plane field of view.}
   {
  We confirm spectroscopic redshifts for all 136 multiple images of 48 source galaxies at $0.9 < z < 6.2$. Within those galaxies, we securely identify 182 multiple images of 66 galaxy components that we use to constrain our lens model. This makes MACS J0416 the cluster with the largest number of confirmed constraints for any strong lens model to date. We identify 116 clumps belonging to background  high-z galaxies; the majority of them are multiple images and span magnitude, size, and redshift intervals of $[-18, -10]$, $[\sim 400-3]$ parsec and $1<z<6.6$, respectively, with the faintest or most magnified ones probing possible single gravitationally bound star clusters. The multiplicity introduced by gravitational lensing allows us, in several cases, to triple the effective integration time up to $\sim$51 hours exposure per single family, leading to a detection limit for unresolved emission lines of a few $10^{-20}$ \ergscm, after correction for lensing magnification. Ultraviolet high-ionization metal lines (and \heii) are detected with S/N~$>10$ for individual objects down to de-lensed magnitudes between $28-30$. The median stacked spectrum of 33 sources with a median M$_{\rm UV} \simeq -17$ and $<z>$ = 3.2 ($1.7<z<3.9$) shows high-ionization lines, suggesting that they are common in such faint sources.}
   {Deep MUSE observations, in combination with existing HST imaging, allowed us to: (1) confirm redshifts for extremely faint high-z sources; (2) peer into their internal structure to
   unveil clumps down to $100-200$ pc scale; (3) in some cases, break down such clumps into star-forming complexes matching the scales of bound star clusters ($<20$ pc effective radius); (4)  double the number of constraints for the lens model, reaching an unprecedented set of 182 bona-fide multiple images and confirming up to 213 galaxy cluster members. These results demonstrate the power of JWST and future adaptive optics facilities mounted on the Extremely Large Telescopes (e.g., European-ELT Multi-conjugate Adaptive Optics RelaY, MAORY, coupled with the Multi-AO Imaging CamerA for Deep Observations, MICADO) or Very Large Telescope (e.g., MCAO Assisted Visible Imager and Spectrograph, MAVIS) when combined in studies with gravitational telescopes.
  } 


   \maketitle
%

\section{Introduction}

The key capabilities of extremely large telescopes (ELTs) for the exploration of the distant Universe will provide unprecedented access to  faint luminosities (thanks to a large collecting area) and angular resolution down to $\sim10$ milliarcsec (mas) thanks to the technology of adaptive optics (AO). In the near future, the imminent launch of the James Webb Space Telescope (JWST) will open up a new wavelength domain redward of the K$-$band that is crucial for capturing rest-frame optical lines well within the reionization epoch.
These future facilities will allow us to routinely analyze the internal structures of high-redshift galaxies at unprecedented small spatial scales. With a typical point spread function (PSF) of $\sim 10$ mas (milli-arcsecond) and a pixel scale of 4 mas per pixel, the high-redshift galaxies that remain unresolved today will finally be dissected into resolution elements of 80 (60) parsec at a redshift of 3 (6), eventually providing constraints down to spatial scales of $\sim 20~(30)$ pc per pixel (e.g., E-ELT/MAORY-MICADO). In this regard, star-forming complexes ($<200$ pc size) at high 
redshift and high mass star clusters (e.g., $<30$ pc radius) will be accessed and compared to local similar star-forming regions, allowing for detailed studies of (1) star-formation modes (e.g., location and spectral signatures of massive stars); (2) the presence of high-ionization lines and the related ionization photon production efficiency \citep[e.g.,][]{bouwens.xsi,amorin17,senchyna17,chevallard18,senchyna19,lam19,senchyna20}; and (3) interactions with the surrounding medium (feedback), including the capacity to modulate the opacity of the interstellar medium to ionizing radiation up to circumgalactic scales (which is key for the escape of ionizing photons, e.g., \citealt{erb15nature,grazian17,vanz_sunburst,he20_ricotti}). These are all key ingredients in the pursuit of answers to two of the most pressing questions in current observational cosmology: i) what sources reionized the Universe \citep[e.g.,][]{robertson15,giallo15,meyer2020,eide20_eor_sim,dayal20}; ii) how globular clusters formed \citep[][]{renzini15,renzini17,pfeffer18,pfeffer19,calura19,bastian.lardo18}; iii) and how these two questions might be related to each other \citep[e.g.,][]{ricotti02,schaerer11, katz13, boylan18,ma20_gc_eor,he20_ricotti}.

Indeed, it has been established in the local Universe that the fraction of forming stars located in gravitationally bound star clusters (also known as cluster formation efficiency, $\Gamma$\footnote{$\Gamma$ is defined as the cluster formation rate (CFR) divided by the star formation rate (SFR) of the hosting galaxy \citep[][]{bastian08}.}), increases as the star-formation rate surface density increases \citep[][]{adamo17,adamo20extreme,adamo20}; this, in turn, relates to the increasing gas pressure in high gas surface density conditions \citep[][]{kruijssen12,CFE_li18}. It has also been proposed that $\Gamma$ positively correlates with redshift, such that, on average, high-density conditions and merger rate in the high redshift Universe would favor $\Gamma > 30-40$\%, whereas it is of a few percent at low redshift ($z<2$) \citep[e.g.,][]{pfeffer18}. Similar arguments, based on the present-day volume density of globular clusters projected back in time, suggest that at $z \gtrsim 5,$ about half of the stellar mass of the Universe was located in star clusters \citep[][]{renzini17}, thus suggesting that a significant fraction of the star formation of the Universe in the first Gyrs took place in these systems.
Therefore, if reionization was mainly driven by star formation -- which is mostly confined to bound star clusters -- then it is plausible that young star clusters played an essential role in this process 
\citep[e.g.,][]{ricotti02,boylan18,bik18,vanz_sunburst,herenz17_ionized_channel}.

For the reasons described above, a census of gravitationally-bound young star clusters at high redshift would represent a big step forward in this investigation. Observationally, such a census requires improvements in angular resolution and depth in the rest frame UV with HST. Observations in the rest frame optical and longer wavelengths (JWST and ALMA) will then be necessary to understand their physical properties in detail.

Even though angular resolution of 10-20  mas in the rest frame UV is currently not attainable overall, significant progress has been made in terms of depth with the VLT, which is performing very deep spectroscopy of the faintest sources in what is currently the deepest field obtained with Hubble \citep[the Hubble Ultra Deep Field, HUDF,][]{HUDF.original,HUDF.extreme,HUDF.koekemoer}. The initial results from an extended integration time ($>30$ hours), obtained with the VLT multi-unit spectroscopic explorer \citep[MUSE,][]{Bacon_MUSE} in the HUDF, have been presented in a series of recent works \citep[e.g.,][]{Bacon17, Inami17, maseda18,maseda20, Lya.glow.MUSE, lya.fraction.MUSE,feltre20}\footnote{The complete list is available here: {\it http://muse-vlt.eu/science/publications/}}, confirming redshifts for galaxies as faint as magnitude-30 \citep[][]{brinchmann17}, including a set of ``HST-dark'' MUSE sources with detected emission lines (typically \lya) with no detection of HST counterparts \citep[][]{Inami17,mary20}.
The VLT/MUSE coverage has revolutionized the study of the high-redshift Universe at $z<6.65$ in the post-reionization epoch \citep[][]{bacon_proc_20}.

A complementary approach
is the use of gravitational lensing magnification ($\mu$) \citep[e.g.,][]{bradley14,atek14,atek15,atek18,Treu_2015,Karman_2015,Karman_2017,Caminha_2019,erb2019,richard_MUSE_HFF_14} provided by clusters of galaxies, which makes background sources brighter and larger on the sky, thus allowing for a much higher effective resolution than in blank fields with the same observational setup. For example, the combination of (1) MUSE integral field spectroscopy; (2) lensed fields; and (3) deep multi-frequency HST imaging (such as the Hubble Frontier Fields, HFF hereafter, \citealt{Lotz_2017HFF,Koekemoer_2014HFF}) has led to the confirmation of an unprecedented number of multiple images per field up to a redshift of $z \simeq 6.7$ \citep[][]{Caminha_macs0416, lagat_19_A370}. These identifications are crucial for constructing high-precision lens models. 
Such observations have given us a first glimpse of what will be accessible with ELTs or extreme AO facilities such as VLT/MAVIS\footnote{{\it http://mavis-ao.org/mavis/} and {\it https://arxiv.org/abs/2009.09242} for the Phase A Science Cases.}  in blank fields.

The sub-kpc spatial resolution provided by lensing magnification revealed spatial variations of, for instance, \lya\ emission along the arcs \citep[e.g.,][]{claeyssens19}, and, in combination with HST deep imaging, it allowed the detection of faint lensed sources \citep[e.g.,][]{mahler18}. As an interesting example,
some star-forming complexes, discovered in the HFFs, have characteristic sizes smaller than 100 pc along with other physical parameters that make them good candidates for globular cluster precursors \citep[][]{vanz16_id11,vanz_paving, vanz_id14}. In fact, along the maximum tangential stretch provided by strong lensing, the effective resolution of HST reaches a few tens of pc, while the boosted signal-to-noise ratio (S/N) enables a morphological analysis that would be impossible in blank fields.
One such a case is the compact object behind MACS~J0416, identified as a young massive star cluster with an intrinsic (i.e., delensed) magnitude of 31.3 and effective radius smaller than 13 pc, which is hosted in a dwarf galaxy at $z=6.149$ \citep[][]{vanz19,calura20}. Similarly, high-redshift star-forming clumps ($<200$ pc size) have been identified in various lensed fields, suggesting that a hierarchically structured star-formation topology emerges whenever the angular resolution increases \citep[see also,][]{livermore15,kawamata15,rigby17,johnson17,mirka17,cava18,zick20}. 
In a recent spectacular case, a highly magnified, finely structured giant arc has been identified as the first example of a young (3 Myr old) massive star cluster at $z=2.37$, directly detected in the Lyman continuum ($\lambda < 912$~\AA) and contributing to the ionization of the IGM \citep[][]{vanz_sunburst,rivera19, chisholm19}.

\begin{figure*}[ht!]
        \centering
        \includegraphics[width=\linewidth]{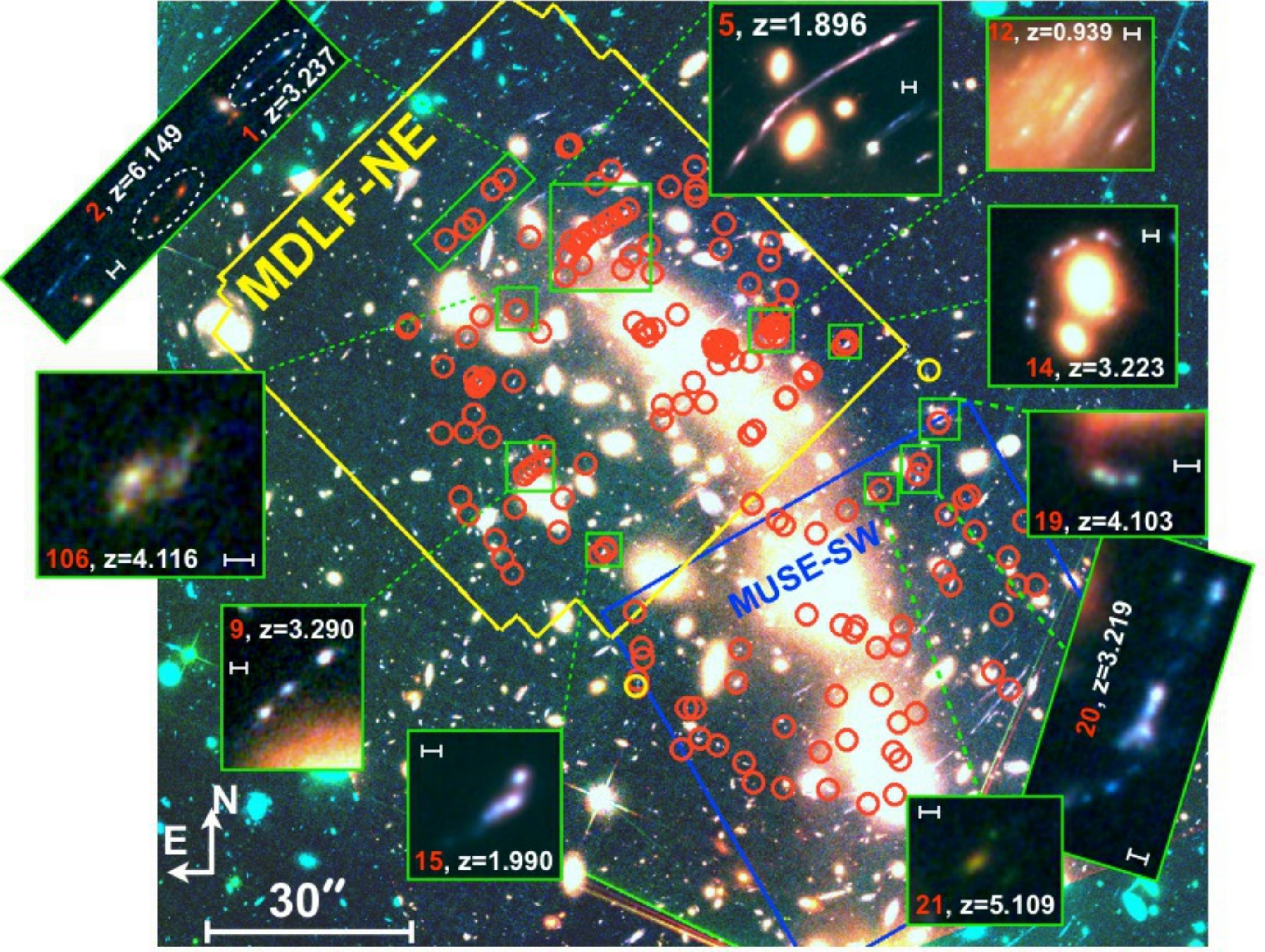}
        \caption{Color image of the HFF galaxy cluster MACS~J0416 shown in the main panel (red, green, and blue as F105W, F814W, and F606W, respectively). The MDLF centered in the north-east part of the cluster is outlined by the yellow contour (17.1h integration time), while the blue contour shows the observation in the south-west (11-hour integration, PI Bauer, 094.A-0525(A)). The red circles mark the positions of the 182 multiple images used to constrain the lens model \citep[][]{bergamini20}. The insets show zoomed examples of strongly lensed objects with detected clumps in the redshift range of $1.5<z<6.5$ (discussed in detail in Sect~\ref{clumps}). In each inset, the catalog ID (red number) and redshift are indicated, while the segment marks the $0\farcs3$ scale. Yellow circles mark two images not covered by MUSE, which are however included in the multiple images sample due to their mirroring lensing properties. }
        \label{fig1}
\end{figure*}

There are two key aspects in the study of the distant Universe in lensed fields driven by MUSE integral field spectroscopy: (1) the $1' \times 1'$ field of view integral field unit (IFU) provides spectroscopic redshifts without any target pre-selection, significantly enlarging the discovery space (the identification of globular cluster precursors and extremely faint sources with intrinsic magnitude $>31$ are two examples among a number of others); and (2) dozens of multiple images can be easily identified by the IFU in a single observation, even when distorted or extended. This yields a vast gain in efficiency compared to traditional target-oriented ``multi-slit spectroscopy,'' at least when a large field of view is not required and when the density of targets is high. 

The dramatic increase in spectroscopically confirmed multiple images is key for producing robust magnification maps with lens models, with a much improved understanding of the systematic uncertainties affecting magnification values and their gradients across the image \cite[e.g.,][]{grillo_2015,grillo_2016,Meneghetti_2017,Caminha_macs0416,atek18,Treu_2016}. This remains a critical step for inferring intrinsic physical properties or the geometry of highly magnified galaxies. 

The confirmation of lensed sources with intrinsic magnitudes in the range of $30-33$ via \lya\ emission shows that sources without HST imaging counterparts are common  in lensed fields and at fluxes fainter than those of similar HST$-$dark MUSE sources found in the HUDF. An example is the high equivalent width ($> 1000$~\AA\ rest-frame) \lya\ arclet at $z=6.629$,  straddling a caustic, which is confirmed in the MDLF. The HST counterpart is barely detected in HST imaging (with an observed magnitude of $\gtrsim 31$ at 2$\sigma$ level), which corresponds to an intrinsic magnitude fainter than 35. This suggests that such star-forming complex possibly hosts extremely metal-poor (or Pop~III) stellar populations \citep[see][for details]{vanz_popiii}. This is perhaps the most compelling example of a blind spectroscopic detection, as it would be impossible to place a slit over such an object based on deep imaging alone.

To push the frontier of integral field spectroscopy of the high redshift Universe,  we present in this work the MUSE Deep Lensed Field over the North-East part of the Hubble Frontier Field galaxy cluster MACS~J0416.1$-$2403 (hereafter MACS~J0416). We show that a total integration of 17.1 hours in a single MUSE pointing on the magnified region of the cluster produces point-like source detection that would require $>100$ ($\mu>2.5$) to $>1000$ ($\mu>7.5$) hours of integration without lensing (where $\mu$ is the magnification factor). With the addition of publicly available data in the south-west region of the same galaxy cluster (with 11-hour integration; see Sect.~\ref{SW}), the number of confirmed multiple images increases to the unprecedented number of 182 in the redshift range of $0.9<z<6.2$. A new lens model based on this set of images is presented in an accompanying paper by \citet{bergamini20}. Soon after this paper appeared on astro-ph, \citet{richard2020} submitted a paper presenting an atlas of MUSE observations of 12 clusters and corresponding lens models. This study included the new MUSE deep observations of MACS~J0416 presented here.

The present work is structured as follows: Section 2 presents the MUSE observations and data reduction.  Section 3 presents the full set of multiple images. Section 4.1 focuses on the sample of star forming clumps identified among the multiple images, and Section 4.2 details the spectral stacking and the most relevant high-ionization lines.
Individual sources are presented in Section 5, highlighting two examples of extremely small objects as potential gravitationally bound star clusters.
We assume a flat cosmology with $\Omega_{M}$= 0.3,
$\Omega_{\Lambda}$= 0.7 and $H_{0} = 70$ km s$^{-1}$ Mpc$^{-1}$.


\section{The MUSE Deep Lensed Field: Observations and data reduction}

Deep MUSE \citep{Bacon_MUSE} observations were allocated in period 100 (Prog.ID 0100.A-0763(A) $-$ PI E. Vanzella) on a single pointing covering the north-east (NE) lensed region of the HFF galaxy cluster MACS~J0416 (Figure~\ref{fig1}). Out of a total of 19 observing blocks (OBs) that were scheduled (22.1h, including overhead), 16 have been successfully acquired with quality A or B (84\%).\footnote{The quality control of OBs executed in service mode is based on the specified constraints in the OB for airmass, atmospheric transparency, image quality and seeing, Moon constraints, twilight constraint, as well as Strehl ratio for Adaptive Optics mode observations (as requested). If all constraints are fulfilled, the OB is marked with the grade "A", while the "B" quality control is assigned if some constraint is up to 10\% violated. The observations with quality control grades A or B are completed, while those with quality control grade "C" (out of constraints) are re-scheduled and may be repeated ({\it www.eso.org}).}
Table~\ref{tab1} lists the log of the observations that were executed in the period between November 2017 till August 2019.
14 OBs out of 16 have been acquired with the assistance of Ground Layer Adaptive Optics (GLAO) provided by the GALACSI module. 
Each exposure was offset by fractions of arcseconds and rotated by 90 degrees to improve sky subtraction. The image quality was very good, spanning the range between $0\farcs4-0\farcs8$, with a median PSF full width at half maximum (FWHM) of $0\farcs6$. The same NE field of the galaxy cluster was observed within a GTO program (Prog.ID 094.A-0115B, PI: J.Richard) in November 2014, for a total of two hours split into four exposures \citep[][]{Caminha_macs0416}. We added the 2014 dataset to our MUSE data, eventually producing a total integration time of 17.1h on-sky with a final optimal image quality of $0\farcs6$. In the following, we refer to this deep pointing as the MUSE Deep Lensed Field (MDLF).

\subsection{Data reduction}

We used the MUSE data reduction pipeline version 2.8.1 \citep{2014ASPC..485..451W} to process the raw data and create the final stacked data-cube.
All standard calibration procedures were applied to the science exposures (i.e.,  bias and flat field corrections, wavelength and flux calibration, etc.).
In order to reduce the remaining instrumental signatures due to slice-to-slice flux variations of the instrument, we  used the self-calibration method.
This method is based on the MUSE Python Data Analysis Framework \citep{2016ascl.soft11003B} and implemented in the last versions of the standard reduction pipeline provided by ESO.
The final astrometry was performed matching sources detected with {\tt SExtractor} \citep{1996A&AS..117..393B} in the white image of the final data-cube and detections in the HFF filter F606W image.
Finally, we applied the Zurich Atmosphere Purge \citep{ZAP16} on the data-cube in order to remove the still remaining sky residuals.

Four OBs, indicated as ``NOAO'' in Table~\ref{tab1}, have been observed with an average natural seeing (i.e., without GLAO) of $0\farcs6$, and simply included in the co-addition of all OBs following the procedure described above. For these datacubes no Raman lines due to the laser are present, especially in the wavelength range of $5800-6000$~\AA. However, the final co-added cube is dominated by OBs obtained with GLAO (14 out of 18).

The final data-cube has a spatial pixel scale of $0\farcs2$, a spectral coverage from 4700~\AA\ to 9350~\AA, with a dispersion of 1.25~\AA/pixel and a fairly constant spectral resolution of 2.6~\AA\ over the entire spectral range.
The total integration time is 17.1h, with an image quality of $0\farcs6$, as measured on two stars available in the field.

%
\begin{table}
\caption{Summary of the MDLF observations}             
\label{tab1}      
\centering          
\begin{tabular}{l c l}     
\hline\hline       
Date & Quality & OB Name \\ 
\hline
 & MDLF & \\
\hline
    22/23-Nov-2017 & A & WFM\_J0416\_NOAO\_1\\  
    10/11-Jan-2018 & A & WFM\_J0416\_NOAO\_2\\
    21/22-Feb-2018 & C & WFM\_J0416\_NOAO\_3\\
    12/13-Mar-2018 & X & WFM\_J0416\_AO\_1 \\
    4/5-Nov-2018   & A & WFM\_J0416\_AO\_10\\
    5/6-Nov-2018   & B & WFM\_J0416\_AO\_1 \\
    5/6-Nov-2018   & A & WFM\_J0416\_AO\_2 \\
    6/7-Nov-2018   & A & WFM\_J0416\_AO\_4 \\
    2/3-Dec-2018   & A & WFM\_J0416\_AO\_11 \\
    4/5-Dec-2018   & A & WFM\_J0416\_AO\_13 \\
    12/13-Dec-2018 & A & WFM\_J0416\_AO\_14 \\
    11/12-Jan-2019 & B & WFM\_J0416\_AO\_5\\
    16/17-Jan-2019 & C & WFM\_J0416\_AO\_6\\
    25/26-Jan-2019 & A & WFM\_J0416\_AO\_6\\
    27/28-Feb-2019 & A & WFM\_J0416\_AO\_17\\
    28-Feb/1-Mar-2019 &A & WFM\_J0416\_AO\_7 \\
    3/4-Mar-2019 & A & WFM\_J0416\_AO\_8\\
    2/3-Aug-2019 & A & WFM\_J0416\_AO\_9\\
    30/31-Aug-2019 & A & WFM\_J0416\_AO\_18\\
\hline     
       & GTO & \\
\hline
    17-Dec-2014 & A & WFM\_J0416\_NOAO\\
    17-Dec-2014 & A & WFM\_J0416\_NOAO\\
\hline
\hline
\end{tabular}
\tablefoot{Log of the observed OBs. The typical exposure time (on sky) of each OB is 3340s. The bottom two rows refer to the previous two hours of observation from the GTO. The column ``OB Name''  indicates the observing mode ("WFM" = wide field mode) and the use (or not) of the adaptive optics "AO/NOAO", meaning "on/off".}
\end{table}

\subsection{Depth of the MDLF}

The performances and the depth achievable with the VLT/MUSE instrument have been well monitored in the past few years from extensive observations, from a few to dozens of hours of integration time \citep[e.g.,][]{Inami17}. In particular, the very deep campaign performed in the Hubble Ultra Deep field, HUDF \citep[e.g.,][and references therein]{Bacon17,maseda18}, suggest a growing S/N that is fully in line with the expected integration time \citep[see also,][]{Bacon15}. Under the assumption of similar observing conditions and data reduction technique, a proper rescaling of the depth reported from deep GTO program \citep[e.g.,][]{Inami17} would suggest a line flux limit for our 17.1h MDLF of $\simeq 2 \times 10^{-19}$ \ergscm\ at $3\sigma$, $\lambda = 7000$~\AA\ and within an aperture of $0\farcs8$ diameter. 

\begin{figure}[ht!]
        \centering
        \includegraphics[width=\linewidth]{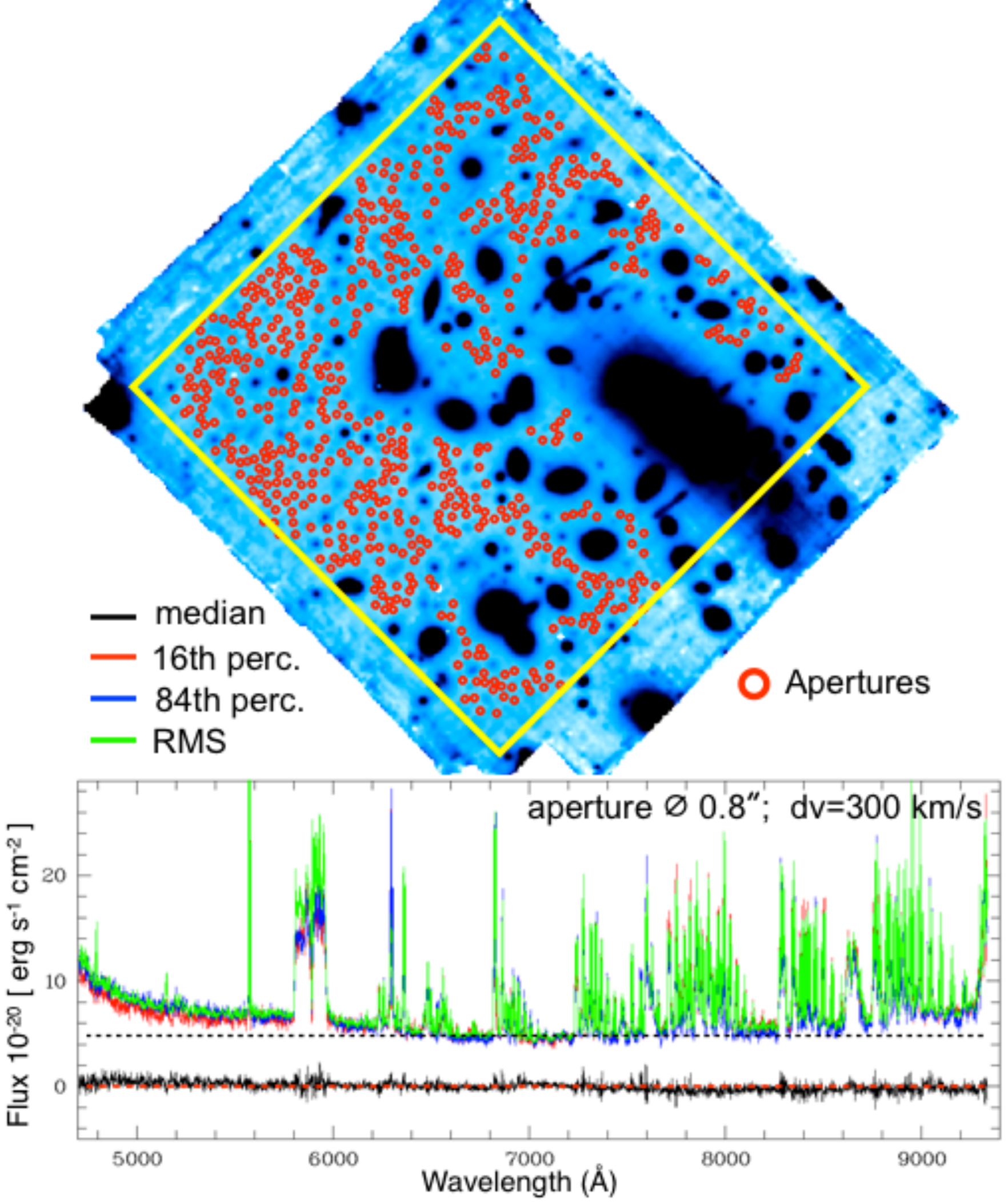}
        \caption{Top panel: White-light image of the MDLF is shown together with the 600 non-overlapping apertures placed in empty zones not intercepting visible objects in the image. The corresponding median value calculated at each wavelength and consistent with the zero level (black line), 16th$-$84th percentiles (red and blue lines), and the rms (green line) are reported below the white image. Apertures have a diameter of $0\farcs8$ and the statistics is computed on each aperture by collapsing slices within dv=300 \kms. The pattern of the sky emission lines is evident. The increased noise in the range of $5800<\lambda<6000$~\AA\ is due to the emission of the laser used for ground layer AO.}
        \label{rms}
\end{figure}

Similarly to what described by \citet{herenz17_MUSE_WIDE}, we then carried out a posteriori checks of the noise fluctuation of the reduced data cube (i.e., after the full data reduction) by placing 
600 non-overlapping apertures (of $0\farcs8$ diameter) on positions visually extracted from the white image (obtained by collapsing the full wavelength range) and not intercepting evident sources. The location of the apertures is plotted over the white image in Figure~\ref{rms}.
We then calculated the flux within each aperture by integrating it over a velocity width of dv=300 \kms\ (kept constant across the full wavelength range), typical of \lya\ emission in high redshift galaxies.  The mean and rms, as well as the median and the 68\% central interval within the 16th and 84th percentiles within the 600 apertures, were extracted at each wavelength with an incremental step of 1.5~\AA. Figure~\ref{rms} shows the median and percentiles as a function of wavelength. The pattern of the sky spectrum clearly emerges, as well as the increased noise in the wavelength range of 5800~\AA$-$6000~\AA\ due to the GLAO sodium-based laser. We derive a $3\sigma$ limit of $1.5 \times 10^{-19}$ \ergscm\ at 7000~\AA\ (where no OH sky lines are present) within an aperture of $0\farcs8$ diameter and collapsed over 300 \kms\ along the wavelength direction.
 
\begin{figure}[ht!]
        \centering
        \includegraphics[width=\linewidth]{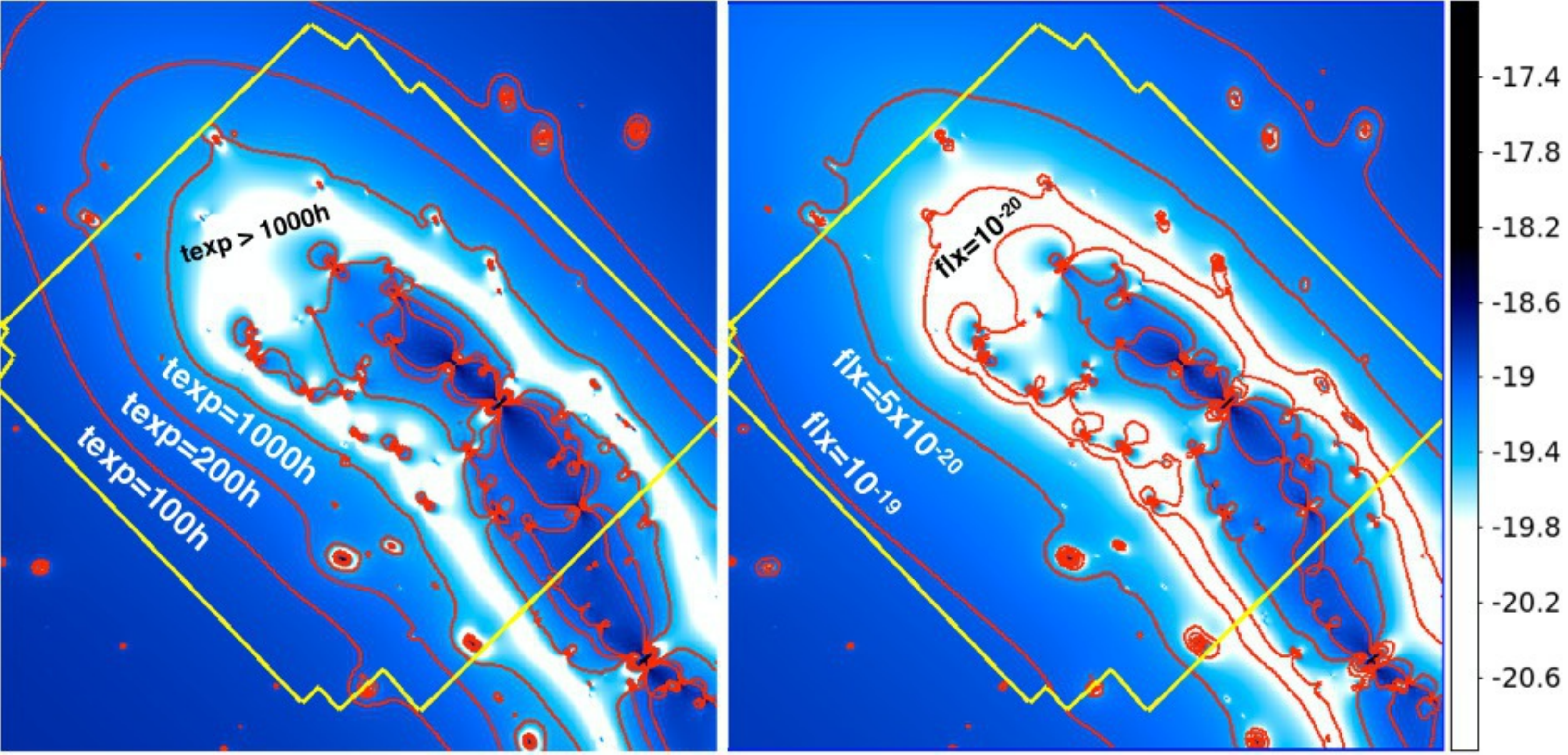}
        \caption{{\it Left:} Equivalent exposure time needed to reach the same S/N probed with the MDLF without lensing, assuming point-like sources at $z=6$.  Contours of iso-exposure are shown (100, 200, 1000 hours equivalent integration). We note that the MDLF is equivalent to 100 h integration in blank fields already with modest magnification, $\mu \sim 2.5$. {\it Right:} Map of the 3-sigma line flux limit adopting a velocity width of 300~\kms\ and a circular aperture of $0\farcs6$ diameter, based on the same assumptions as in the left panel. The color-bar indicates the log of flux values in~\ergscm, whereas the three contours correspond to the flux (flx) of $10^{-19}$, $5 \times 10^{-20}$, and $10^{-20}$~\ergscm.}
        \label{equivalent}
\end{figure}

\begin{figure}[ht!]
        \centering
        \includegraphics[width=\linewidth]{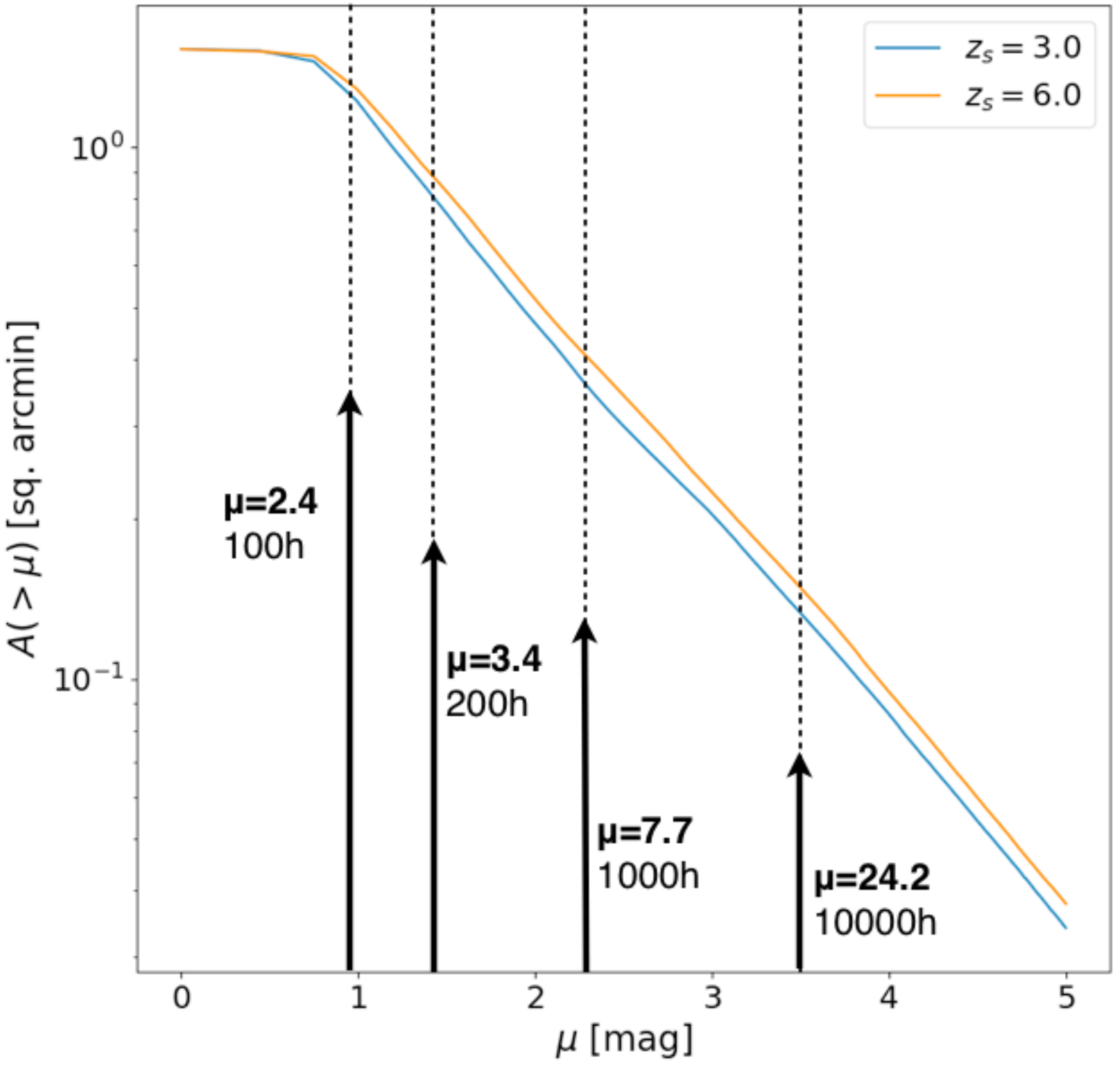}
        \caption{Cumulative surface area of the MDLF in the image  plane as a function of magnification in magnitude units ($\mu[{\rm mag}] = 2.5\log_{10}(\mu$)) for redshift 6 (orange line) and 3 (blue line). Arrows mark the magnifications corresponding to the exposure time (=100, 200, 1000, 10000 hours) as shown in Figure~\ref{equivalent}, necessary for obtaining the same depth of the MDLF without lensing.}
        \label{surface}
\end{figure}

The magnification across the field provided by the gravitational lensing effect further decreases the detectable line flux limit in the MDLF when compared to the Hubble Ultra Deep Field \citep{Bacon17}.
Assuming a point-like emitting source and that, at first order, the magnification is the ratio between the observed flux and the de-lensed(intrinsic) one, $\mu = F_{obs} / F_{intr}$, the equivalent integration time (texp) in absence of lensing required to obtain the same S/N achievable in lensed fields is obtained by rescaling the MDLF integration to $\mu^2$.  T(MDLF) $\times~\mu^2$, where T(MDLF)=17.1h.
Figure~\ref{equivalent} shows the texp map needed to obtain the same depth of MDLF without lensing. It is widely known that strong lensing boosts the detection of faint sources and represents a complementary approach to observations in blank fields, however, deep observations like the MDLF allow us to reach equivalent texp~$\simeq 100$~h even in regions where magnification is modest, $\mu \sim 2-3$. The 90\% of the MDLF field of view is equivalent, in terms of depth, to $>100$~h of integration in non-lensed fields (with the most magnified regions pushing texp up to 1000~h where $\mu > 7.7$). 
The same figure also shows the equivalent 3-sigma line flux limit after rescaling to texp. Line fluxes down to a few $10^{-20}$ \ergscm\ can be probed in regions with large magnification (texp $> 200$ hours). Such a depth allows us to detect high-ionization lines on individual objects with intrinsic magnitudes $27-30$ (see Sect.~\ref{stack.section}).
The outer regions of the galaxy cluster at relatively low-$\mu$ have the advantage to be relatively free from contamination by galaxy cluster members and are less affected by large uncertainties on the magnification, being far from the critical lines. 
The major drawback of strongly lensed fields is the smaller intrinsic area probed behind the lens when compared to the non-lensed fields. An illustration of this effect is shown in Figure~\ref{surface}, which shows the cumulative surface area on the lens plane probed by the MDLF as a function of magnification (in magnitude units). The surface area decreases rapidly with $\mu$ reaching half of its original coverage when $\mu \sim 6.3$.

\subsection{The MUSE pointing in the South-West: MUSE-SW}
\label{SW}
Relatively deep observations in the SW region of the same galaxy cluster (J0416) were carried out under the  ID 094.A-0525(A) program (PI:  F.E.  Bauer).  This includes  58  exposures  of approximately 11 minutes each, executed over the period October 2014 $-$ February 2015. We use the same reduced data-cube described in \citet{Caminha_macs0416}. Despite a  relatively long exposure in the SW pointing (formally 11h integration),  the S/N of the spectra does not scale according to expectations, resulting to an equivalent integration of $\sim 4$ hr only. \citet{Caminha_macs0416} attribute this inconsistent depth to the significantly worse seeing of the SW pointing ($1''$ vs. typically $0\farcs6$) and the large number of short exposures used which, due to residual systematics in the background subtraction, did not yield the expected depth in the coadded data-cube.
As discussed in \citet{bergamini20}, the depth provided by the MDLF produces a major gain in the number of bona fide multiple images, as discussed in the next section.

\section{Catalog of multiple images}
\label{multiple}

In combining MUSE-SW and the initial two hours of integration in the north-east from GTO observations, \citet{Caminha_macs0416} identified 37 galaxies producing 102 multiple images in the redshift range of $1<z<6.2$. The MDLF and a careful identification of confirmed additional lensed families led to an unprecedented set of 182 multiple images, spanning the same redshift range of $0.9<z<6.2$. To search for new sources, we followed the procedure described in \citet{Caminha_macs0416}. Firstly, using our lens model we looked in the vicinity of the predicted positions of multiple images of families partially lacking spectroscopic information.
This led us to complete the spectroscopic information of several lensed systems.
Secondly, new sources have been identified by exploring narrow-band continuum subtracted cubes and analyzing spectra extracted at the position of candidate multiply imaged objects. This process was based on (a) visual inspection of color images, (b) the assistance by the lens model which was progressively refined \citep[][]{bergamini20}, (c) the ASTRODEEP photometric redshift catalog \citep{castellano16, merlin16}. Different versions of continuum-subtracted cubes were generated varying the width of the central window within which slices are collapsed (with typical dv=300-500 \kms) and the redward and blueward regions used to estimate the continuum level, typically with widths of $20-30$~\AA.

\begin{figure}
        \centering
        \includegraphics[width=\linewidth]{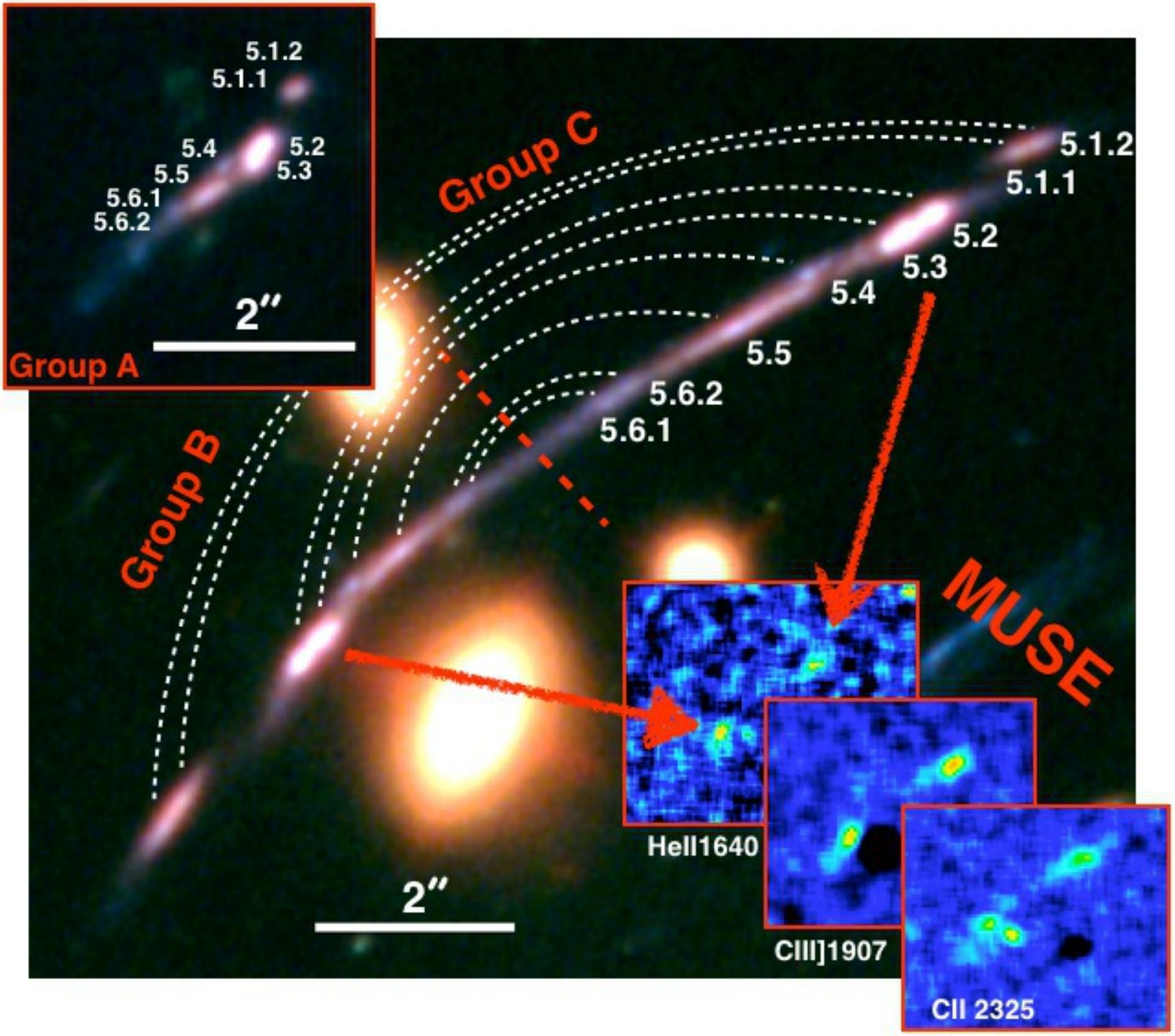}
        \caption{Eight clumps identified for source 5 are marked with IDs 5.1.2, 5.1.1, 5.2, 5.3, 5.4, 5.5, 5.6.2, 5.6.1 and indicated on both sides of the critical line (red dashed line), labeled as group C (5c) and B (5b). The top-left inset shows on the same scale the least magnified image 5a, in which all the corresponding clumps can be recognized. The bottom-right insets show the continuum-subtracted narrow-band images extracted from MUSE around the wavelength of the most prominent high ionization emission lines.}
        \label{sys5}
\end{figure}

The MDLF observations allowed us to increase significantly the number of multiple images in the NE region of the cluster and triple the S/N of the previous 2h exposure data-cube from GTO. 
Individual sources contain in some cases multiply imaged clumps (see, e.g., Figure~\ref{sys5} and Sect.~\ref{clumps}), in which more than one family can be part of the same high-z galaxy.\footnote{A family is defined as a set of multiple images of the same background object. An object can be either a single galaxy or a single sub-component of the same galaxy (e.g., a clump). For example, source 5 has 6 families with three multiple images each (a,b,c): 5.1(abc), 5.2(abc), 5.3(abc), 5.4(abc), 5.5(abc), and 5.6(abc), which have been used to constrain the lens model.
In some cases, HST single-band imaging reveals that individual families are further split into two knots, which are labeled with an extra digit, e.g., 5.1.1 and 5.1.2 or 5.6.1 and 5.6.2. These extra knots are not used as constraints in the lens model. 
Source 1, made of multiple images 1a, 1b, and 1c corresponds to only one family.} In particular, the total number of families increases to 66, with the new ones covering the redshift range of $0.9<z<6.2$. As a result, the number of individual high redshift galaxies generating the set of multiple images is lower than 66, amounting to 48 independent sources. The arclet at $z=6.629$ has not been included among the constraints of the model because no clear HST counterparts have been currently identified \citep[see][]{vanz_popiii}.

\begin{figure}
        \centering
        \includegraphics[width=\linewidth]{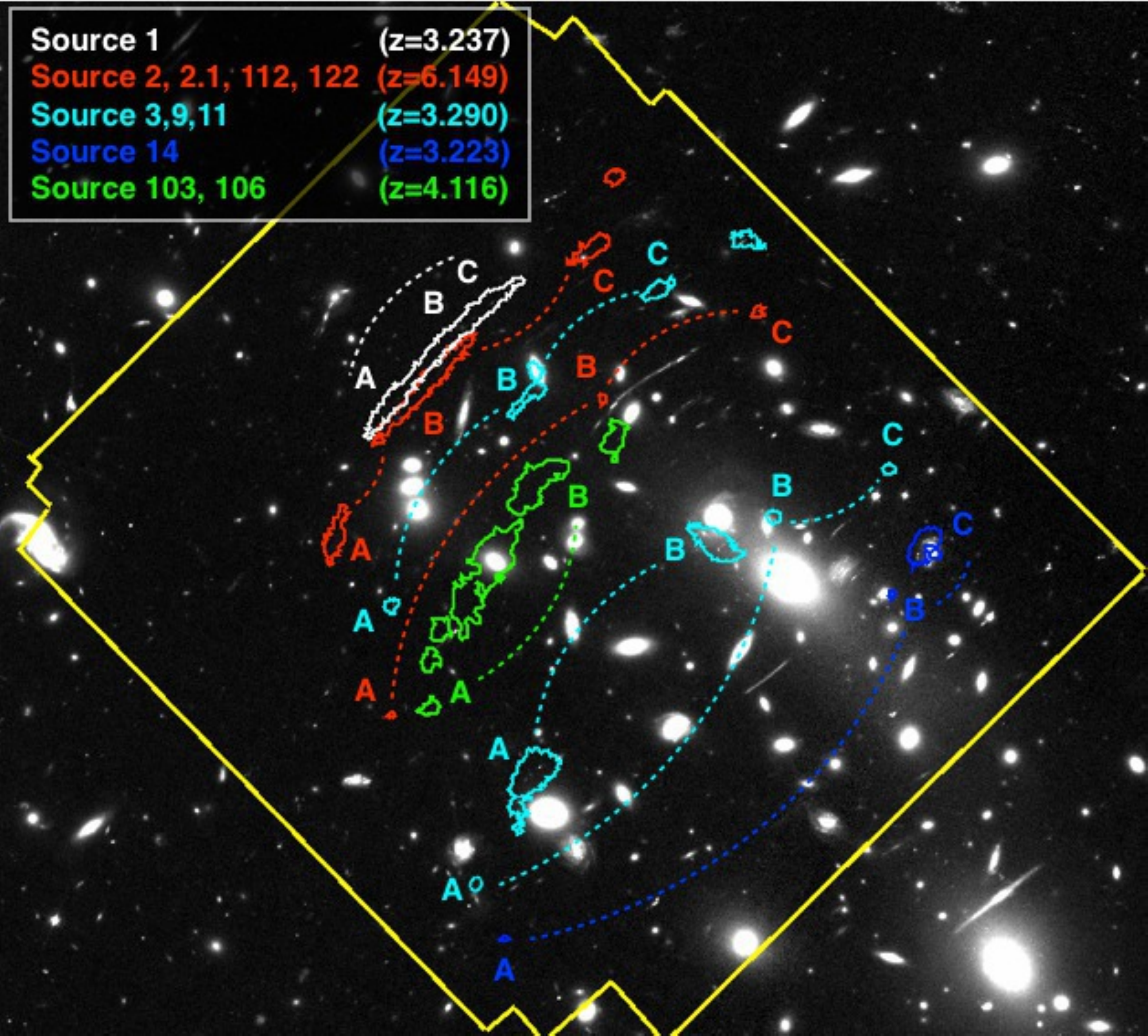}
        \caption{Examples of the most prominent multiply imaged \lya\ emitting regions spatially resolved in the MDLF. Multiple images are indicated as $A$, $B$ and $C$ (connected with dotted lines) with color-coded contours at the 2$\sigma$ level calculated on the continuum-subtracted narrow-band MUSE images, centered on \lya\ emission. In the case of source 14, the \civalone~$\lambda1548$ emission is shown instead of \lya\ (being deficient in \lya\ emission (see also Figure~\ref{fig_id14} and discussion in \citealt{vanz_id14}). For more details about sources 103 and 106, see Figure~\ref{sys106}.}
        \label{nebulae}
\end{figure}

As discussed in Sect.~\ref{clumps}, a close inspection of the confirmed multiple images in deep HST data reveals a significant fraction of multiply imaged clumps emerging from each high-z galaxy. Those that are firmly identified are included as constraints in the lens model. The number of clumps typically increases where magnification increases, eventually making them individually recognizable (enhanced spatial scale) and detectable (enhanced S/N). The inclusion of multiple clumps is particularly useful for better constraining the position of critical lines and the high magnification values in these regions. Such examples are families 5 at $z=1.8961$ (see Figure~\ref{sys5}) and source 12 at $z=0.9390$, in which the large magnification close to the critical lines is better sampled by a high spatial density of local constraints corresponding to star-forming clumps (see \citealt{bergamini20} for more details).

At the end of this process, the spectroscopic confirmation of new high-z galaxies and the addition of individual clumps increase the total number of multiple images used for the lens model to 182 (66 families),  spanning the redshift range of $0.9<z<6.2$.  
\citet{bergamini20} present the details of the lens model, which is currently the one exploiting the largest number of spectroscopically confirmed constraints for any galaxy cluster. It includes 80 additional multiple images compared to the previous model and 213 confirmed galaxy cluster members (20 more than in the previous model), by reproducing the positions of all 182 multiple images with an rms accuracy of only $0\farcs40$.
In Appendix~\ref{multiple_appendix}, we present the details of all multiple images, showing, for each of them, the HST cutouts and MUSE narrow band continuum-subtracted imaging at the wavelength position of the most relevant emission lines. 

The \lya\ emission is often spatially resolved and extends beyond the HST counterpart down to the very faint fluxes permitted by lensing magnification. A dedicated analysis of the spatially extended \lya\ emission and intrinsic spatially-varying profiles (e.g., relative intensities of the blue and red peaks) will be presented in a future work. An example showing the most prominent cases in the MDLF is reported in Figure~\ref{nebulae}, where multiple images of \lya\ nebulae at $z=3-6$ extend along the tangential direction and possibly include even fainter clustered sources (currently not detected on HST images, e.g., \citealt{ribas16}) contributing to the \lya\ emission. One of them, source 9, shows a spatially-varying multi-peak \lya\ emission and nebular high-ionization lines emerging from three well-recognized knots
(this system was already presented in \citealt{vanz_id9} using  much shallower MUSE observations).
 
 \begin{figure*}
        \centering
        \includegraphics[width=\linewidth]{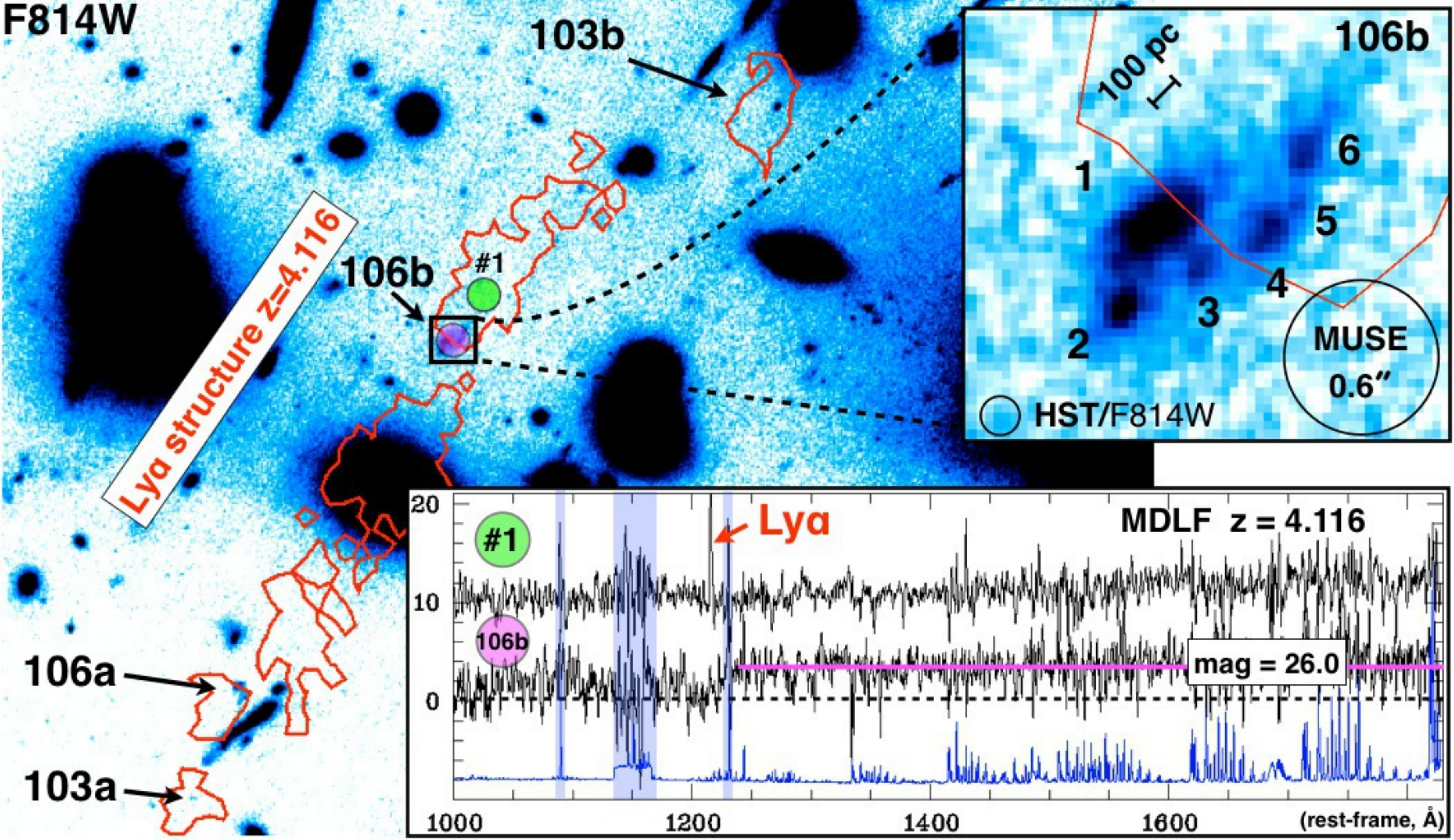}
        \caption{\lya\ nebulae (red contours at 2-$\sigma$ level) including images 106(a,b) and 103(a,b), belonging to the same physical structure at z=4.116, are shown in the main panel (HST ACS/F814W band). In the top-right inset, the lensed galaxy 106b broken in six clumps is shown; the smaller ones ($2-6$) have intrinsic UV magnitude $\simeq 30.5-31$ (corresponding to M$_{\rm UV} \simeq -15$) and intrinsic sizes of the order of (or smaller than) 100 pc along the tangential stretch, as indicated by the ruler. The bottom inset shows two one-dimensional spectra extracted from the MDLF. One is on galaxy 106b (within a circular aperture of $1.2''$ diameter, see magenta circle). We note the ultraviolet continuum of 106b with F814W = $26.01 \pm 0.03$ is well-detected (magenta line) above the zero level (black dashed line), and its continuum-break confirmed at z=4.116. The top spectrum is extracted from a nearby region showing \lya\ emission without HST counterpart (green circle with $1.2''$ diameter, labeled as \#1), vertically shifted at position 10 for illustrative purposes. The error spectrum is shown in blue at the bottom to highlight the location of the sky emission lines. The shaded band marks the wavelength region affected by the GLAO sodium-based laser. }
        \label{sys106}
\end{figure*}
 
The depth of the MDLF also allows us to confirm sources without \lya\ emission, down to magnitude $\simeq 26$. One example is source 106b at $z=4.116$ (see Figure~\ref{sys106}), for which the continuum-break is clearly detected, its redshift is measured by cross-correlating the spectrum with high-z templates\footnote{Redshifts have been measured using the {\tt Eazy} package within the  Pandora environment \citep[][]{garilli10}.}, and found consistent with the spatially offset \lya\ nebula. Interestingly, the spectroscopic redshift is also in very good agreement with the photometric redshift derived from ASTRODEEP, $z_{phot}=4.20$ \citep[][]{castellano16}. As discussed in the Sect.~\ref{clumps}, source 106b is also a good example of how a galaxy can be resolved into several sub-components by strong lensing (at least six star-forming regions of $\sim 100-200$ parsec size).

\subsection{The full MUSE spectroscopic catalog}

In addition to the set of multiple images specifically used by \citet{bergamini20} to constrain the lens model, we also released a version of the MUSE spectroscopic catalog that includes all the sources we identified in the MUSE datacubes. By combining the MDLF and MUSE-SW pointing, this catalog contains 424 individual objects, spanning the redshift interval up to z=6.7, thus extending the sample of 182 multiple images (48 objects). Faint sources with observed magnitude down to $m_{1500}> 28$ have been confirmed, corresponding to intrinsic $m_{1500}>29-30$ in the case of $\mu \simeq 3-6$. The typical error at this magnification regime is less than 20\%, implying that the error on the intrinsic magnitude is mainly dominated by the photometric uncertainty (for the given cosmology). 

\begin{figure*}
        \centering
        \includegraphics[width=\linewidth]{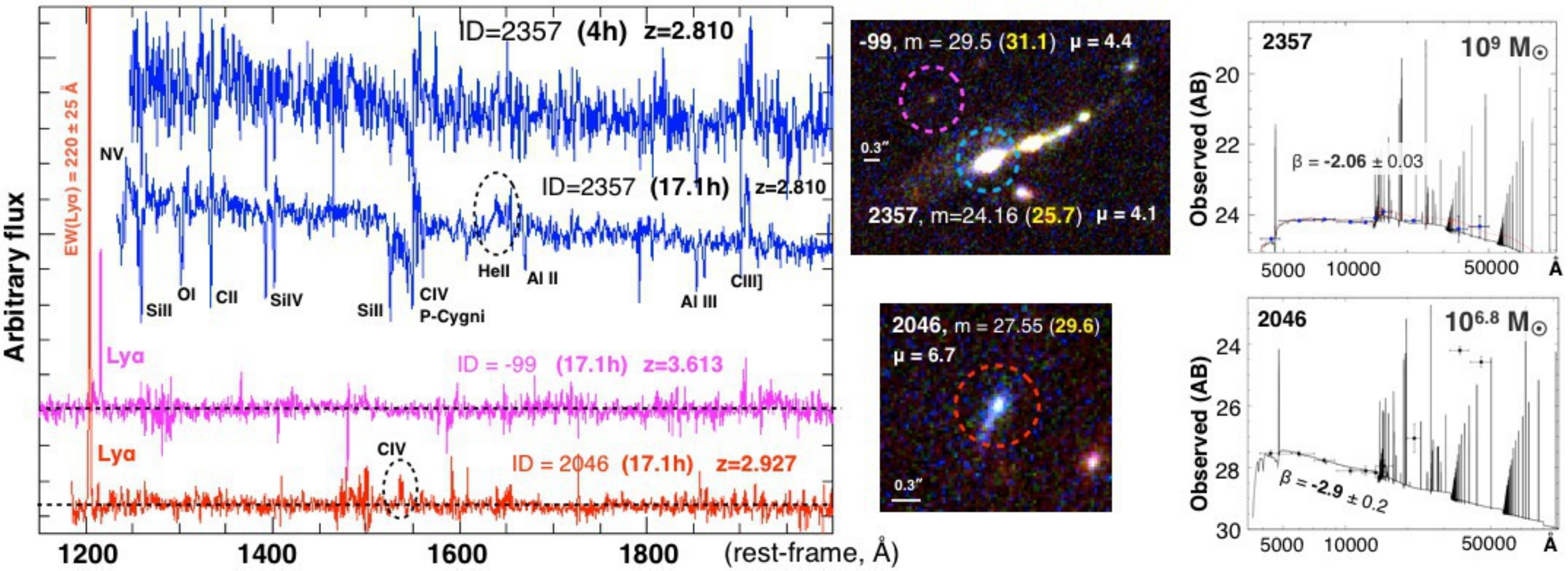}
        \caption{Examples of MUSE spectra of moderately magnified sources not producing multiple images. In the left panel the one dimensional spectra are color coded accordingly to the dashed circles marking the sources in the HST color images (middle panels). The spectrum of source 2357 from four hours of integration without AO is compared with the one from the MDLF full depth (17.1 hours) including AO (the P-Cygni profile of the \civalone\ is affected by the increased noise due to laser AO correction). The deep spectrum shows many features in absorption, whereas in emission, the \ciiidoub\ and broad \heii\ (marked with a dashed ellipse) are clearly detected, as well as a possible P-Cygni of the \nvdoub, close to the blue edge of the spectral coverage. Two other faint objects, 2046 and -99, are shown with magenta and red colors. The observed(intrinsic) magnitude is shown in the middle panels, with the intrinsic one reported with a yellow color. Source 2046, with $m_{1500} = 31.1 \pm 0.3$, shows a large \lya\ equivalent width of $220\pm25$~\AA~rest-frame and a very steep ultraviolet slope, $\beta = -2.9 \pm 0.2$. Nebular \civ\ doublet is also detected for this object (see dashed ellipse on the red spectrum). The rightmost panels show the SED fits and magnitudes as performed by \citet{castellano16}. The red spectrum is blueshifted by 12~\AA\ with respect to the other spectra for illustrative purposes.}
        \label{field}
\end{figure*}

Figure~\ref{field} shows three examples of the aforementioned cases. In particular, the spectra of two of these confirmed very faint sources at $z=3.613$ (ID = -99) and $z=2.927$ (ID = 2046) (in magenta and red colors respectively) have intrinsic magnitudes $m_{1500}$ = 31.1 and 29.6, with an error of 0.3 mag (including the magnification uncertainty). ID=2046 also shows an extremely blue ultraviolet slope with a relatively small error, as estimated from the HST F606W, F814W and F105W photometric bands, $\beta = -2.9 \pm 0.2$ (F$_\lambda \simeq \lambda^{-\beta}$, \citealt{castellano12}). Interestingly, the same object also shows a very large equivalent width of the \lya\ ($220 \pm 25$~\AA) and the presence of faint nebular \civ\ doublet, associated to an object with an estimated stellar mass of a few million solar masses ($10^{6.8}$ \msun).

Another example in Figure~\ref{field} shows that deep MUSE observations of (intrinsically) relatively bright, moderately magnified galaxies ($\mu \simeq 3-6$) reveal or consolidate spectral features clearly associated with the presence of massive stars. Source 2357 is the brightest clump of a complex system at $z=2.810$ showing various knots. Its observed magnitude of 24.16 (25.7 intrinsic) makes it relatively bright, however, the increased S/N provided by the MDLF reveals multiple spectral features (if compared to the initial four-hour integration), such as the broad \heii\ and \civalone\ P-Cygni profile, both indicating the presence of strong stellar winds arising from massive O and WR stars, with a possibly further signature of P-Cygni of \nv\ indicative of ages younger than 5 Myr \cite[e.g.,][]{senchyna_massive_stars,vanz_sunburst,chisholm19}.
Appendix~\ref{multiple_appendix} presents the spectroscopic catalog of all high-z objects, including some that are not multiple images.

%
\begin{table}
\caption{Summary of the MDLF observations}             
\label{tab1}      
\centering          
\begin{tabular}{l c l}     
\hline\hline       
Date & Quality & OB Name \\ 
\hline
 & MDLF & \\
\hline
    22/23-Nov-2017 & A & WFM\_J0416\_NOAO\_1\\  
    10/11-Jan-2018 & A & WFM\_J0416\_NOAO\_2\\
    21/22-Feb-2018 & C & WFM\_J0416\_NOAO\_3\\
    12/13-Mar-2018 & X & WFM\_J0416\_AO\_1 \\
    4/5-Nov-2018   & A & WFM\_J0416\_AO\_10\\
    5/6-Nov-2018   & B & WFM\_J0416\_AO\_1 \\
    5/6-Nov-2018   & A & WFM\_J0416\_AO\_2 \\
    6/7-Nov-2018   & A & WFM\_J0416\_AO\_4 \\
    2/3-Dec-2018   & A & WFM\_J0416\_AO\_11 \\
    4/5-Dec-2018   & A & WFM\_J0416\_AO\_13 \\
    12/13-Dec-2018 & A & WFM\_J0416\_AO\_14 \\
    11/12-Jan-2019 & B & WFM\_J0416\_AO\_5\\
    16/17-Jan-2019 & C & WFM\_J0416\_AO\_6\\
    25/26-Jan-2019 & A & WFM\_J0416\_AO\_6\\
    27/28-Feb-2019 & A & WFM\_J0416\_AO\_17\\
    28-Feb/1-Mar-2019 &A & WFM\_J0416\_AO\_7 \\
    3/4-Mar-2019 & A & WFM\_J0416\_AO\_8\\
    2/3-Aug-2019 & A & WFM\_J0416\_AO\_9\\
    30/31-Aug-2019 & A & WFM\_J0416\_AO\_18\\
\hline     
       & GTO & \\
\hline
    17-Dec-2014 & A & WFM\_J0416\_NOAO\\
    17-Dec-2014 & A & WFM\_J0416\_NOAO\\
\hline
\hline
\end{tabular}
\tablefoot{The log of the observed OBs is reported. The typical exposure time (on sky) of each OB is 3340s. The bottom two rows refer to the previous 2 hours observation from the GTO. The column ``OB Name''  indicates the observing mode ("WFM" = wide field mode) and the use (or not) of the adaptive optics "AO/NOAO," meaning "on/off".}
\end{table}

\section{Clumpy high-z galaxies}
\label{clumps}

A common morphological property of high redshift star-forming galaxies is the presence of clumps \citep[][]{zanella15,zanella19}, that seem to emerge whenever the angular resolution increases. Strong gravitational lensing reveals such clumps down to a $\sim 100$ pc scale \citep[][]{livermore15, rigby17, cava18} that further continue fragmenting down, approaching the sizes of massive stellar clusters ($ \lesssim 30$ pc) in high magnification regimes, $\mu > 10$ \citep{vanz19, vanz_sunburst, johnson17}.

\begin{figure*}[ht!]
        \centering
        \includegraphics[width=\linewidth]{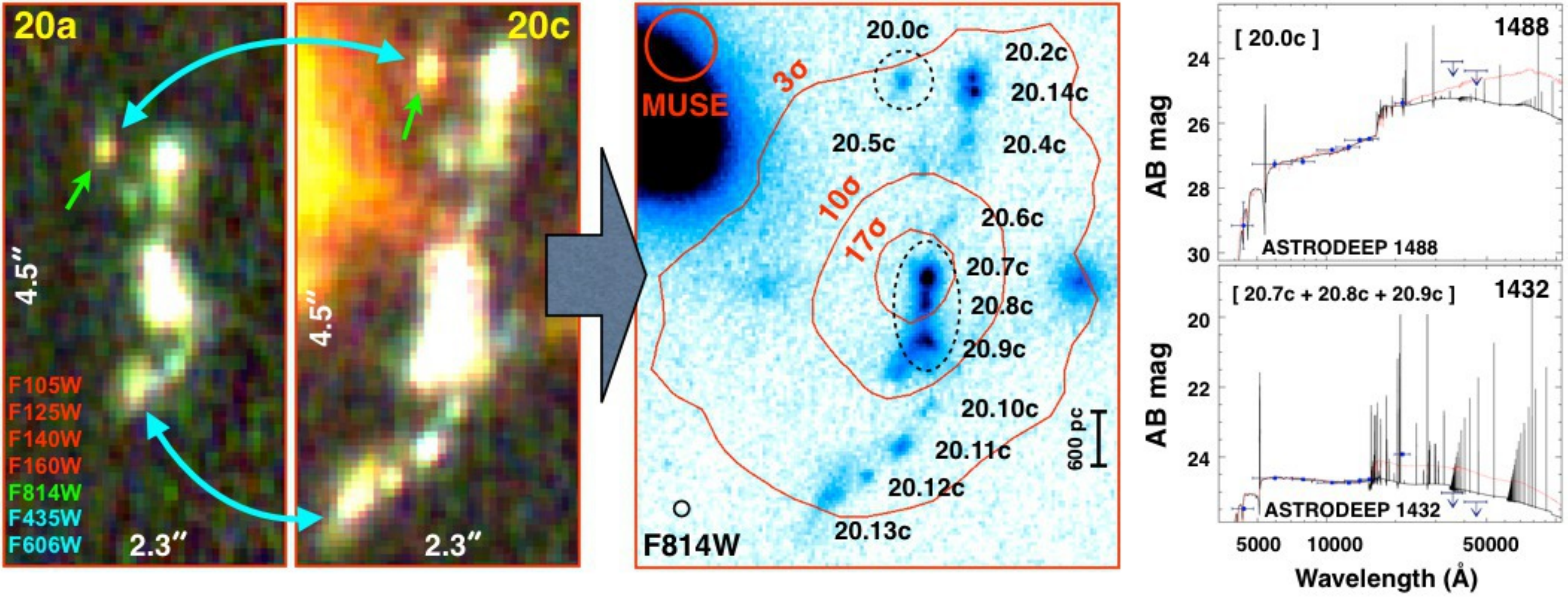}
        \caption{Example of a magnified $z=3.222$ galaxy (source 20). Deep color RGB images of components 20a and 20c are shown in the two leftmost panels; the knot showing a redder color with respect to the rest of the galaxy is marked with a green arrow. Cyan arrows mark the two extremes of the structure, indicating the corresponding physical regions. In the middle panel, the F814W blue color-code HST image details the most magnified component 20c, in which at least 13 clumps are identified, across a region of 8 physical kpc on the source plane. The contours are drawn from the MUSE \lya\ emission. The MUSE and ACS/F184W PSF sizes are indicated with a red (top-left) and black (bottom-left) circles. The SED-fits (from the ASTRODEEP photometric catalog, \citealt{castellano16}) of two regions marked with dashed black ellipses are shown in the rightmost panels. The mirrored symmetry between images 20a and 20c confirms that all clumps belong to the galaxy, including knot 20.0 showing a clearly different color (see SED fits). The physical scale is reported on image F814W, bottom-right (1 HST pixel corresponds roughly to 60 pc along the vertical extension of the galaxy in the source plane).}
        \label{system20}
\end{figure*}

The identification of star-forming clumps in the secure multiple images discussed here has been visually performed by looking at HST/ACS and WFC3 images and their RGB color version, in addition to taking into account the mirroring and parity properties introduced by strong lensing (see Appendix~\ref{clumps_appendix}). The latter reinforces the identification of extremely faint clumps (e.g., observed magnitude $>29-30$) that are otherwise elusive even for deep spectroscopy; this represents a unique advantage provided by lensing. Figure~\ref{system20} shows an example where at least 13 clumps associated to source 20 at $z=3.222$ are identified, including very faint or isolated knots which display in some cases different colors (see also source 5, Figure~\ref{sys5}). Other examples are shown in Appendix~\ref{clumps_appendix}. 

All high-$z$ multiple images have been visually inspected and the consistency with their parity properly checked. Among the 66 families spanning the redshift range of $1<z<6.2$, we identify structured clumps in the majority of the high-z galaxies (more than 60\%).
Appendix~\ref{clumps_appendix} describes the sample of clumps, reporting for each of them the HST cutouts and MUSE spectra.
Despite lensing magnification acting to magnify (and distort) galaxies, the identification of clumps is typically not performed by automatic tools of source extraction \citep[e.g., SExtractor package,][]{1996A&AS..117..393B} since a delicate trade-off between de-blending and detection threshold segmentation is needed. Indeed, the majority of the clumps discussed here are not present in the ASTRODEEP \citep{castellano16} or HFF Deep Space \citep{deepspace18} catalogs of HFF~J0416. Moreover, the presence of bright cluster galaxies in the field makes faint object detection and photometry (contamination) difficult. In order to characterize their magnitude distribution and homogenize measurements, we made use of the the APHOT tool \citep{merlin19} and we performed photometric measurements on each of them over the same images used to build the ASTRODEEP color catalog.
To estimate their magnitudes, we adopted 2~$\times$~FWHM diameter apertures and measure their local background through a sigma clipping procedure in annuli of 10 pixel radius, at 1.2 times the Kron radii around each source (see \citealt{merlin19} for details). The figures in Appendix~\ref{clumps_appendix} reveal that such clumps have rather compact sizes, several of them are marginally resolved or entirely unresolved and slightly elongated. The inferred magnitude can therefore be somewhat affected, however, we did not apply any correction in this work since our scope is focused on the characterization of the new parameter space opened by these observations -- specifically, the size and luminosity at the faint end.

\begin{figure}[ht!]
        \centering
        \includegraphics[width=\linewidth]{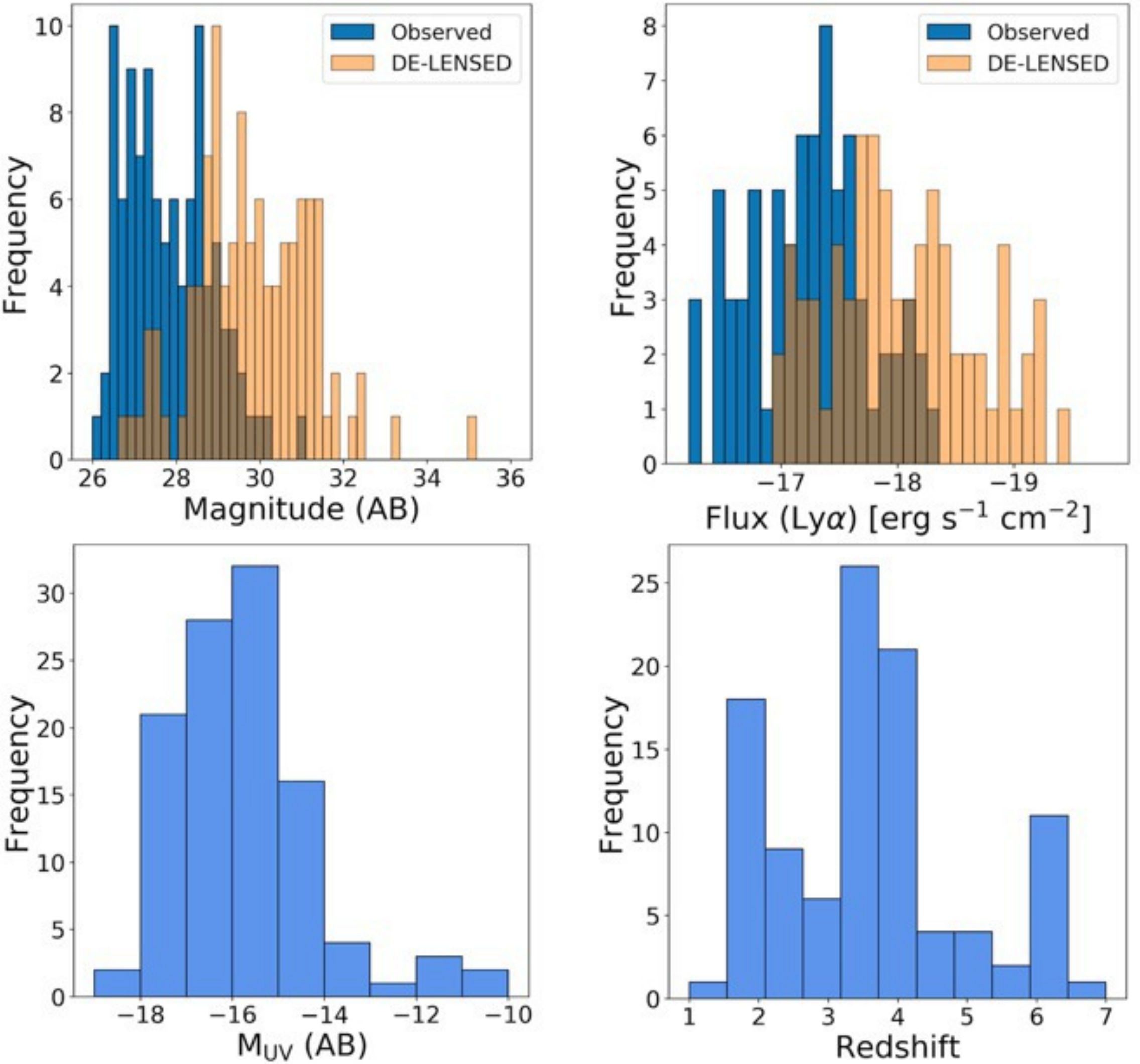}
        \caption{Observed (de-lensed) 1500~\AA~magnitude distributions (top-left) and observed (de-lensed) \lya\ flux distribution (top-right) for the sample of clumps (individual sources).  The bottom panels show the absolute magnitude (left) and redshift (right) distributions. No multiple images are included. The median absolute magnitude of the sample is $-16$ corresponding to an AB magnitude $\sim 30$.}
        \label{distr}
\end{figure}

Figure~\ref{distr} shows the observed/intrinsic magnitude distribution of all clumps at $z<4.8$ ($z>4.8$) extracted from the HFF HST/F814W (F105W) band, as well as the observed or intrinsic \lya\ fluxes. The absolute magnitude spans the range of [$-18$, $-10$] with a median of $-16$, over a redshift range of [$1-6.7$], with a median of $z=3.5$. 
The distribution of the \lya\ fluxes is shown in the same figure. Fluxes were extracted from a fixed aperture of $0\farcs8$ diameter; we did not attempt to tune apertures to capture the different morphology of the emitting regions, which are also shaped by lensing distortion. Unfortunately, the MUSE PSF (FWHM=$0\farcs6$) prevents us from extracting spectra for the majority of the clumps, which are blended because of the lower angular resolution with respect to HST. With this caveat in mind, it is worth stressing that for compact \lya\ emitters the measured fluxes extend down to a few $10^{-19}$ \ergscm, with the faintest tail approaching $10^{-20}$ \ergscm,
as in the case discussed by \citet{vanz_popiii} at $z=6.629$ straddling the caustic, implying extremely faint and small sizes of the emitting regions.
High-ionization emission lines (typically emerging from much smaller regions than those producing scattered \lya, and typically aligned with the HST stellar continuum) are also captured at the faintest luminosities in single sources
and with high S/N ratios on the stacked spectrum, as described in Sect.~\ref{stack.section}. 

\begin{figure*}[ht!]
        \centering
        \includegraphics[width=\linewidth]{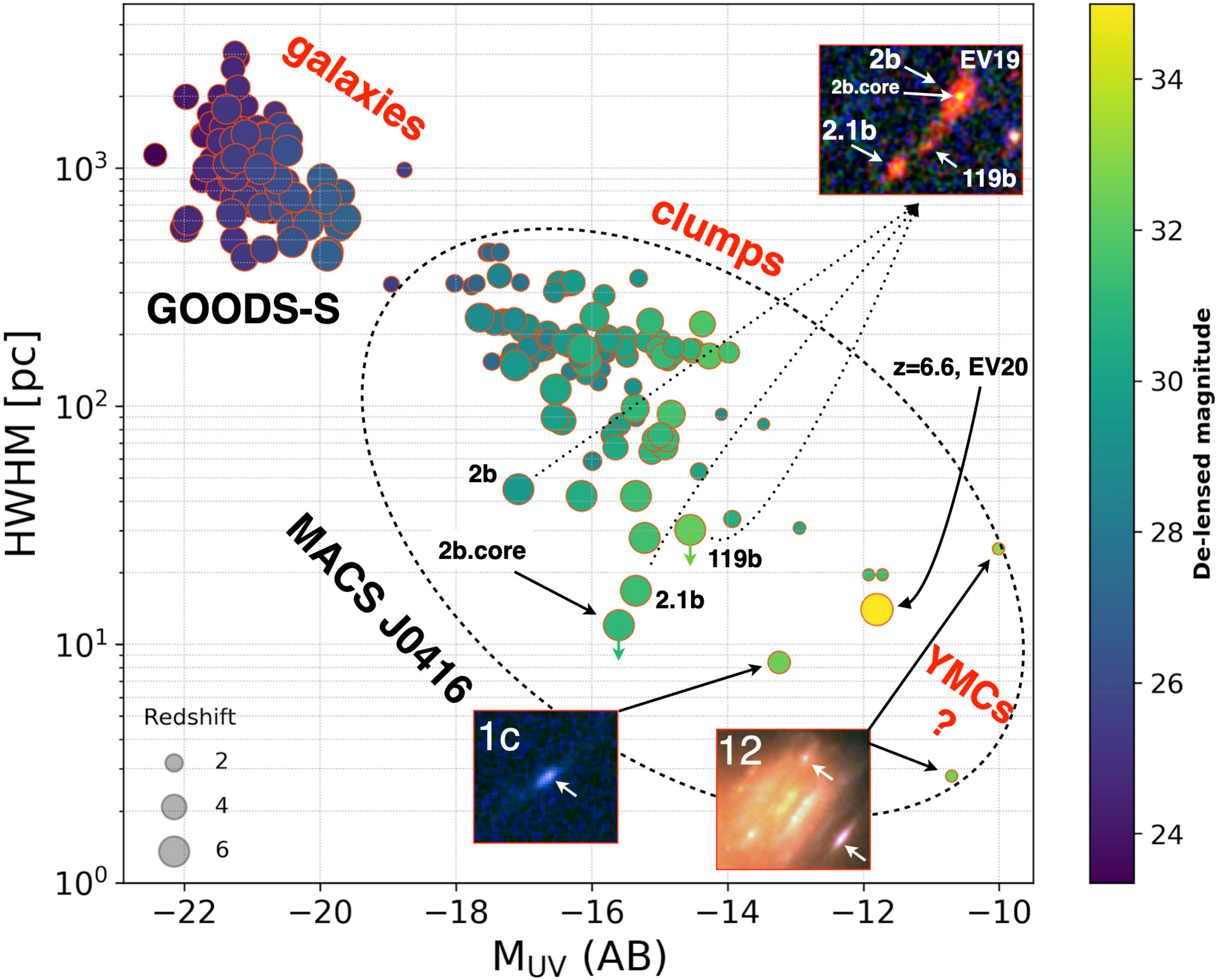}
        \caption{Intrinsic size (HWHM) as a function of absolute magnitude for all the clumps or single compact (isolated) sources identified in the MDLF. The points are color-coded according to their intrinsic magnitude (color bar on the right) and size-coded with the redshift value. The cloud of points in the top-left of the diagram represents non-lensed objects belonging to the GOODS-South field (see text). The insets show the color images of three examples with large magnification and possibly hosting extremely small objects,  compatible with being single star clusters, such as the source dubbed ``D1core'' \citep[EV19,][]{vanz19} or the star complex possibly hosting Pop~III stellar population \citep[EV20,][]{vanz_popiii}. Images 1c and 12 are discussed in Sect~\ref{grav}. From top-left to the bottom-right of the diagram, luminosity from $-22$ to $-10$ and sizes from kpc to a few parsec scale embrace galaxies, clumps, and star clusters. The error on M$_{\rm UV}$ depends on the magnification and photometric uncertainties, and is typically $\lesssim 0.5$ magnitude for the lensed sources. The typical error on the HWHM also depends on magnification and, conservatively, it is within the 50\% of the reported value. }
        \label{scatter_clumps}
\end{figure*}

An accurate estimate of the size of each object (e.g., the effective radius) will be part of a future work. Here, we perform a first analysis by computing the physical size that the HST PSF would have if it had been placed at the same locations, using the magnification maps from out new lens model. Clearly this is a simple assumption and would overestimate the effective radius for compact point-like objects (for which the PSF deconvolution would lead to radii even smaller than the pure PSF, comparable to a single HST pixel, e.g., \citealt{vanz19}). Conversely, it would underestimate the size in the case of extended objects (see how clumps appear in Appendix~\ref{clumps_appendix} and figures therein). Figure~\ref{scatter_clumps} shows the half width at half maximum (HWHM) as a function of the intrinsic absolute magnitude, redshift, and intrinsic magnitude for the whole sample.

Several clumps appear as faint as (or fainter than) those reported by \citet{maseda18} (or \citealt{feltre20}) from the MUSE deep observations performed in the Hubble Ultra Deep Field, where magnitudes $30-31$ are probed with S/N~$ <5$. In the present case, and not surprisingly, strong lensing allows us to probe physical scales out of reach in blank fields (e.g., $<100$ pc) and access comparable flux limits with high S/N or even faint sources that have been totally missed in non-lensed fields (e.g., intrinsic magnitude fainter than 31).
We note, in fact, that several such tiny star-forming regions (e.g., M$_{\rm UV} > -16$) are well-detected with $S/N > 10 $, thus allowing a morphological and SED-fitting analysis even on single sources.
As an example, a point-like object with an intrinsic magnitude of 29.5 will move down to a magnitude of 26.5 with a magnification of $\mu = 10$; such a magnitude is typically measured with S/N~$> 20$ at the HFF depth. Similarly, MDLF$-$like observations will probe emission lines at unprecedented faint flux levels (see Sect.~\ref{stack.section}). 

In order to highlight the gain provided by strong lensing, Figure~\ref{scatter_clumps} also includes a sample of galaxies extracted from non-lensed fields at $3.5<z<6.5$ (from the GOODS-South, \citealt{vanz09,giava04}). High-z galaxies studied in non-lensed fields have typical sizes of kpc (or sub-kpc) scale and magnitudes typically brighter than M$_{\rm UV} = -18$. In lensed fields and in this work, the same class of sources can be decomposed into clumps of 100-200 pc size at typical magnitude M$_{\rm UV} = -16$ as the angular resolution increases.
These clumps includes the most extreme cases for which single star clusters can be probed, down to M$_{\rm UV} = -15$ with sizes smaller than 50 parsec \citep[e.g.,][]{zick20,bouwens17, kawamata15}, including globular cluster precursors \citep[][]{vanz_paving,vanz19}.
Concerning the very faint-end of the magnitude-size distribution, sources that are barely detected $-$ even when assisted by lensing magnification $-$ correspond to intrinsic magnitudes in the range of $33-35$, with extreme cases even fainter than 35 \citep{vanz_popiii}. It is worth stressing that unresolved objects (smaller than $40-50$ pc) showing prominent \lya\ emission at $z=3.5$ and suggesting a high ionization field provided by young stellar populations, with magnitudes fainter than $m_{1500}=31$ (${\rm M_{UV}}=-14.8$), correspond to stellar masses of $\lesssim 10^{6}$\msun\ in the instantaneous burst assumption and are weakly dependent on metallicity or IMF \citep[][]{leitherer14}. Irrespective of the nature of such objects,
they are more likely to belong to the realm of star forming complexes or even massive star clusters with ${\rm M_{UV}}=-13$ or fainter \citep[][]{atek15,alavi14,alavi16,atek18,bouwens17,livermore17}. Thus, the current demography of the faint-end of the ultraviolet luminosity functions of ``high-z galaxies'' may be contaminated or even perhaps dominated by these low-mass star systems \citep[][]{pozzetti19,boylan18,elmegreen12}, implying that the term ``galaxy'' for this class of faint sources does not seem appropriate.

\begin{figure*}[ht!]
        \centering
        \includegraphics[width=\linewidth]{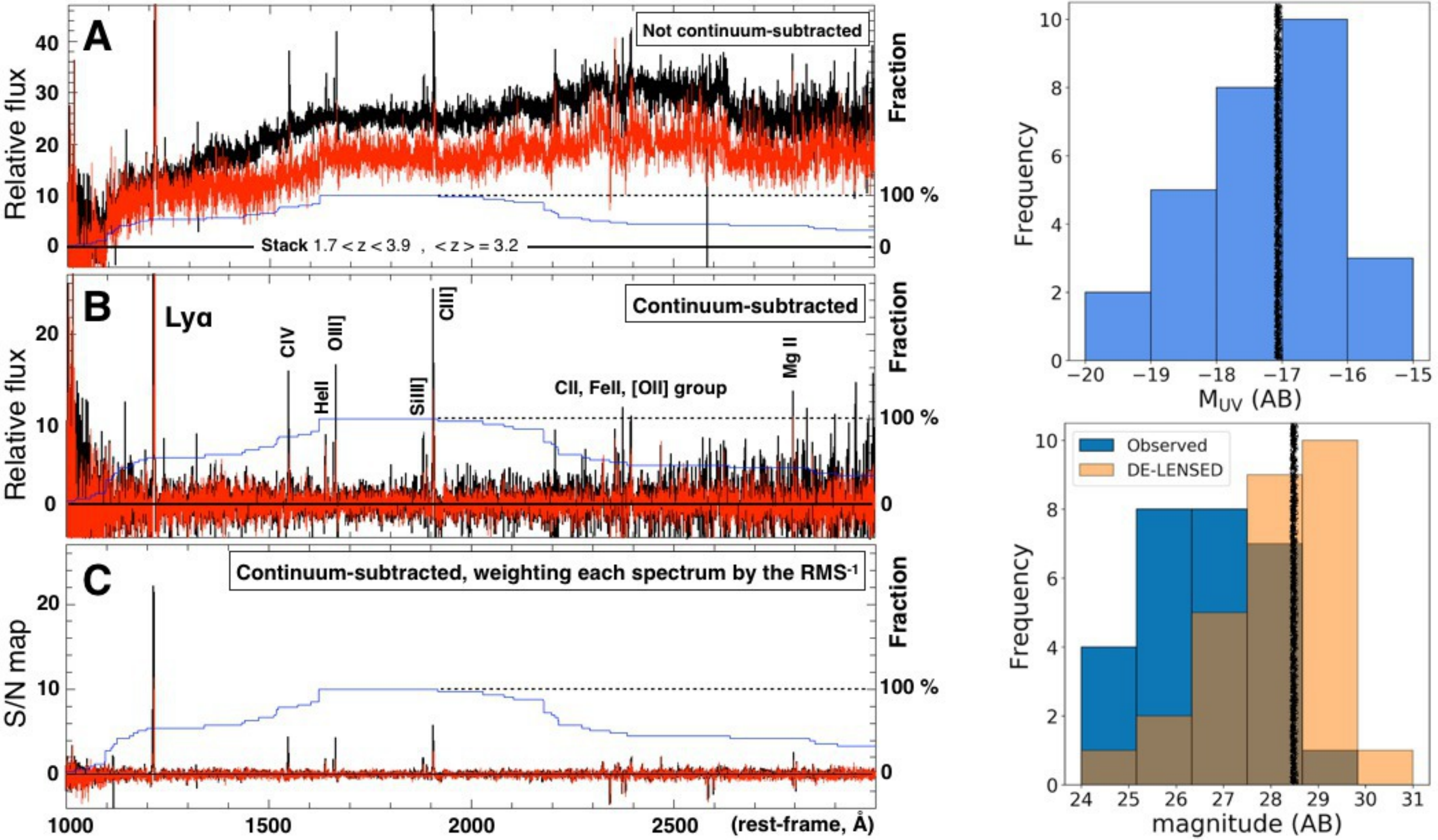}
        \caption{{\it Left panels:} Mean(median) stacked spectrum of 33 sources at redshift $1.7 < z < 3.9$ shown in black(red). Panel A shows the raw stacking of sources without any continuum subtraction. The contamination from the foreground galaxy cluster is evident. In the middle panel B, the spectra are stacked after subtracting the continuum. Panel C shows the result of weighting by the inverse of the noise the continuum subtracted spectra (see Figure~\ref{stack} for a zoomed version), which we use as an emission line detection map. The blue curve in panels A, B, and C shows the percentage of input spectra (out of 33) entering in the stacking. {\it Right panels:} Top-right histogram shows the delensed absolute magnitudes of the sources in the stack; the bottom-right panel shows the observed (blue) and intrinsic (orange) magnitudes. The vertical black stripes indicate the median values of the distributions.}
        \label{stack_3versions}
\end{figure*}

\section{Spectral stacking and high-ionization nebular lines detected on individual sources to M$_{\rm UV}\sim-16$}
\label{stack.section}

 Before discussing the method used to coadd spectra from a set of sources, it is worth mentioning two main differences between lensed and non-lensed fields. First, in lensed fields, the high-redshift background sources are contaminated by the intracluster light and by generally red galaxy cluster members, especially in the innermost regions of the galaxy cluster. Therefore, spectral features due foreground cluster galaxies may remain imprinted in the final stacked spectrum if not subtracted properly. Second, the presence of multiple images allows us to increase the effective total integration time for a single family. For example, when three multiple images with similar levels of magnification (e.g., comparable magnitudes) and free from foreground contamination are available, the total integration time for the single source increases to 51.3 hours ($17.1 \times 3$). Naturally, when only one image is available (for whatever reason), the integration time reduces to the original integration of 17.1 h for the MDLF (a similar argument applies to the SW pointing).

\begin{figure*}[ht!]
        \centering
        \includegraphics[width=\linewidth]{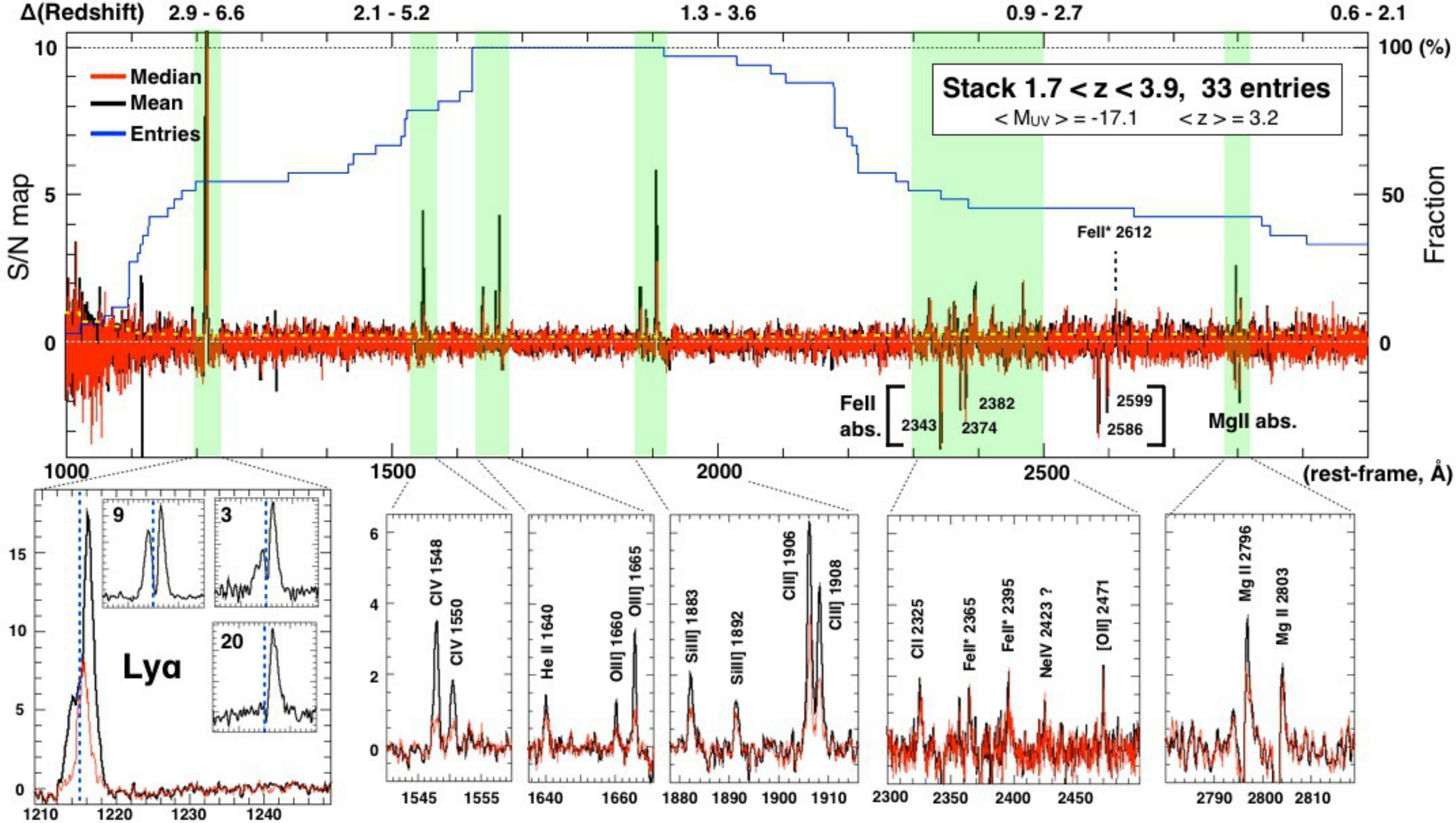}
        \caption{Mean (median) S/N stacked detection map of sources lying in the redshift range of $1.7<z<3.9$ is shown in black (red) on the top panel. The number of entries included in the stack is shown in blue and the fraction reported in the right Y-axis, where 100\% corresponds to 33 entries. On the top axis, the corresponding redshift interval probed by MUSE is indicated for the rest-frame wavelengths reported on the X-axis. The yellow dashed line represent the 1-sigma error of the stack, increasing at the edges of the wavelength range. The green transparent stripes indicate the positions of the most relevant groups of high-ionization lines, whose zoomed versions are shown in the bottom panels. On the bottom leftmost panel, the zoomed \lya\ line is shown and is a result of a variety of different line profiles; as an example, three different cases are reported in the insets of the same panel, extracted from sources 9, 3, and 20 shown in relative flux units, with the systemic redshift indicated with the vertical blue dashed lines.}
        \label{stack}
\end{figure*}

We mitigate the first issue by stacking continuum-subtracted spectra. This procedure implies that we miss the final continuum slope of the coadded spectrum and it tends to wash out the absorption lines as well, even though some signature of absorption lines still persist (see below). In this section, we focus on the detection of emission lines. 

We adopted the following strategy to compute the stacked spectrum: (1) For spectra in the redshift range of $1.7<z<3.9,$ the \ciiidoub\ line wavelength is captured by MUSE. The systemic redshifts have been measured from at least one of the following nebular high-ionization emission lines: \civ, \heii, \oiiiuv,\ and \ciii, which often are detected on individual spectra (as also the median stacked spectrum demonstrates, see below). The redshift from the \lya\ line is used if no other lines are present. Here, we decided to exclude the sample at $z>3.9$, for which the high-ionization lines mainly lie in the forest of sky emission lines. 

\noindent (2) Each one-dimensional spectrum is continuum-subtracted by using a smoothing-spline and successively weighted by the inverse of the corresponding error spectrum provided by the MUSE pipeline.
The resulting continuum-subtracted S/N spectra have more regular sky residuals and can be considered as S/N detection maps. The measurements of line ratios are, however, performed on the continuum-subtracted stack.
(3) Spectra belonging to multiple images of the same family have been combined by computing a weighted average, where the weights are assigned after a visual inspection of each multiple image, based on the observed magnitudes, the magnifications factors, and presence of possible contaminants (e.g., by excluding the cases outshone by nearby foreground objects).

In this way, we selected 61 (out of 66) individual objects, excluding five of them due to redundant information -- namely,  close clumps that are undistinguished by the MUSE extraction aperture of  $0\farcs8$ in diameter and that would enter more than one time in the stacking); 33 out of 61 satisfy the condition $1.7 < z < 3.9$.
The average weighted exposure time for the 61 objects is 33 hours (ranging between 17.1 to 51.3 hours) and the equivalent total weighted integration time for the stacked spectrum in the wavelength range of \lya $-$ \ciii\ spans between 600 to 1000 hours, without including the amplification $\mu$.
By adopting an average $\mu=4$, the equivalent integration time needed to obtain a similar depth in unlensed fields would add up to $\gtrsim 10000$ hours.

Figure~\ref{stack_3versions} illustrates the stacking steps. The raw mean/median stack without continuum subtraction is shown in panel A, highlighting the smooth red pattern emerging from the foreground cluster contamination. The mean/median stack of continuum subtracted spectra is reported in panel B. The S/N detection map obtained after inversely weighting each continuum-subtracted spectrum by its error spectrum is shown in panel C. The latter is the best probe for the presence of faint emission lines (including some absorption lines).

Figure~\ref{stack} zooms in on panel C of Figure~\ref{stack_3versions}. The stacked median (mean) S/N detection map clearly reveals the presence of high-ionization emission lines from \lya\ up to \mgii, for sources with 
magnitude spanning the range of $[-15,-19]$ and a median absolute magnitude M$_{\rm UV}=-17.1$. Magnitude distribution of the objects entering the stack are shown in the right panels of Figure~\ref{stack_3versions}. They include 28 sources for which reliable photometry could be obtained.
The line ratios among key nebular emission lines discussed by \citet{feltre16} calculated from the continuum-subtracted stacked spectrum (panel B of Figure~\ref{stack_3versions}) suggest that, on average, the ionizing source is dominated by stellar emission, rather than being powered by AGNs. In particular, the line ratios Log$_{10}$(\civ\ / \heii) = $0.294 \pm 0.026$ and Log$_{10}$(\oiiiuv\ / \heii) = $0.381 \pm 0.043$ lie well within the area populated by star-forming regions \citep[e.g.,][]{feltre16, gutkin16, mainali17, vanz_id14}. The same conclusion is reached based on the \ciiidoub, \civ,\ and \heii\ line ratios.
It is also worth noting that such nebular lines are at best marginally resolved at the MUSE spectral resolution (R $\simeq 3000$, d$\lambda = 2.6$~\AA), implying line widths $\sigma_v~(\rm or~FWHM) \lesssim 50~(120)$ \kms. Indeed, some cases subsequently observed with VLT/X-Shooter at higher spectral resolution of ${\rm R=8900}$, for instance, source 14 discussed here, show that such nebular lines can be as narrow as $\sigma_v~(\rm or~FWHM) \lesssim 15~(33)$ \kms, being marginally resolved also in the X-Shooter data \citep[][]{vanz_id14, vanz16_id11}.

Unlike in \citet{feltre20}, where no lines are individually detected at S/N~$>3$ for individual objects, the combination of lensing and deep MUSE observations allows us to detect several high-ionization lines individually, even for objects with de-delensed magnitudes as faint as $\sim 28-30$. In fact the intrinsic fluxes of such lines are in the range of a few $10^{-20} - 10^{-18}$ \ergscm\ in single sources (three examples are shown in Figure~\ref{examples}). 
In Appendix~\ref{stack_appendix}, we show stacks of a subset of faint one-dimensional spectra for which high-ionization lines have been detected individually.

The sample of lensed sources observed with the MDLF confirms the results obtained by \citet{feltre20} (and \citealt{maseda18}) on the HUDF, extending the luminosity range down to M$_{\rm UV}=-16$ and increasing the wavelength coverage up to $\lambda \sim 2800$\AA. High-ionization lines are common in very low-luminosity regimes (confirmed even for single $m>28-30$ objects), given their presence in the median stack over the full sample (see Appendix~\ref{stack_appendix} for a comparison between mean and median stack for a subset of sources). 
While such nebular emission lines will be modeled individually elsewhere, we note here that the presence of nebular emission doublet at the \civ\ wavelengths emerging from the ionized gas is indicative of  very low (${\rm Z<0.002}$) interstellar metallicity \citep[][]{senchyna17,senchyna19,vidal17civ}, especially for the fraction of sources where such nebular emission is most prominent and dominates the averaged stacked spectrum.
Not surprisingly, the sources probed in this work at the faintest luminosity regime sample the tail of the very low stellar mass objects ($10^{6-8}$\msun), for which a low metallicity would be expected by extrapolating the mass-metallicity relation at such masses and redshift \citep[see][for a review]{maiolino.mannucci.rev2019}.
The stacked spectrum shown in Figure~\ref{stack} complements in terms of luminosity and stellar mass the stacked spectrum derived from the Project {\tt MEGaSaURA} (the Magellan Evolution of Galaxies Spectroscopic and Ultraviolet Reference Atlas, \citealt{rigby18_I,rigby18_II}). In that study, a composite spectrum of 14 highly magnified star-forming galaxies at $1.6<z<3.6$, with stellar mass $\gtrsim 10^9$ \msun\ and median sub-solar metallicity (37\% Z$_{\odot}$), reveals numerous weak nebular emission lines, stellar photospheric absorption lines and strong absorption from interstellar medium on a high signal-to-noise detected continuum. Unlike the {\tt MEGaSaURA} stacked spectrum, the spectrum reported here includes much fainter objects and shows evident narrow nebular emission lines (including the \heii) that are not (or only marginally) present in the brighter galaxy sample used in \citet{rigby18_II}.

It is worth noting that in several cases nebular high-ionization lines emerge from single clumps, as shown in Figure~\ref{examples} for sources 1, 9, and 14. In particular, source 9 shows three distinct clumps, $9.1a$, $9.2a$ and $9.3a$, each one barely resolved in HST images implying effective radii smaller than $100-200$ pc. The bluest of the three (and the most nucleated one, 9.2a) shows the strongest \heii\ emission. Gravitational lensing allows us to identify such small clumps and (in this case) to extract spectra for each of them. Another (and most extreme) example is the Sunburst arc, in which the very large magnification allowed us to recognize a single 3 Myr old star cluster, 
showing evident P-Cygni profiles of \nvalone, \civalone\ and broad \heii\ arising from O-type and Wolf-Rayet stars \citep[][]{vanz_sunburst}.

\begin{figure*}
        \centering
        \includegraphics[width=\linewidth]{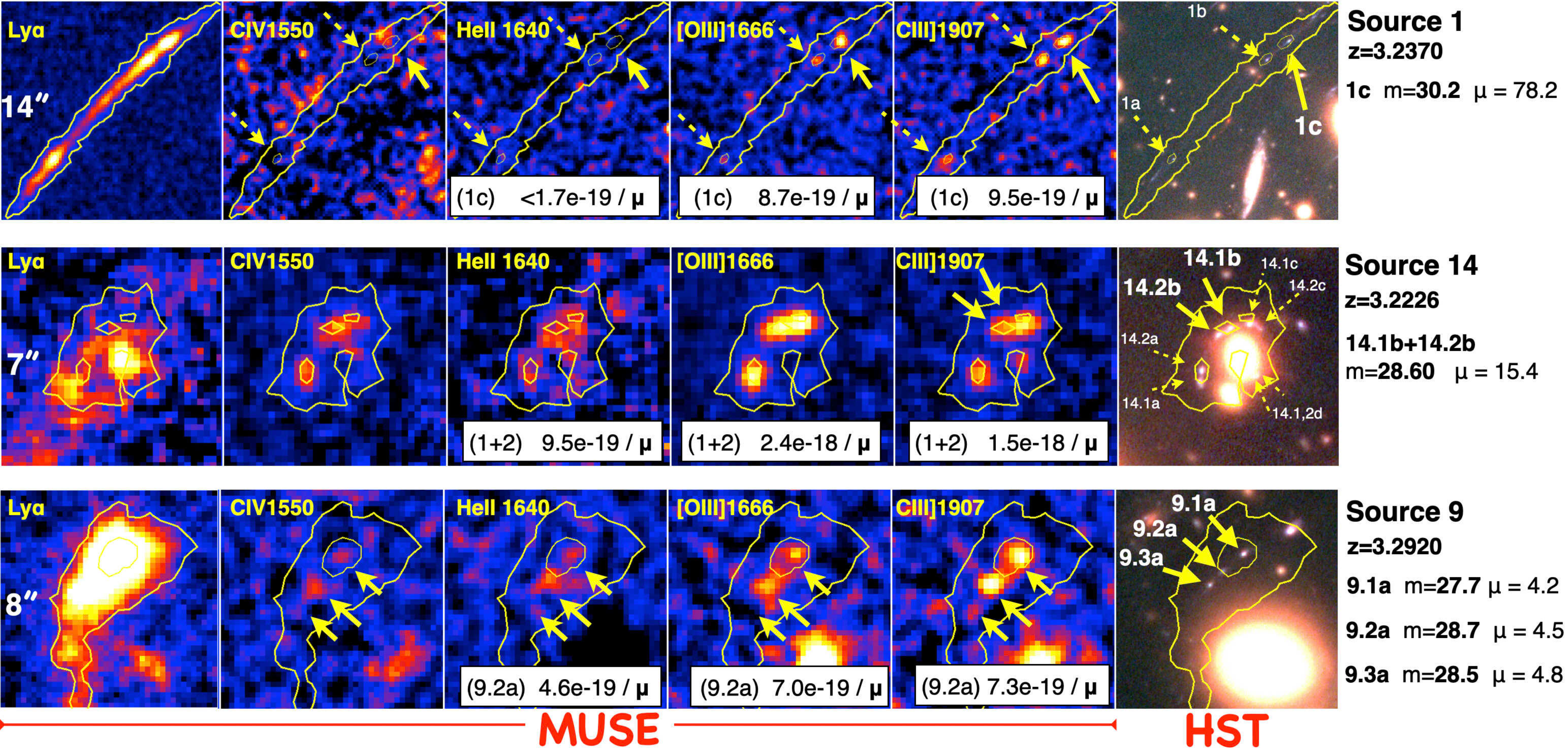}
        \caption{Three examples for which high-ionization emission lines are detected individually. From left to right, the most relevant atomic transitions are shown (labeled on each panel). Yellow contours outline \lya\ (\civalone) in the top and bottom (middle) rows at 2- ad 10-$\sigma$ levels. The MUSE continuum-subtracted narrow-band images ($dv=300$ \kms) have been smoothed with a $2\times2$ boxcar filter along the spatial dimension. On the right, the color HST image (red = F105W, green = F814W, blue = F606W) with arrows showing relevant sources. Solid arrows refer to individual sources, while the dashed ones to the associated multiple images, when present. Measured line fluxes are also reported for \heiialone, \oiiialone\ and \ciiialone; the intrinsic values are obtained using the  $\mu$ values quoted on the right, together with redshifts and de-lensed magnitudes. Magnifications for sources 1 and 14 are also reported in Table~\ref{ratios}. We note the case of source 9 (bottom) where high-ionization lines emerge differently from the three individual clumps (9.1a, 9.2a, 9.3a). The \lya\ emission in the leftmost panels show a typical arc$-$like shape (top, \citealt{vanz_paving}), a deficit of \lya\ emission if compared to carbon emission (middle, see one-dimensional spectrum in Figure~\ref{individual14} and \citealt{vanz_id14}), and a \lya\ nebula (bottom, \citealt{vanz_id9})
        }
        \label{examples}
\end{figure*}

\section{Gravitationally bound star clusters at cosmological distance and prospects for future AO-assisted instrumentation}
\label{grav}

In this section, we discuss the interpretation of the clumps in terms of gravitationally bound star clusters, in the context of current observational limits and future AO-assisted instrumentation. 
The typical uncertainty on the amplification factor $\mu$ (including systematics) at high magnification regimes, $\mu>5$, is on the order of $20-40$\% and is discussed in Appendix~\ref{muerr} for a subset of sources discussed in this work.

\subsection{Looking for bound star clusters at high redshift}
\label{general}

A way to assess whether a stellar cluster is gravitationally bound is to calculate its dynamical age ${\rm \Pi}$, defined as the ratio between the age and the crossing time ${\rm T_{CR}}$, 
${\rm \Pi = Age / T_{CR}}$.  The crossing time expressed in Myrs is defined as 
${\rm T_{CR} = 10 \times (R_{\tt eff}^{3} / GM)^{0.5}}$, where M and ${\rm R_{eff}}$ are the stellar mass and the effective radius, respectively, and ${\rm G \approx 0.0045~pc^{-3} \msun^{-1} Myr^{-2}}$ is the gravitational constant. Stellar systems evolved for more than a crossing time have ${\rm \Pi > 1}$, suggestive of being bound (\citealt{gieles11}; see also discussion by \citealt{adamo20}). This criterion has been used extensively for the identification of star clusters in the local Universe  \citep[e.g.,][]{calzetti15,adamo17,ryon17}. The criterion is valid under the assumptions that the system is in virial equilibrium, follows a Plummer density profile and the light traces the underlying mass.

\begin{figure}
        \centering
        \includegraphics[width=\linewidth]{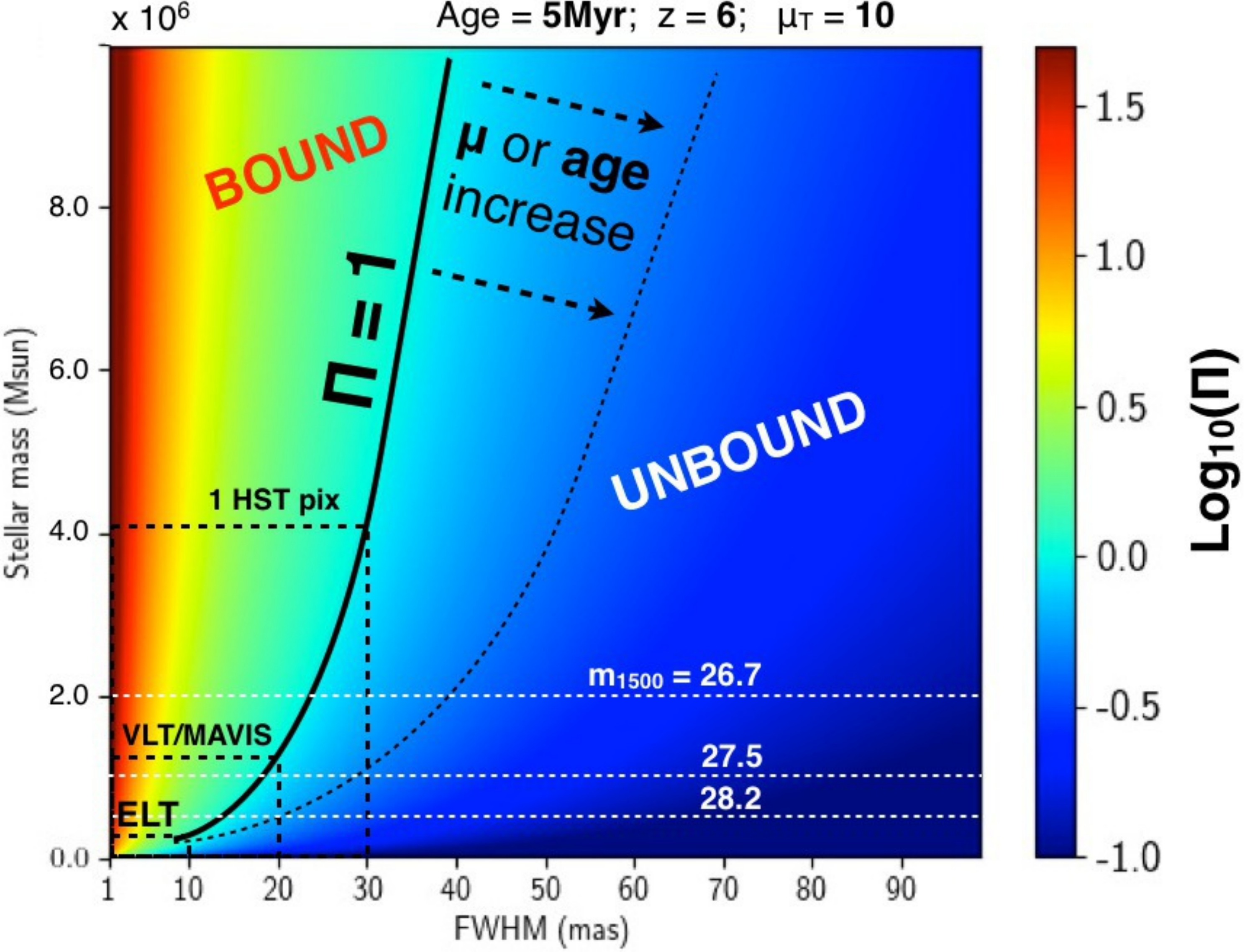}
        \caption{Dynamical age in logarithmic units (color-coded) as a function of stellar mass and angular resolution (milliarcsecond, mas), for a 5-Myr-old star cluster placed at z=6 and observed at tangential magnification $\mu_{T}=10$. The solid black curve marks the locus of $\Pi = 1$ separating bound ($\Pi > 1$) from unbound ($\Pi < 1$) systems. The cases of HST (1 pixel, 30 mas), VLT/MAVIS with an expected FWHM of 20 mas and E-ELT/MAORY-MICADO with FWHM of 10 mas are shown. VLT/MAVIS and E-ELT/MAORY-MICADO can probe star clusters down to a few $10^{5}$\msun, provided that the S/N and the knowledge of the PSF allow a proper morphological
        analysis and/or PSF deconvolution. The slope of the $\Pi = 1$ curve flattens if the magnification and/or the age of the system increase (dotted curve). UV apparent magnitudes at 1500~\AA, corresponding to 2, 1 and $0.5 \times 10^{6}$ \msun\ objects, are indicated as horizontal white dotted lines.}
        \label{angular}
\end{figure}

Figure~\ref{angular} shows the required angular resolution needed to distinguish among bound and unbound star clusters as a function of stellar mass. For this exercise, the age of the cluster is fixed at 5 Myr, $z=6$ and magnification $\mu_{T}=10$ 
(the stretch along the tangential direction over which the size is probed). Instruments like E-ELT/MAORY-MICADO and VLT/MAVIS will reach 10 and 20 mas resolution in the near infrared and optical wavelengths, respectively, formally allowing for the identification of bound star clusters down to a few $10^{5}$ \msun\ (with the adopted $\mu$). Clearly, 
the discerning power depends on the S/N and the knowledge of the PSF over the field of view.
The S/N, in turn, depends on the magnification factor. For illustration, we compute the image plane magnitude of a star cluster with stellar mass {\rm M} as a function of $\mu$.
Assuming an instantaneous burst and Salpeter IMF, a 5 Myr old star cluster has an absolute magnitude M$_{1500} \approx -16.7$ \citep[][]{leitherer14}, which corresponds to a reference magnitude m$_{1500}^{{\rm ref}} = 30.0(28.8)$ at redshift 6(3). Therefore, the lensed apparent magnitude, $m$ can be written as:

\begin{equation}
    m = m_{1500}^{{\rm ref}} -2.5\log_{10}(\mu) - 2.5\log_{10}(10^{-6}\, {\rm M}/\msun).
    \label{eq1}
\end{equation}

The equation implies that a young $5\times 10^{5}$ \msun\ star cluster, magnified by $\mu=10$ at redshift 6(3), has a magnitude 28.2(27.0).  
It is worth noting that the very compact size of star clusters (e.g., ${\rm R_{eff}} < 30$ pc, \citealt{adamo20}) will favor the detection in deep imaging, in comparison to extended sources \citep[e.g., Figure~4 of ][]{bouwens_impact_size}. 
While dedicated simulations using realistic AO-based PSFs are needed to 
quantify the size reconstruction as a function of the S/N, we note that magnitudes 
$\lesssim 28$ are plausibly within reach of big telescopes, especially considering that relatively massive star clusters with ${\rm M > 10^{6}}$ \msun\ will be even brighter,  $m<27.5$. Moreover,  E-ELT/MAORY-MICADO, with a moderate magnification of $\mu=3$, will easily identify massive ${\rm M \simeq 10^{7}}$ \msun\ star clusters expected to have $m=26.3$ (from Eq.~\ref{eq1}), while still probing a physical scale of 11 pc/pix, considering the MICADO pixel scale of 4 mas and $\mu_T = 2$ (assuming $\mu$ = 3 = $\mu_{T} \times \mu_{R}$, i.e., the tangential stretch slightly dominates over the radial one, as typically happens for the MDLF). By reaching star-cluster like sizes with modest magnifications, the ELTs will pave the way for the exploration of much larger volumes than those currently accessible with 8-10m telescopes that require high magnification (e.g., $\mu>10$ in Figure~\ref{surface}).

\begin{figure*}
        \centering
        \includegraphics[width=\linewidth]{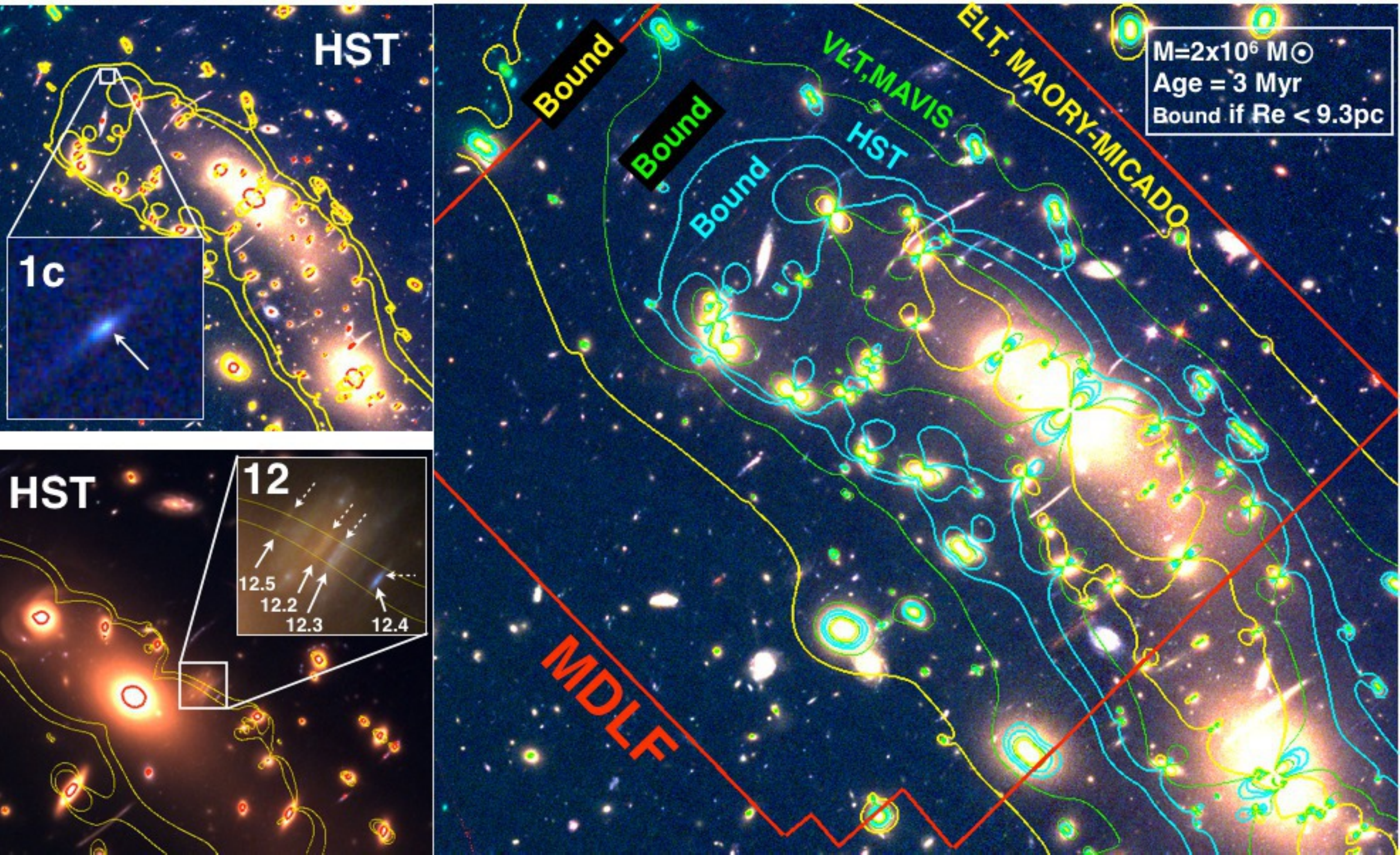}
        \caption{{\it Left panels:} Locus of dynamical age $\Pi=1$ at the redshift of image 1c ($z=3.237$, top) and 12 ($=0.939$, bottom) is highlighted with yellow contours; a star cluster with an effective radius matching one HST pixel (30 mas) lying within such contours is gravitationally bound if its age is larger than 3 Myr or it has a stellar mass larger than $2\times 10^{6}$ \msun. The compact knots detected in image 1c and 12 are candidate gravitationally-bound star clusters under the above assumptions. The insets show the zoomed source 1c and 12, with arrows indicating the most magnified knots. {\it Right panel:} Same contours of $\Pi=1$ at the redshift of source 1 adopting the HST 30 mas/pix scale (cyan line), VLT/MAVIS (7.5 mas/pix, green line), and E-ELT/MAORY-MICADO (4 mas/pix, yellow line) are superimposed onto the same HST color image shown in the left panels (red = F105W, green = F814W and blue = F606W). The layout of the MDLF is shown in red. It is worth noting that ELT can probe gravitationally bound young massive clusters even with moderate magnification $2 < \mu < 4$, a regime in which $\mu$ is free from systematic errors and by foreground contamination by galaxy cluster members.}
        \label{dyn}
\end{figure*}

Currently, deep HST imaging on lensed fields and PSF deconvolution down to the single HST pixel (30 mas) on MUSE-confirmed sources is producing intriguing candidate star clusters. In order to explore the potential of HST observations, we calculate the lensed dynamical-age-cross-section starting from the magnification maps of J0416 extracted from the latest \citet{bergamini20} lens model. As an exercise, Figure~\ref{dyn} shows the contours of dynamical age $\Pi = 1$ assuming a star cluster age of 3 Myr, a stellar mass of $2\times 10^{6}$ \msun\ and assuming the object is not (or it is marginally) resolved down to an effective radius ${\rm R_{eff} = 1}$ HST pixel of 30 mas (the same Figure also shows the case for E-ELT/MAORY-MICADO and VLT/MAVIS). Such a limit has been recently reached with, for example, {\tt Galfit} \citep[][]{Peng_2010} after a proper PSF deconvolution \citep[e.g.,][]{vanz19,zick20}. Contours at $\Pi = 1$ have been calculated at the redshift of source 1 ($z=3.237$) and source 12 ($z=0.939$), which seem to host very small knots. Figure~\ref{dyn} shows the resulting area within which $\Pi > 1$ under the above assumptions (i.e., the region within which 1 HST pixel probes $<9.3$ pc, which is the corresponding  ${\rm R_{eff}}$ at $\Pi=1$),
with the positions of the observed images 1c and knots of source 12. 
The very large tangential magnification coupled with their very nucleated appearance is a suggestion that their sizes are extremely small. 

\noindent {\it Source 1.} Image 1c shows an effective radius of 2.5 pixels \citep[][]{vanz_paving} that would correspond to $< 10$ pc along the tangential stretch ($\mu_{T} \simeq 69$). Specifically, the same tangentially elongated image also shows a nearly point-like spatially offset knot (indicated with a white arrow in Figure~\ref{dyn}, see also bottom-right panel of Figure~2 in \citealt{vanz_paving}).
The effective radius of such a knot is even smaller than the entire image 1c, conservatively not larger than 2 pixels with a size smaller than 7 pc.
Under the assumption that the knot hosted on image 1c is not younger than 3 Myr and with a stellar mass not smaller than $10^{6}$ \msun, a ${\rm R_{eff}} < 7$ pc would imply ${\rm \Pi > 1}$, matching the condition for a gravitationally-bound star cluster. The distribution of possible $\Pi$ depends on the solutions for the stellar mass and ages within certain confidence levels, given the magnification uncertainly, and it is not calculated here. However, it represents a good candidate-bound star cluster that is likely dominating the \lya\ and high-ionization line emission (see image 1c on Figure~\ref{examples}) and that will need further exploration, for instance, by adding near-infrared spectroscopic observations to constrain the aforementioned age and stellar mass. Such an object is also reminiscent of similar local star clusters dominating the ionization field and the \lya\ emission \citep[e.g.,][]{bik18}. 
A very similar object showing a spatially offset knot hosted in a more elongated image has been discussed by \citet{zick20}, however, with a lower magnification regime that allows them to put constraints down to 40 pc physical scale.

\noindent {\it Source 12.} Source 12 is a spiral galaxy at $z=0.939$ that is straddling the corresponding critical line. Its proximity to several nearly point-like knots hosted in 12
(12.2, 12.3, 12.4, and 12.5) suggests magnification values in the range of $20-100$, strongly stretch along the tangential 
direction \citep[][]{bergamini20} and corresponding to a spatial scale of $10-1$ pc/pixel, respectively.  Knots 12.2, 12.3, 12.4, and 12.5 are shown in Figure~\ref{dyn} (also Figure~\ref{MI1}). Assuming a HWHM of $0.06''$ for the HST ACS/F435W PSF (of $0\farcs12$, \citealt{merlin16}), a rough estimate of the sizes span the 
range of $4-20$ pc along the tangential direction. Under the above assumptions (age and stellar masses), the knots would touch the boundaries where ${\rm \Pi > 1}$ (Figure~\ref{dyn}), especially the object 12.4b(,c) that is extremely close to the critical line with a plausible size 
smaller than 6 pc. Performing a detailed mass, age, and size estimation of such extreme cases it is not the scope of this work, however Figure~\ref{dyn} shows that relatively rare (due to the required magnification) gravitationally bound star clusters can be identified at cosmological distance with HST imaging on lensed fields (see also the analysis of the {\em Sunburst} arc in  \citealt[][]{vanz_sunburst}). 

\subsection{A possible pair of massive star clusters at z=3.223 }
\label{14}

Source 14 is exceptional, given that it is magnified by the galaxy cluster that produces three multiple images and a couple of cluster members, which further splits one of the images into four.
In total, source 14 generates six images, $14a,b,c,d,e,f$ 
(\citealt{Caminha_macs0416},
see also \citealt{bergamini20}). \citet{vanz_id14} based on two initial hours of MUSE integration confirmed five out of six multiple images, with the sixth and the least magnified one ($14f$) being tentatively identified via photometric redshift. Here, we confirm the five previously identified images and revisit the identification of the sixth $14f$, now confirmed with the MDLF. In particular, image $14f$ corresponds to the ASTRODEEP source ID=1127 with a magnitude of F814W = $27.78\pm0.07$ (see Appendix~\ref{individual14}).
 The magnification at the location of $14f$ is $\mu=2.1 \pm 0.1$, implying an intrinsic magnitude for source 14 of 28.6.
 
 Surprisingly, the most magnified version of source 14 (e.g., $14a$ or $14b$) shows that the spatially unresolved image $14f$ is made of two distinct and much smaller knots (labeled as ``1'' and ``2''), which do not appear at position $f$ (Figure~\ref{fig_id14}). The two very magnified knots at position $a$ (or $b$) have very similar ultraviolet magnitude ($\simeq 26.5$, \citealt{vanz_id14}) and are separated by $\sim 390$ pc in the source plane. Assuming that each knot  contributes equally to the observed magnitude of 28.6, the intrinsic magnitude of each of them is of the order of $\sim 29.4$. This value may be a lower limit to the brightness if the host galaxy contributes any flux.

The updated magnification, inferred by comparing the observed fluxes and using the improved lens model, implies that source 14 is made of  a pair of compact knots having ${\rm R_{eff}} \lesssim 30$ pc each, with both having a de-lensed magnitude of $\gtrsim 29.4$, or M$_{\rm 1500} \gtrsim -16.3$ (see Appendix~\ref{individual14}). From the  SED-fitting performed by \citet{vanz_id14}, we know that their ages span the range of $10-30$ Myr and stellar masses are in the range of $(1-10)\times 10^{6}$ \msun. Intriguingly, the combination of these quantities (e.g., ${\rm R_{eff} \sim 30}$ pc, M = $5\times 10^{6}$ \msun\ and Age = 20 Myr) produces a dynamical age of $\Pi \simeq 2$, supporting the hypothesis that the two knots are indeed a pair of gravitationally bound massive stellar systems separated by 390 pc on the source plane and approaching the definition of young massive star clusters.
Radii of the order of 30 pc appear quite large for local star clusters \citep[e.g.,][]{bastian.lardo18}. However, more typical values of $\lesssim 25$ pc are still within the uncertainties of the present data.

From the perspective of ELT performance, an instrument such as MAORY-MICADO will probe a spatial scale of 50 pc at the redshift of image $14f$ magnified by $\mu = 2.1$, allowing for the identification of the two knots (unresolved by HST), although each one will be not spatially resolved. Remarkably, MAORY-MICADO with 10 mas PSF resolution on images $14.1a$, $14.2a$, $14.1b$, $14.2b$ ($\mu \simeq 15$) will probe 6.7 pc (or 2.7 pc / pix, adopting 1 pix = 4 mas), along the direction of the maximum stretch ($\mu_{T} \sim 11$). If MAORY-MICADO will probe 2.7 pc/pix in the rest-frame optical wavelengths, VLT/MAVIS will cover the rest-frame ultraviolet down to $\sim 5$ pc/pix on the same images (adopting 7.5 mas/pix)\footnote{It is worth noting that the very limited sky-coverage offered by MUSE in the narrow field mode configuration make the observation of such objects prohibitive.}. This will be a dramatic step forward in the study of these kinds of objects, allowing us to calculate what fraction of the stars in the galaxy have formed in gravitationally bound star clusters (the cluster formation efficiency, $\Gamma$). It is worth noting that if at least one of the two knots is a gravitationally bound star cluster and the host is marginally contributing to the emerging ultraviolet light (as the most magnified images seem to imply), then it would suggest a large $\Gamma \sim  50\%$ in this system. In other words, more than half of the ultraviolet light comes from stars bounded in a star cluster. If they are a physical pair of massive clusters, then $\Gamma$ could be well above 50\%. Rare and large $\Gamma$ values ($\gtrsim 80\%$) in the local Universe under extreme environment conditions (starburst galaxies) have been observed with masses as high as $10^{7}$ \msun. Recently, \citet[][]{adamo20extreme} described such cases for a sample of six galaxies within 80 Mpc distance from the Earth, suggesting that such large values of $\Gamma$ and high truncation mass of the star cluster mass function would be more common in the high redshift Universe. State-of-the-art instruments will allow us to begin exploring these properties in greater detail.

\section{Conclusions}

In this work, we present the MUSE Deep Lens Field (MDLF) with a total integration time of 17.1 h over a single pointing, targeting one of the best cosmic telescopes, HFF~MACS~J0416 at $z=0.396$, and providing line flux limits down to 
$2\times 10^{-19}$ \ergscm\ within 300 \kms\ and continuum detection down to magnitude 26, both at three sigma level at $\lambda = 7000$~\AA. 
While the effective area probed in lensed fields rapidly decreases with the magnification $\mu$, when compared to non-lensed fields (Figure~\ref{surface}), the combination of a long exposure (17.1 hours) and amplification allow us to probe very faint fluxes, which would require well above $100$ hours in blank fields. Specifically, about 90\% of the MDLF field of view is equivalent to $>100$ h integration without lensing, assuming point-like emission (see Figure.~\ref{equivalent}). 
By combining deep MUSE spectroscopy with deep HST multi-band imaging, we obtain the following initial results:

\begin{enumerate}
    \item{We increased the number of multiple images to 182 in the redshift range of $0.9<z<6.2$, emerging from 66 families extracted from 48 background individual sources. These multiple images, including multiple clumps detected around the critical lines, are used to constrain the new lens model presented by \citet{bergamini20} in an accompanying paper. This unprecedented number of spectroscopically confirmed images enhance significantly the reliability of the magnification maps for high redshift studies \citep[e.g.,][]{johnson_sharon16,Caminha_rxc2248}.
    }
    \item{The majority of the multiple images show star-forming clumps over a wide redshift range, as discussed in Sect~\ref{clumps} and Appendix~\ref{clumps_appendix}. Strong lensing geometry coupled to MUSE spectroscopy allow us to confirm very compact and faint objects, including sub-components that
    would be beyond reach also at the MDLF depth. In a future work, lensing magnification of such systems will enable individual analysis (e.g. SED fitting) and in some cases to perform localized spectroscopy, with an effective resolution of 100-200 pc physical scale (as shown in Figures~\ref{sys106},~\ref{system20},~\ref{sys5}, and discussed in Appendix~\ref{clumps_appendix}).}
    \item{High ionization metal lines of \civ, \oiiiuv, \siiiuvem\ and \ciiidoub\  (including  \heii)\, have been detected with S/N in the range of $>5-30$ on individual objects down to intrinsic magnitude 28-30, with de-lensed line fluxes of $10^{-20} - 10^{-19}$ \ergscm, including several with sizes smaller than $<200$ pc (see examples in Figures~\ref{examples} and~\ref{fig_id14}). Such lines emerge very clearly in the mean and median stacked spectra (Figure~\ref{stack}). At a median redshift $z=3.2$, the high-ionization lines seem to persist down the faintest limits probed by the MDLF, for instance, M${\rm_{UV}} \simeq -16$, thus extending the results of \citet{feltre20} to fainter luminosity regimes. In particular, the cases showing the most prominent nebular lines are indicative of a low-metallicity regime}.
    \item{Candidates for gravitationally bound star clusters with sizes smaller than 30 pc have been identified at cosmological distance (Sect.~\ref{grav}), including a doubly imaged likely physical pair young massive star cluster separated by $\sim 400$ pc in the source plane (source 14). 
    Dynamical-age-cross sections have been calculated and prospects for future AO-assisted instrumentation discussed in Sect.~\ref{grav}. In particular, future instruments with resolutions of $10-20$ mas (e.g., E-ELT/MAORY-MICADO or VLT/MAVIS) will be able to identify young gravitationally-bound star clusters with ages smaller than $< 5$ Myr and stellar masses $\gtrsim 10^{5}$ \msun\  up to the reionization epoch. 
    }
\end{enumerate}

The MDLF gives us a first glimpse of the high redshift universe at luminosity and resolution that would have been impossible just a few years ago. It demonstrates very clearly the power of gravitational telescopes in complementing the physical parameter space accessible with the deepest blank fields, in which reaching MLDF depths would  require 100-1000 hr and the resolution would be unattainable. The main limitation of the MLDF is the progressively smaller volume probed in highly magnified regions. This limitation can be overcome by a concerted campaign of deep MUSE follow-up observations of lensing clusters.

\begin{acknowledgements}
      We thank the anonymous referee for the careful reading and constructive comments. This project is partially funded by PRIM-MIUR 2017WSCC32 ``Zooming into dark matter and proto-galaxies with massive lensing clusters''. We acknowledge funding from the INAF 
      for ``interventi aggiuntivi a sostegno della ricerca di main-stream'' (1.05.01.86.31).  PB acknowledges financial support from ASI though the agreement ASI-INAF n. 2018-29-HH.0. MM acknowledges support from the Italian Space Agency (ASI) through contract ``Euclid - Phase D" and from the grant MIUR PRIN 2015 ”Cosmology and Fundamental Physics: illuminating the Dark Universe with Euclid”. FC acknowledges support from grant PRIN MIUR 2017 $-$ 20173ML3WW$\_$001. CG acknowledges support by VILLUM FONDEN Young Investigator Programme through grant no. 10123. KC and GBC acknowledge funding from the ERC through the award of the Consolidator Grant ID~681627-BUILDUP. EV thanks Davide Vanzella for helping collecting data from the literature shown in Figure~\ref{scatter_clumps}. TT acknowledges support by the National Science Foundation through grant NSF-1810822 "COLLABORATIVE RESEARCH: The Final Frontier: Spectroscopic Probes of Galaxies at the Epoch of Reionization". MG was supported by by NASA through HST-HF2-51409. This research made use of the following open-source packages for Python and we are thankful to the developers of these: Matplotlib \citep{matplotlib2007}, MPDAF \citep{MPDAF2019}, PyMUSE \citep{PyMUSE2020}, Numpy \citep[][]{NUMPY2011}.
\end{acknowledgements}

%
%

\bibliographystyle{aa}
\bibliography{bibliography}

\begin{appendix} 
\section{The MUSE spectroscopic catalog}
\label{multiple_appendix}

\begin{figure*}
\centering
\includegraphics[width=17.5cm]{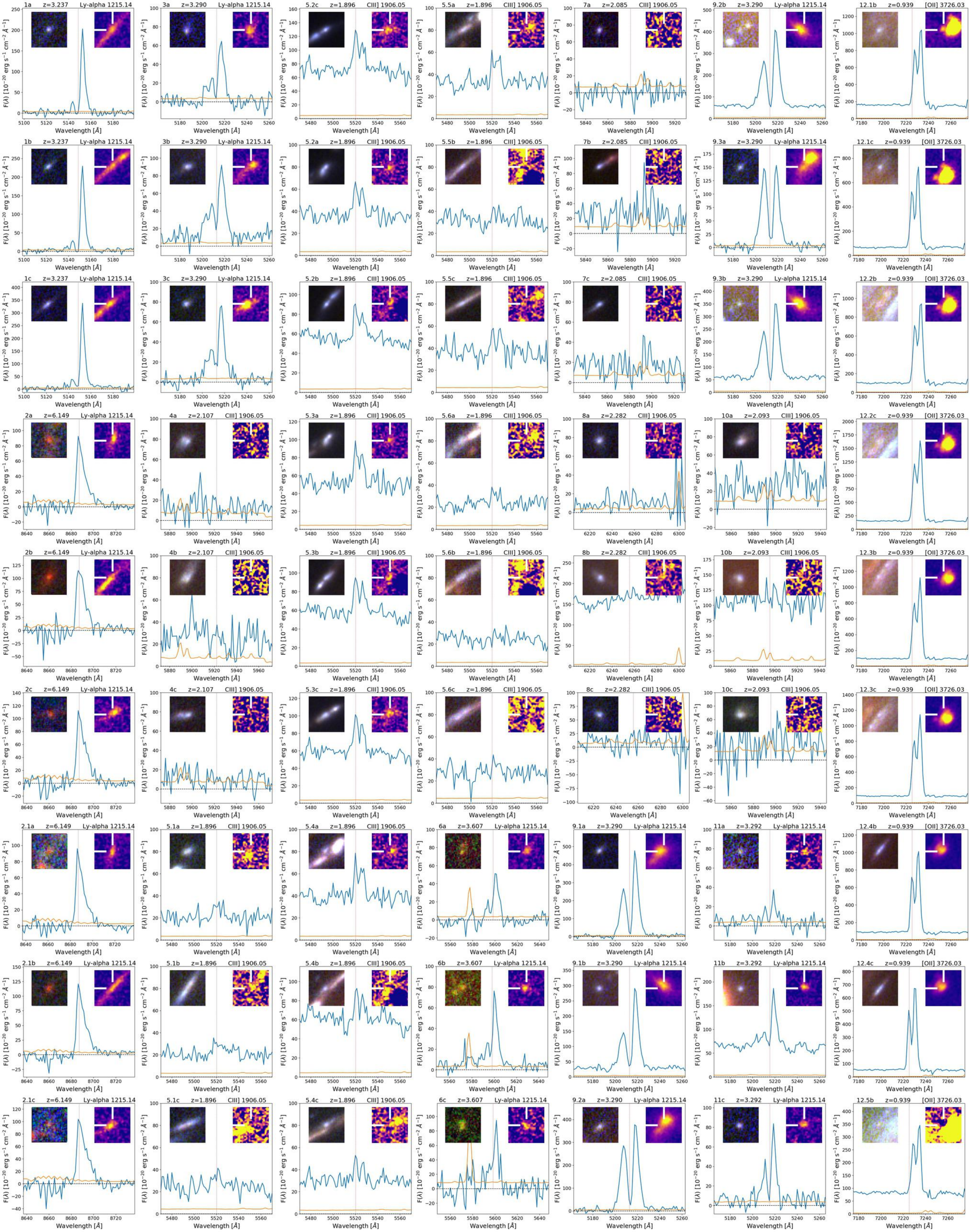}
\caption{Panels show the one-dimensional MUSE spectrum (blue) for the 182 multiple images used to build the lens model. Spectra are centered over the two most relevant atomic transitions: \lya\ or \ciiidoub, depending on redshift. The orange line is the error spectrum, and the red vertical line at the center of each cutout marks the wavelength of \lya\ (or \ciiiblue) at the given redshift. The spectra have been extracted from circular apertures with $0\farcs8$ diameter. Each thumbnail reports in the top, from left to right, the ID, redshift and the line transition. In each cutout, the inset on the top-left shows the HST RGB image corresponding to red = F814W, green = F606W, blue = F435W bands if $z<5.2$, or red = F105W, green = F814W, blue = F606W bands if $z>5.2$. The inset on the top-right is the continuum subtracted MUSE image of the same object, collapsed along the wavelength direction over 10(20)\AA\ for \ciiialone~(\lya) line.}
\label{MI1}
\end{figure*}

\begin{figure*}
\centering
\includegraphics[width=17.5cm]{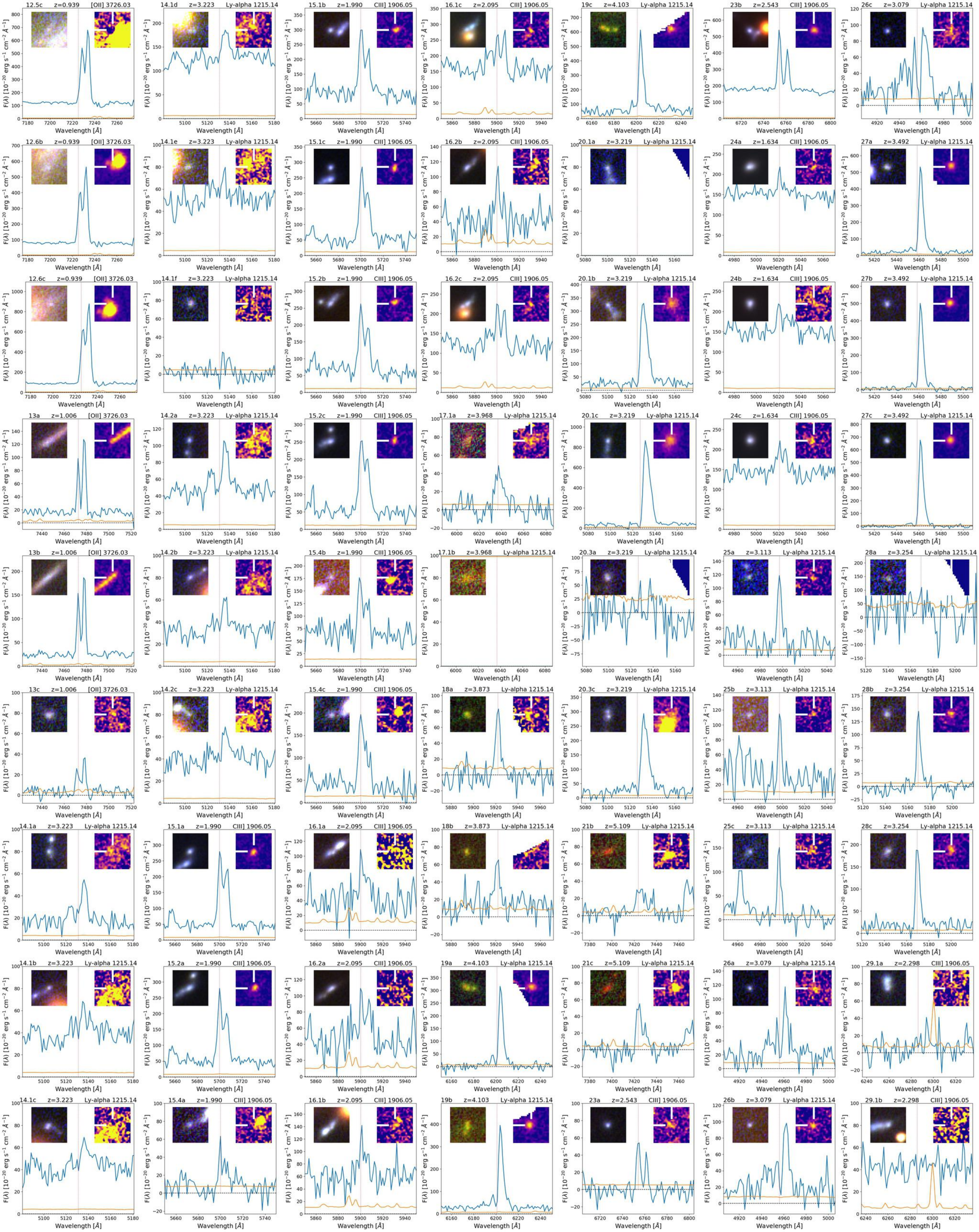}
\caption{Details in Figure~\ref{MI1}.}
\label{MI2}
\end{figure*}

\begin{figure*}
\centering
\includegraphics[width=17.5cm]{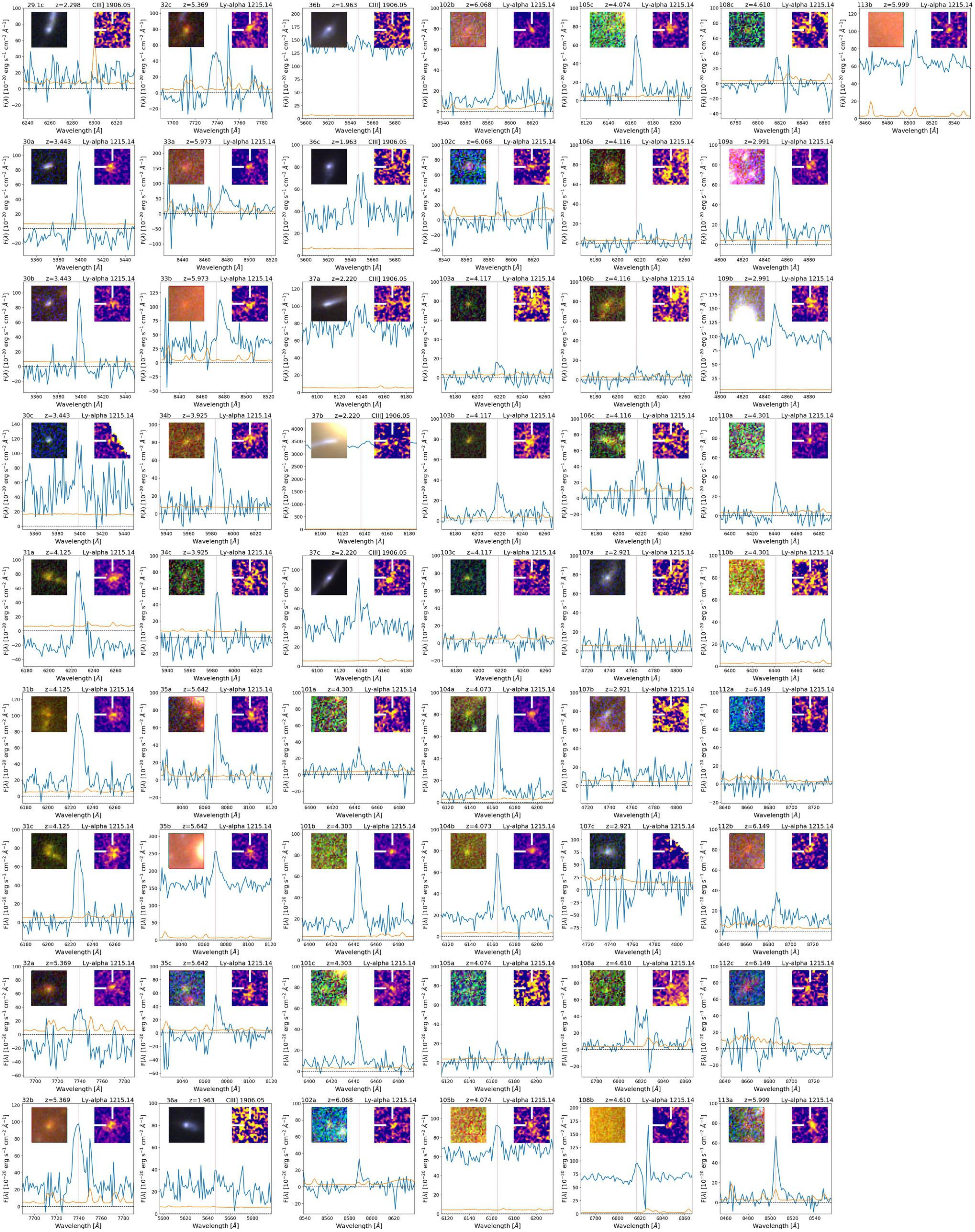}
\caption{Details in Figure~\ref{MI1}.}
\label{MI3}
\end{figure*}

Figures~\ref{MI1},~\ref{MI2}, and~\ref{MI3} show the 182 multiple images extracted from the HST/WFC3 F814W or F105W bands in the case the redshift is lower(higher) than $z=5.2$. Each thumbnail includes  a one-dimensional MUSE spectrum zoomed at the location of the most relevant emission lines: \lya\ or \ciiidoub, specifically at the position of the blue component of the doublet (1906.05~\AA). Redshifts are measured by cross-correlating the spectra with templates of high redshift galaxies using the Pandora
software package, within the Easy-Z environment \citep[][]{garilli10}.
Redshifts have also been secured from different emission lines or absorption lines in the cases where neither \lya\ or \ciiidoub\ are present. The list of multiple images, coordinates RA, DEC and redshifts are reported in Tables~\ref{tab_MI1} and~\ref{tab_MI2}.

In addition to the set of multiple images, we also release the first version of the full MUSE catalog collecting also sources not showing multiple images. An extracted example of the catalog is reported in Table~\ref{full_cat}, while the full list has been published online.

\begin{table*}[h]
\caption{List of confirmed multiple images used in the lens model of \citet{bergamini20}.}             
\label{tab_MI1}      
\centering          
\begin{tabular}{c c c c | c c c c}     
\hline\hline       
ID & RA & DEC & Redshift & ID & RA & DEC & Redshift \\ 
\hline
1a       &      64.049084        &      -24.062862       &      3.2370   & 9.2b   &      64.039996        &      -24.066651       &      3.2899 \\
1b       &      64.046959        &      -24.060797       &      3.2370   & 9.3a   &      64.045504        &      -24.072672       &      3.2899 \\
1c       &      64.046449        &      -24.060397       &      3.2370   & 9.3b   &      64.039925        &      -24.066616       &      3.2899 \\
2a       &      64.050874        &      -24.066542       &      6.1485   & 10a    &      64.044564        &      -24.072092       &      2.0930 \\
2b       &      64.047842        &      -24.062059       &      6.1485   & 10b    &      64.039576        &      -24.066623       &      2.0930 \\
2c       &      64.043586        &      -24.059005       &      6.1485   & 10c    &      64.034336        &      -24.063734       &      2.0930 \\
2.1a     &      64.050804        &      -24.066410       &      6.1485   & 11a    &      64.046841        &      -24.075385       &      3.2920 \\
2.1b     &      64.048175        &      -24.062403       &      6.1485   & 11b    &      64.038515        &      -24.065965       &      3.2920 \\
2.1c     &      64.043408        &      -24.058915       &      6.1485   & 11c    &      64.035223        &      -24.064731       &      3.2920 \\
3a       &      64.049232        &      -24.068174       &      3.2900   & 12.1b  &      64.036838        &      -24.067456       &      0.9392 \\
3b       &      64.045269        &      -24.062763       &      3.2900   & 12.1c  &      64.036504        &      -24.067024       &      0.9392 \\
3c       &      64.041556        &      -24.059997       &      3.2900   & 12.2b  &      64.036658        &      -24.067316       &      0.9392 \\
4a       &      64.048126        &      -24.066957       &      2.1070   & 12.2c  &      64.036592        &      -24.067231       &      0.9392 \\
4b       &      64.047468        &      -24.066039       &      2.1070   & 12.3b  &      64.036567        &      -24.067368       &      0.9392 \\
4c       &      64.042209        &      -24.060541       &      2.1070   & 12.3c  &      64.036496        &      -24.067272       &      0.9392 \\
5.1a     &      64.047367        &      -24.068671       &      1.8961   & 12.4b  &      64.036283        &      -24.067485       &      0.9392 \\
5.1b     &      64.043479        &      -24.063523       &      1.8961   & 12.4c  &      64.036267        &      -24.067462       &      0.9392 \\
5.1c     &      64.040783        &      -24.061609       &      1.8961   & 12.5b  &      64.036904        &      -24.067201       &      0.9392 \\
5.2c     &      64.041083        &      -24.061802       &      1.8961   & 12.5c  &      64.036833        &      -24.067101       &      0.9392 \\
5.2a     &      64.047462        &      -24.068823       &      1.8961   & 12.6b  &      64.036608        &      -24.067572       &      0.9392 \\
5.2b     &      64.043075        &      -24.063084       &      1.8961   & 12.6c  &      64.036292        &      -24.067157       &      0.9392 \\
5.3a     &      64.047483        &      -24.068851       &      1.8961   & 13a    &      64.039245        &      -24.070383       &      1.0055 \\
5.3b     &      64.043021        &      -24.063021       &      1.8961   & 13b    &      64.038301        &      -24.069728       &      1.0055 \\
5.3c     &      64.041162        &      -24.061848       &      1.8961   & 13c    &      64.034234        &      -24.066016       &      1.0055 \\
5.4a     &      64.047583        &      -24.068884       &      1.8961   & 14.1a  &      64.034492        &      -24.066956       &      3.2226 \\
5.4b     &      64.042908        &      -24.062865       &      1.8961   & 14.1b  &      64.034188        &      -24.066485       &      3.2226 \\
5.4c     &      64.041479        &      -24.061979       &      1.8961   & 14.1c  &      64.034000        &      -24.066439       &      3.2226 \\
5.5a     &      64.047650        &      -24.068971       &      1.8961   & 14.1d  &      64.033967        &      -24.066901       &      3.2226 \\
5.5b     &      64.042762        &      -24.062771       &      1.8961   & 14.1e  &      64.035171        &      -24.067919       &      3.2226 \\
5.5c     &      64.041704        &      -24.062128       &      1.8961   & 14.1f  &      64.046063        &      -24.076785       &      3.2226 \\
5.6a     &      64.047737        &      -24.069012       &      1.8961   & 14.2a  &      64.034467        &      -24.066860       &      3.2226 \\
5.6b     &      64.042571        &      -24.062628       &      1.8961   & 14.2b  &      64.034304        &      -24.066543       &      3.2226 \\
5.6c     &      64.042071        &      -24.062319       &      1.8961   & 14.2c  &      64.033900        &      -24.066493       &      3.2226 \\
6a       &      64.047808        &      -24.070164       &      3.6070   & 15.1a  &      64.041804        &      -24.075731       &      1.9904 \\
6b       &      64.043657        &      -24.064401       &      3.6070   & 15.2a  &      64.041833        &      -24.075826       &      1.9904 \\
6c       &      64.037676        &      -24.060756       &      3.6070   & 15.4a  &      64.042096        &      -24.075976       &      1.9904 \\
7a       &      64.047098        &      -24.071105       &      2.0850   & 15.1b  &      64.035250        &      -24.070988       &      1.9904 \\
7b       &      64.040664        &      -24.063586       &      2.0850   & 15.1c  &      64.030771        &      -24.067126       &      1.9904 \\
7c       &      64.039795        &      -24.063081       &      2.0850   & 15.2b  &      64.035171        &      -24.071002       &      1.9904 \\
8a       &      64.044624        &      -24.071488       &      2.2820   & 15.2c  &      64.030771        &      -24.067217       &      1.9904 \\
8b       &      64.040485        &      -24.066330       &      2.2820   & 15.4b  &      64.035008        &      -24.070843       &      1.9904 \\
8c       &      64.034256        &      -24.062997       &      2.2820   & 15.4c  &      64.030996        &      -24.067308       &      1.9904 \\
9.1a     &      64.045104        &      -24.072345       &      3.2899   & 16.1a  &      64.033596        &      -24.069500       &      2.0955 \\
9.1b     &      64.040079        &      -24.066738       &      3.2899   & 16.2a  &      64.033525        &      -24.069446       &      2.0955 \\
9.2a     &      64.045350        &      -24.072512       &      3.2899   & 16.1b  &      64.032600        &      -24.068616       &      2.0955 \\
\hline
\end{tabular}
\tablefoot{The identifiers (with "a,b,c..." indicating the corresponding multiple images), the observed positions RA, DEC and redshift are listed.}
\end{table*}

\begin{table*}[h]
\caption{List of confirmed multiple images used in the lens model of \citet{bergamini20}.}             
\label{tab_MI2}      
\centering          
\begin{tabular}{c c c c | c c c c}     
\hline\hline       
ID & RA & DEC & Redshift & ID & RA & DEC & Redshift \\ 
\hline    
16.1c    &      64.032446        &      -24.068435       &      2.0955   & 32c    &      64.022988        &      -24.077265       &      5.3691 \\
16.2b    &      64.032650        &      -24.068659       &      2.0955   & 33a    &      64.032017        &      -24.084230       &      5.9730 \\
16.2c    &      64.032413        &      -24.068414       &      2.0955   & 33b    &      64.030821        &      -24.083697       &      5.9730 \\
17.1a    &      64.040496        &      -24.078397       &      3.9680   & 34b    &      64.027632        &      -24.082609       &      3.9246 \\
17.1b    &      64.035108        &      -24.073855       &      3.9680   & 34c    &      64.023731        &      -24.078477       &      3.9246 \\
17.1c$^{\star}$  &      64.027163        &      -24.068238       &      3.9680   & 35a    &      64.033729        &      -24.085702       &      5.6417 \\
18a      &      64.040177        &      -24.079872       &      3.8734   & 35b    &      64.028662        &      -24.084216       &      5.6417 \\
18b      &      64.033937        &      -24.074565       &      3.8734   & 35c    &      64.022125        &      -24.077279       &      5.6417 \\
19a      &      64.040129        &      -24.080313       &      4.1030   & 36a    &      64.031614        &      -24.085762       &      1.9626 \\
19b      &      64.033667        &      -24.074766       &      4.1030   & 36b    &      64.028339        &      -24.084553       &      1.9626 \\
19c      &      64.026596        &      -24.070494       &      4.1030   & 36c    &      64.024074        &      -24.080895       &      1.9626 \\
20.1a    &      64.040350        &      -24.081474       &      3.2190   & 37a    &      64.029809        &      -24.086363       &      2.2196 \\
20.1b    &      64.032162        &      -24.075098       &      3.2190   & 37b    &      64.028610        &      -24.085973       &      2.2196 \\
20.1c    &      64.027571        &      -24.072671       &      3.2190   & 37c    &      64.023345        &      -24.081580       &      2.2196 \\
20.3a    &      64.040325        &      -24.081228       &      3.2190   & 101a   &      64.048082        &      -24.074314       &      4.3029 \\
20.3c    &      64.027454        &      -24.072209       &      3.2190   & 101b   &      64.039685        &      -24.064269       &      4.3029 \\
21b      &      64.030775        &      -24.074169       &      5.1093   & 101c   &      64.036549        &      -24.063271       &      4.3029 \\
21c      &      64.029292        &      -24.073327       &      5.1093   & 102a   &      64.048412        &      -24.073606       &      6.0680 \\
23a      &      64.035668        &      -24.079920       &      2.5435   & 102b   &      64.040998        &      -24.064084       &      6.0680 \\
23b      &      64.032638        &      -24.078508       &      2.5435   & 102c   &      64.036405        &      -24.062218       &      6.0680 \\
24a      &      64.035833        &      -24.081321       &      1.6341   & 103a   &      64.048181        &      -24.070890       &      4.1169 \\
24b      &      64.031039        &      -24.078953       &      1.6341   & 103b   &      64.042892        &      -24.063898       &      4.1169 \\
24c      &      64.026239        &      -24.074337       &      1.6341   & 103c   &      64.037669        &      -24.061026       &      4.1169 \\
25a      &      64.038073        &      -24.082404       &      3.1127   & 104a   &      64.043922        &      -24.075066       &      4.0730 \\
25b      &      64.030366        &      -24.079015       &      3.1127   & 104b   &      64.037232        &      -24.069674       &      4.0730 \\
25c      &      64.025446        &      -24.073648       &      3.1127   & 105a   &      64.046427        &      -24.076733       &      4.0735 \\
26a      &      64.037722        &      -24.082388       &      3.0786   & 105b   &      64.035986        &      -24.067871       &      4.0735 \\
26b      &      64.030484        &      -24.079222       &      3.0786   & 105c   &      64.033727        &      -24.065794       &      4.0735 \\
26c      &      64.025186        &      -24.073575       &      3.0786   & 106a   &      64.047744        &      -24.068648       &      4.1162 \\
27a      &      64.037469        &      -24.083657       &      3.4920   & 106b   &      64.045866        &      -24.065809       &      4.1162 \\
27b      &      64.029409        &      -24.079889       &      3.4920   & 106c   &      64.037746        &      -24.059831       &      4.1162 \\
27c      &      64.024946        &      -24.075021       &      3.4920   & 107a   &      64.046032        &      -24.068796       &      2.9209 \\
28a      &      64.038350        &      -24.084126       &      3.2542   & 107b   &      64.044766        &      -24.066694       &      2.9209 \\
28b      &      64.028322        &      -24.079004       &      3.2542   & 107c   &      64.036203        &      -24.060649       &      2.9209 \\
28c      &      64.026330        &      -24.076705       &      3.2542   & 108a   &      64.046513        &      -24.076163       &      4.6100 \\
29.1a    &      64.036696        &      -24.083910       &      2.2980   & 108b   &      64.036659        &      -24.068027       &      4.6100 \\
29.1b    &      64.028408        &      -24.079743       &      2.2980   & 108c   &      64.033508        &      -24.065017       &      4.6100 \\
29.1c    &      64.026054        &      -24.077252       &      2.2980   & 109a   &      64.043756        &      -24.073669       &      2.9912 \\
30a      &      64.033628        &      -24.083185       &      3.4426   & 109b   &      64.037761        &      -24.068837       &      2.9912 \\
30b      &      64.031251        &      -24.081904       &      3.4426   & 110a   &      64.042733        &      -24.072187       &      4.3008 \\
30c      &      64.022699        &      -24.074595       &      3.4426   & 110b   &      64.039160        &      -24.069769       &      4.3008 \\
31a      &      64.035486        &      -24.084679       &      4.1246   & 112a   &      64.049288        &      -24.070949       &      6.1487 \\
31b      &      64.029234        &      -24.081813       &      4.1246   & 112b   &      64.043300        &      -24.062949       &      6.1487 \\
31c      &      64.023412        &      -24.076125       &      4.1246   & 112c   &      64.038892        &      -24.060640       &      6.1487 \\
32a      &      64.035054        &      -24.085504       &      5.3691   & 113a   &      64.045972        &      -24.074033       &      5.9990 \\
32b      &      64.028403        &      -24.082993       &      5.3691   & 113b   &      64.039850        &      -24.066907       &      5.9990 \\
\hline
\end{tabular}
\tablefoot{Continue from Table~\ref{tab_MI1}.\\
\tablefootmark{$\star$}{The object is not covered by the MUSE field of view, however is confirmed because of the evident parity introduced by strong lensing.}\\
}
\end{table*}

\begin{table*}[h]
\caption{Full VLT/MUSE spectroscopic catalog on HFF~J0416.}             
\label{full_cat}      
\centering          
\begin{tabular}{c c c c c c c c}     
\hline\hline       
ID & RA & DEC & Redshift & QF$^{\star}$ & Multiplicity & error-mult. & ID-name \\ 
\hline
-29  & 64.032157 & -24.075108 & 3.2175 & 3 &     1  &    0.0000  &  MUSE\_20b  \\
-56  & 64.033936 & -24.074583 & 3.8710 & 9 &     1  &    0.0000  &  MUSE\_1327 \\
-59  & 64.027763 & -24.073143 & 3.2175 & 9 &     1  &    0.0000  &  MUSE\_1426 \\
-66  & 64.028374 & -24.082983 & 5.3650 & 9 &     1  &    0.0000  &  MUSE\_643 \\
\hline
\end{tabular}
\tablefoot{This is a portion of the full catalog that is to be published online. The columns from left to right are: ID, RA, DEC (J2000), redshift, quality of the redshift measurement (QF), multiplicity (i.e., the number of independent redshift measures), error among redshift estimates if multiplicity is $>1$, the internal ID (ID-name), and the reference associated to the measured redshift  (ref.), if present.\\
\tablefootmark{$\star$}{QF = 2, likely; QF = 3, secure; QF = 9, single-line.}\\
}
\end{table*}

\section{High-z clumps in the MDLF}
\label{clumps_appendix}

The identification of clumps or sub-structures on confirmed and most magnified high-z galaxies (those producing multiple images, see Appendix~\ref{multiple_appendix}) has been visually performed by looking at HST images and their composite RGB color version. 
Indeed, the search for clumps or sub-structures in lensed high-z galaxies is currently better performed via visual inspection. An example is shown in Figure~\ref{system20}, in which the identification of clumps on source 20c is supported by the identification of a very similar mirrored image on the other side of the galaxy cluster (20a), a task which would have been very difficult to accomplish with automatic tools. 

\begin{figure*}[ht!]
        \centering
        \includegraphics[width=\linewidth]{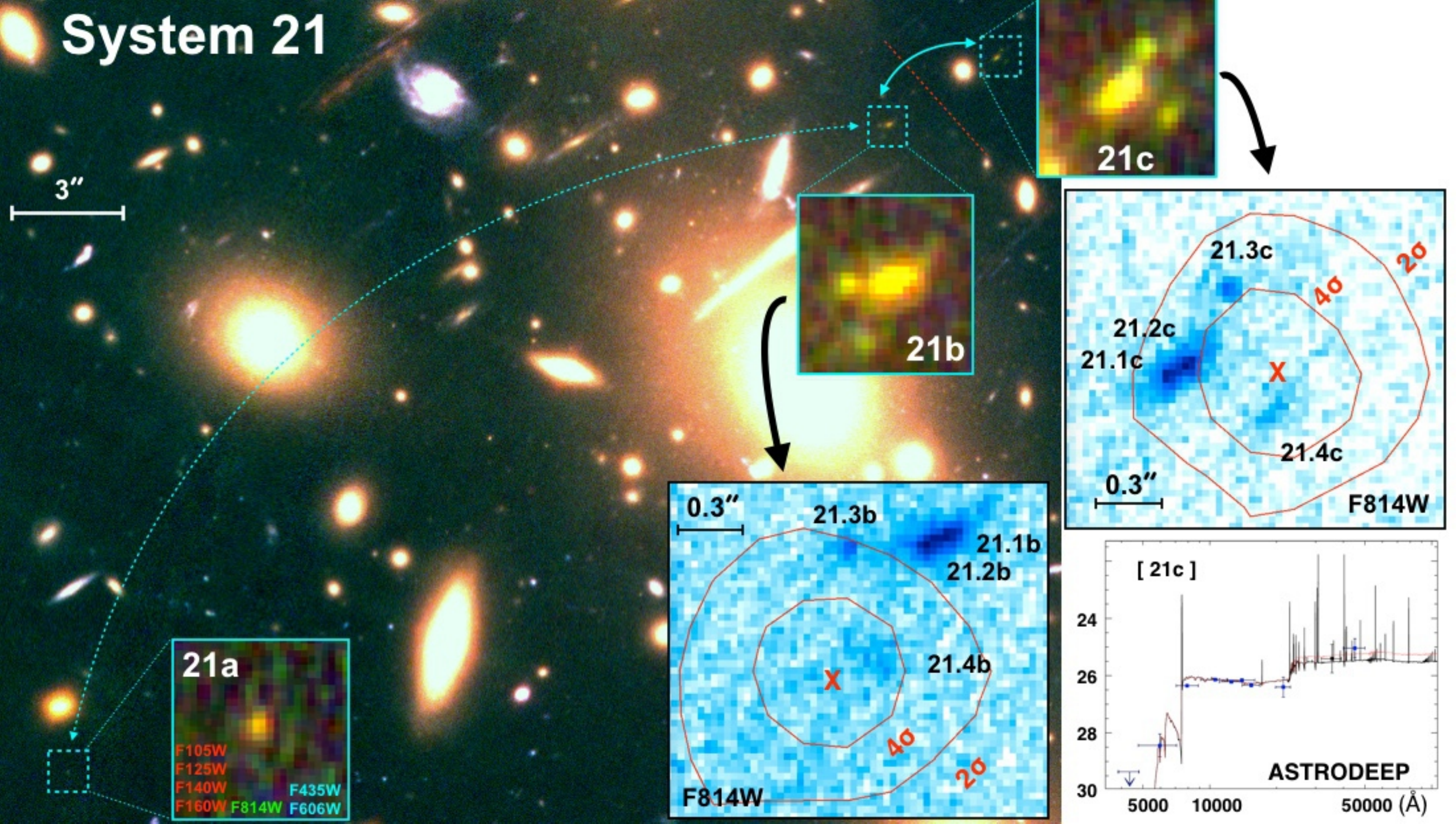}
        \caption{Three images (a, b, c) of source 21 are shown over the color image of the cluster. The insets show the deep RGB composite (F105W+F125W+F140W+F160W, F814W, F435W+F606W) of the multiple images. Image 21a is currently not covered by spectroscopy, however, we rely on the fact it is the only source with photometric redshift ($z_{phot}=5.36$), which is consistent with the spectroscopic one (z=5.109) within an area of $10''\times 10''$ from the predicted position provided by the lens model ($1.2''$ away from the source 21a (ID(ASTRODEEP)=678). On the right, the F814W images (negative blue) with superimposed \lya\ contours (red) at 2 and 4 $\sigma$ are shown with the SED-fitting performed with  ASTRODEEP photometry. The peak of the \lya\ emission is nearly located on top of clump 21.4b and 21.4c.}
        \label{system21}
\end{figure*}

It is worth nothing that even though some of the clumps have inconclusive redshift measurements due to, for instance, the faintness, then the redshift inferred from the brighter parts of the system coupled with the mirroring introduced by strong lensing guarantees that all the clumps belong to the same physical region.
In the example above, without the mirrored image, clump 20.0c would have been identified as isolated and hardly ascribable to the main system, given its different color. Another evident example is shown in Figure~\ref{system21}, where image 21b and 21c on opposite sides of the critical line show the tiny mirrored knots with observed magnitudes fainter than 28.5 (21.3b $\leftrightarrow$ 21.3c and 21.4b $\leftrightarrow$ 21.4c). In particular, 21.3c would have been out of reach also for the MDLF, being the \lya\ emission of the whole system mainly arising from clump 21.4b (or 21.4c). All high-z multiple images have been visually inspected following this approach, assuring that the consistency among multiple images is guarantee (where present), and ultimately validated by the lens model \citep[][]{bergamini20}. 

Figures~B1$-$B11 show all 116 individual clumps or tiny sources identified following the procedure described above. 
For each clump, the HST cutout in the F814W(F105W) band at $z<5.2(z>5.2)$ is shown with reported the ID ad other relevant quantities, such as the total magnification, redshift, absolute magnitude and the physical scale along the maximum stretch (parsec/pixel). A rough guess of the de-lensed stellar mass and star formation rate are also reported by assuming an instantaneous burst and fixed age of 10 Myr, based on starburst99 models \citep[][]{leitherer14}. 
We defer the reader to a dedicated work on this subject.
Together with each cutout, the one-dimensional MUSE spectra zoomed on the most relevant atomic transitions are shown, extracted from circular apertures of $0\farcs8$ diameter. The inset within each of them shows the continuum-subtracted MUSE narrow-band images of the same relevant line, collapsed over 20~\AA\ for \lya\ and 10~\AA\ for the other lines.

\begin{figure*}
\centering
\includegraphics[width=15.0cm]{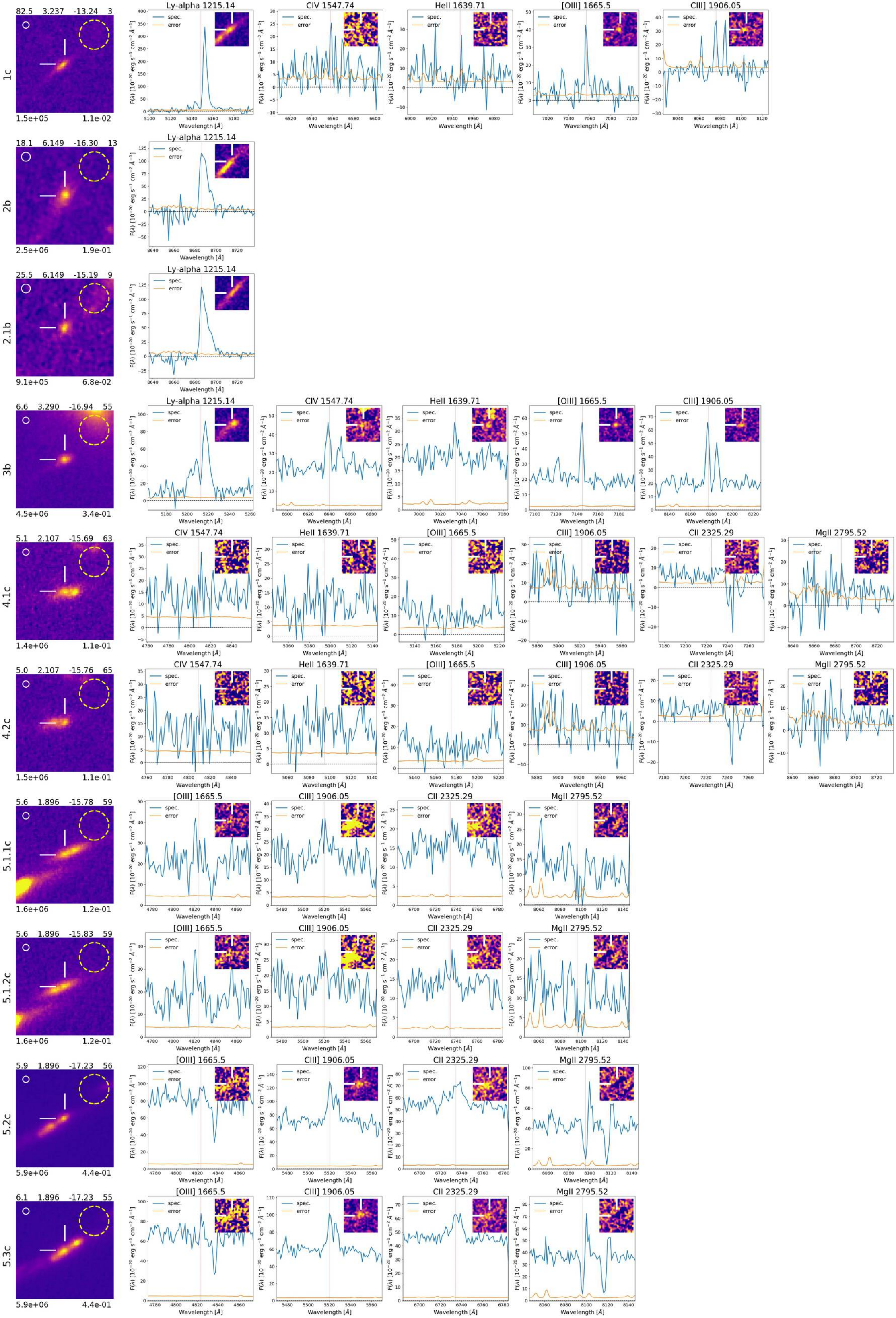}
\caption{Summary of the individual star-forming clumps identified among the multiple images. Each row reports from left to right the square HST cutout of $1''$ size and the extracted one-dimensional spectra zoomed at the positions of the most relevant atomic transitions probed by MUSE. The HST cutout  shows the clump ID on the left, and clockwise, the magnification, redshift, absolute magnitude, parsec per 30 milliarcsecond (1 HST pix), the SFR (in units of M$_{\odot}~yr^{-1}$), and the stellar mass M (M$_{\odot}$) (see text for details about the calculation of SFR and M). The circle in the top-right reports the F814W(F105W) PSF in case of $z<5.2(>5.2)$, while the yellow dashed indicates the MUSE PSF. MUSE spectra (blue) and error (orange) are shown for the most relevant lines depending on redshift. The insets are the corresponding MUSE NB images of the lines (as described in Figure~\ref{MI1}).}
\label{MI_3}
\end{figure*}

\begin{figure*}
\centering
\includegraphics[width=15.0cm]{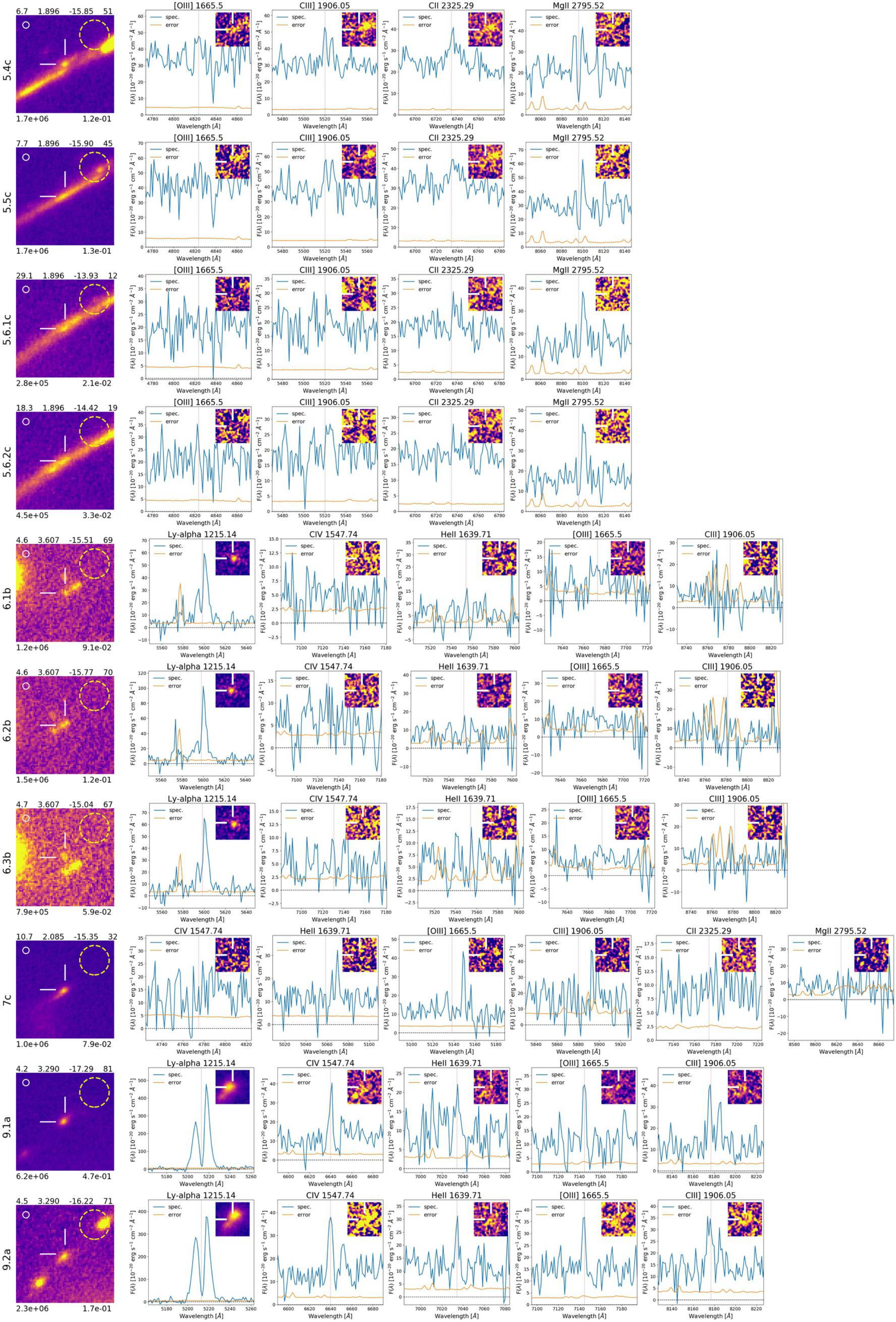}
\caption{Continued.}
\label{MI_3}
\end{figure*}

\begin{figure*}
\centering
\includegraphics[width=14.0cm]{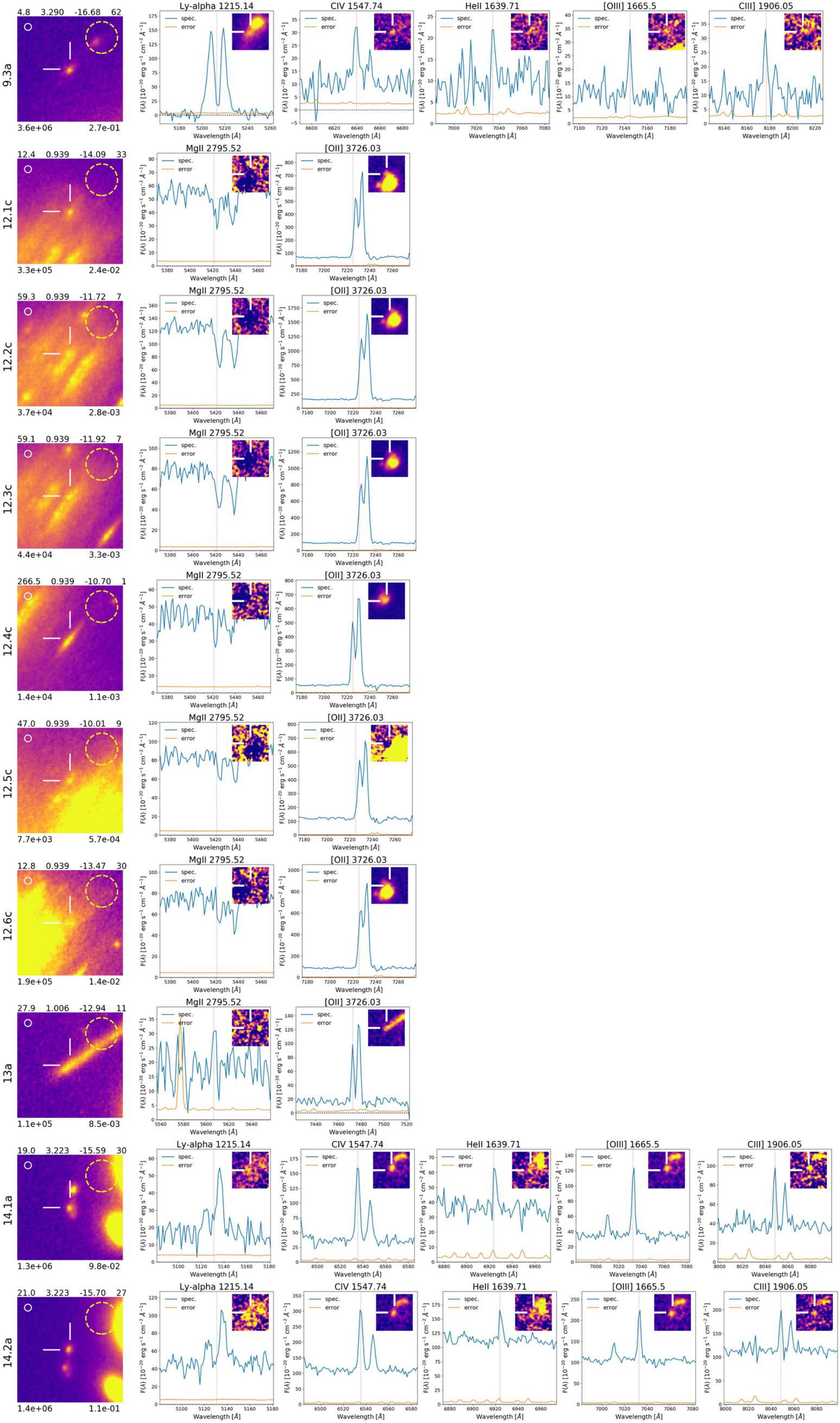}
\caption{Continue.}
\label{MI_3}
\end{figure*}

\begin{figure*}
\centering
\includegraphics[width=16.5cm]{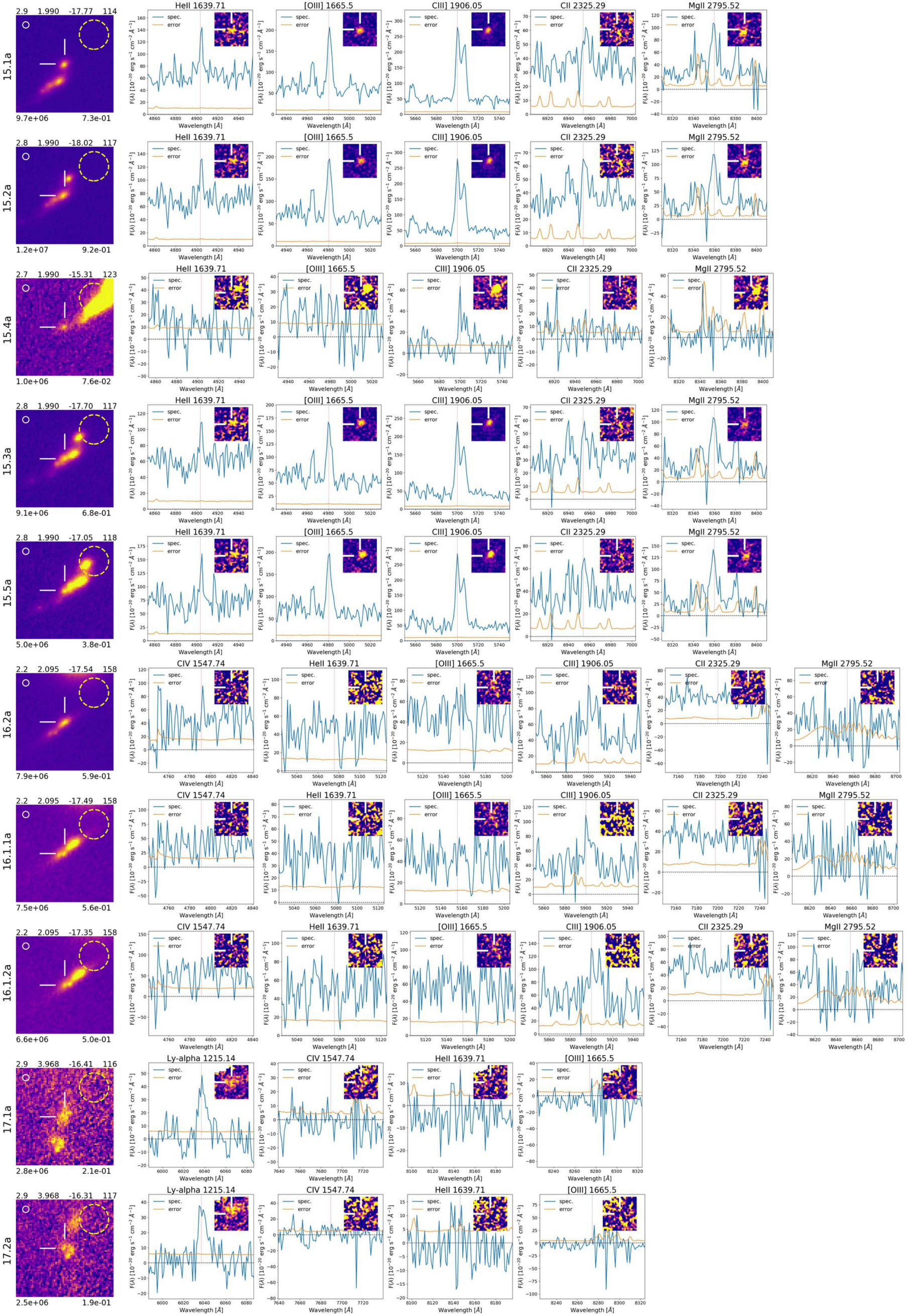}
\caption{Continue.}
\label{MI_3}
\end{figure*}

\begin{figure*}
\centering
\includegraphics[width=14.0cm]{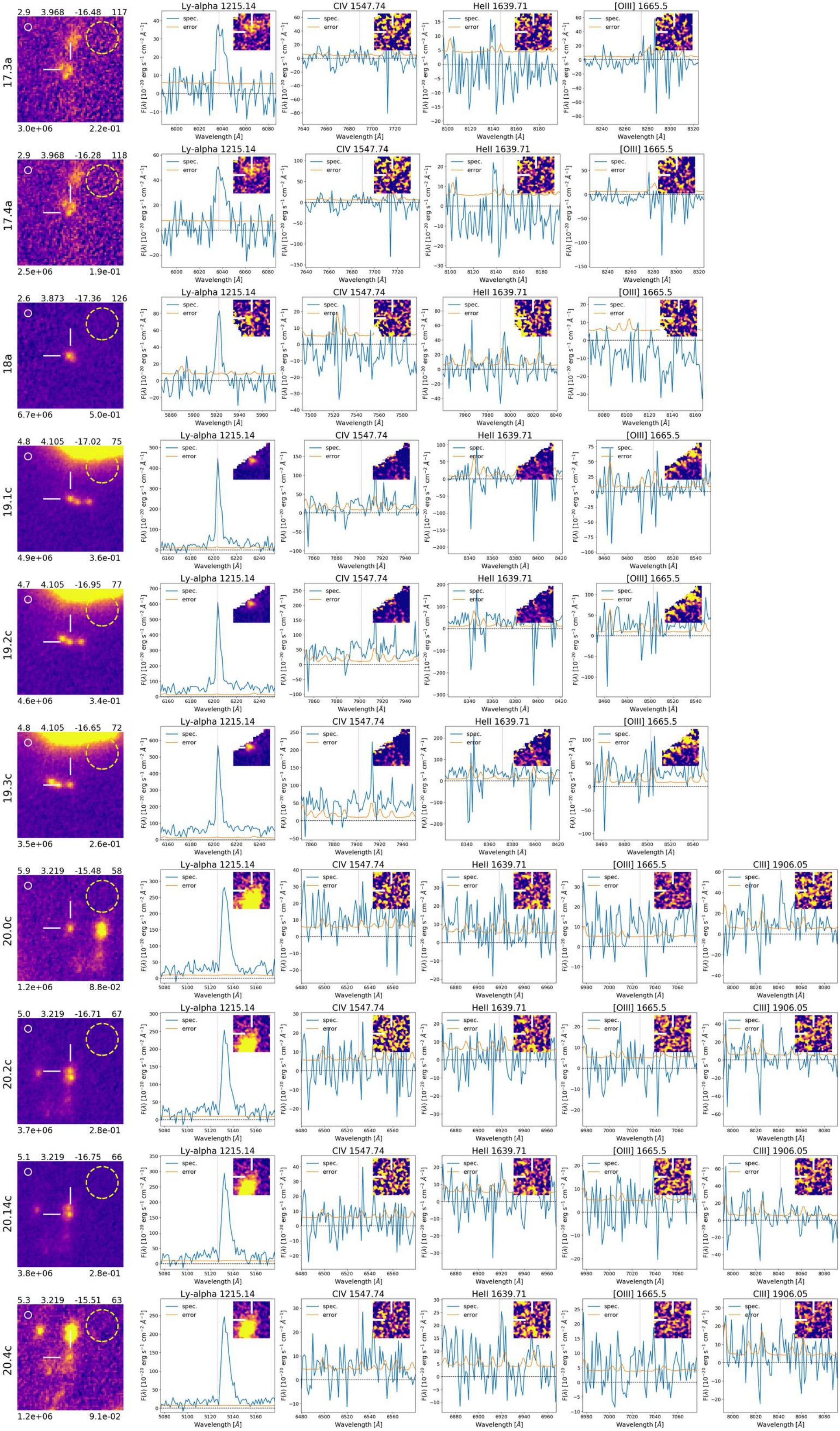}
\caption{Continued.}
\label{MI_3}
\end{figure*}

\begin{figure*}
\centering
\includegraphics[width=14.0cm]{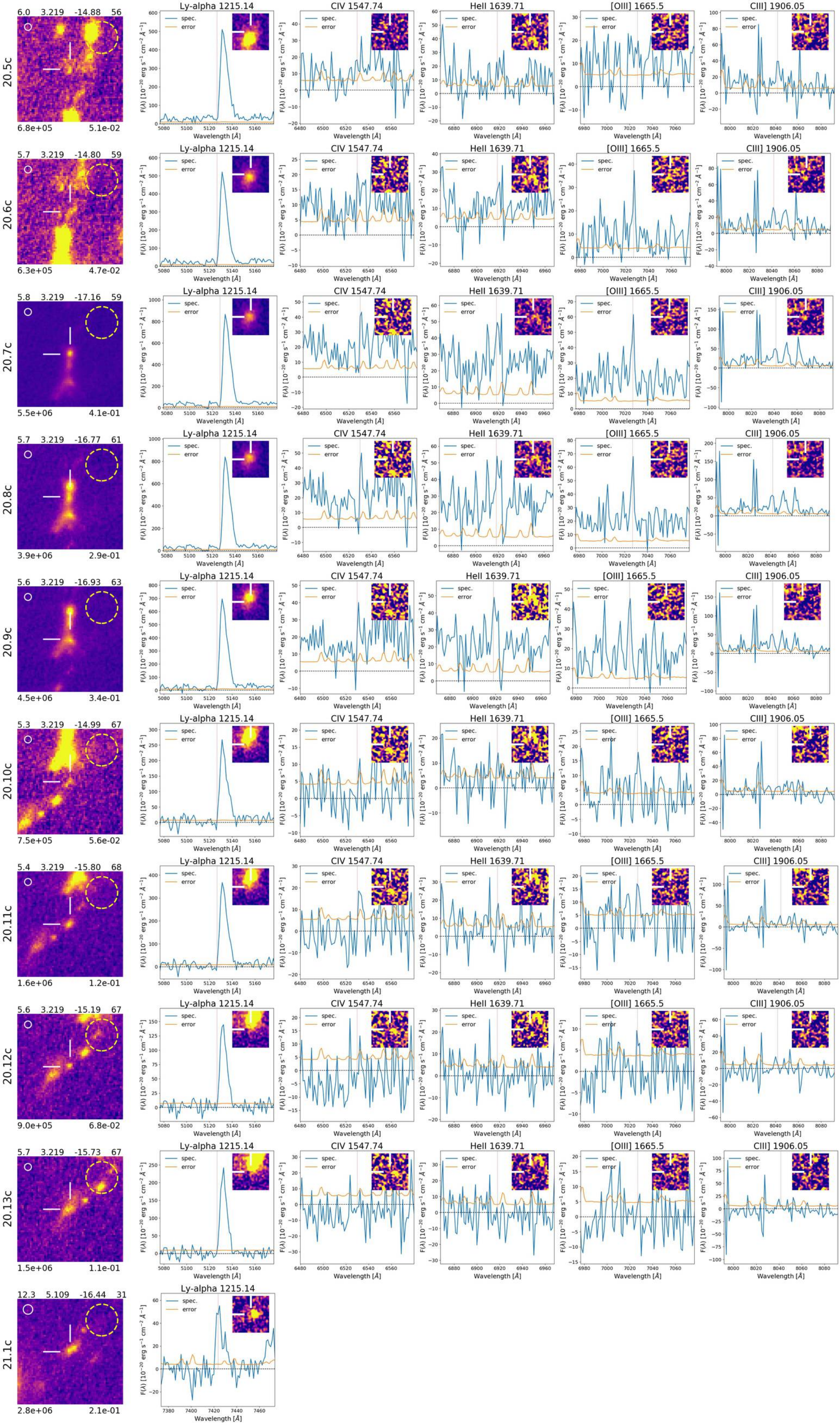}
\caption{Continued.}
\label{MI_3}
\end{figure*}

\begin{figure*}
\centering
\includegraphics[width=16.0cm]{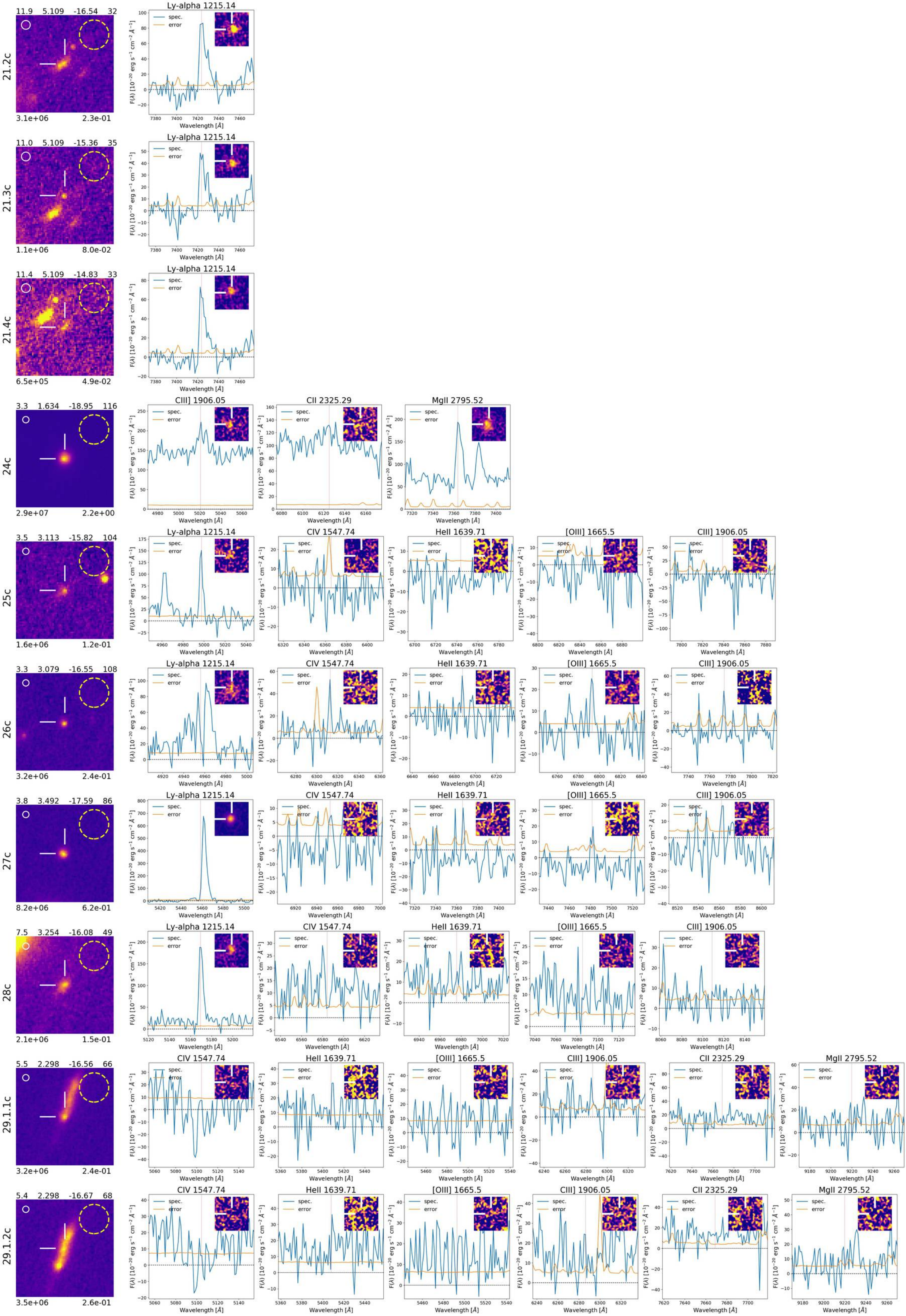}
\caption{Continued.}
\label{MI_3}
\end{figure*}

\begin{figure*}
\centering
\includegraphics[width=16.3cm]{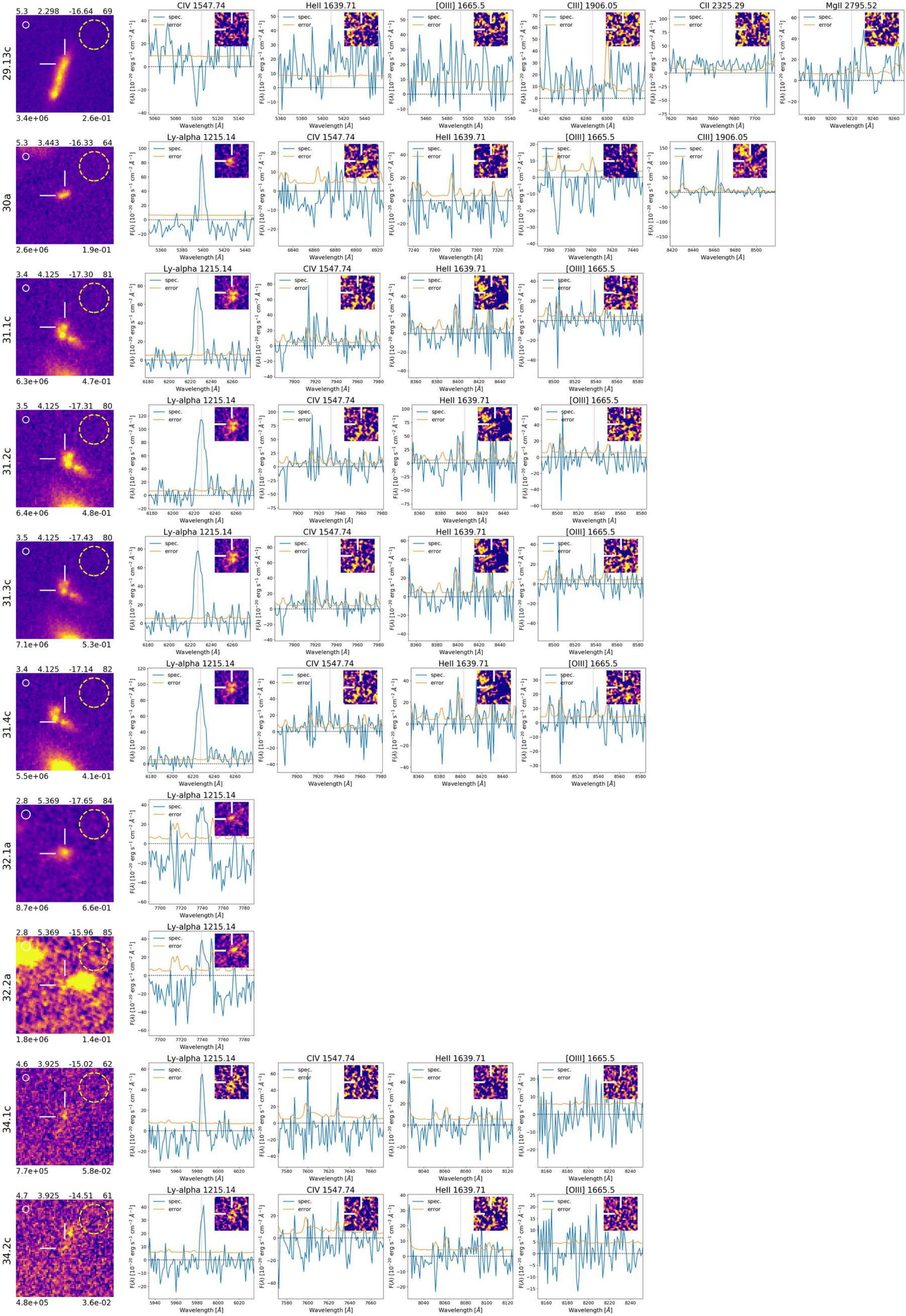}
\caption{Continued.}
\label{MI_3}
\end{figure*}

\begin{figure*}
\centering
\includegraphics[width=16.0cm]{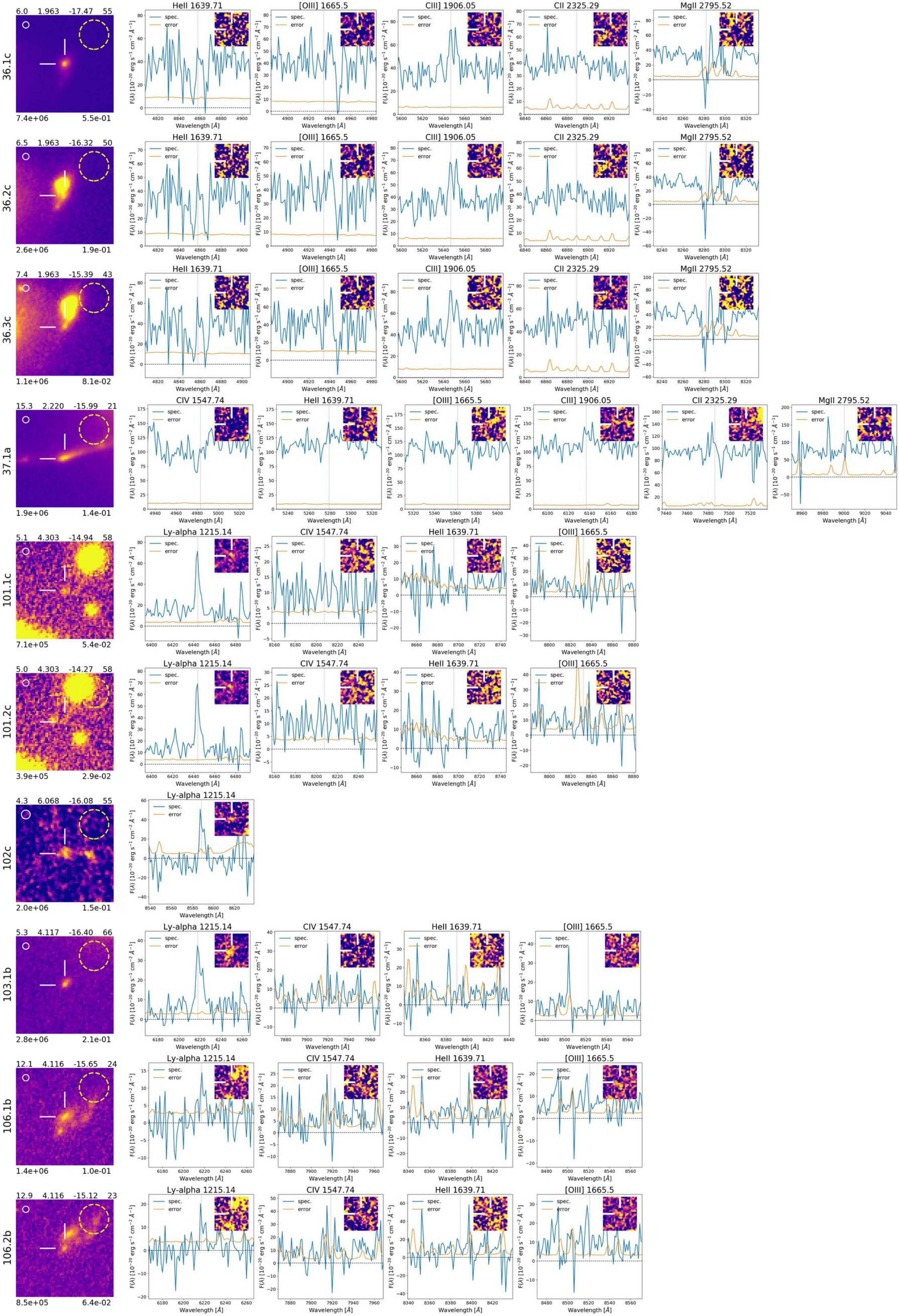}
\caption{Continued.}
\label{MI_3}
\end{figure*}

\begin{figure*}
\centering
\includegraphics[width=16.5cm]{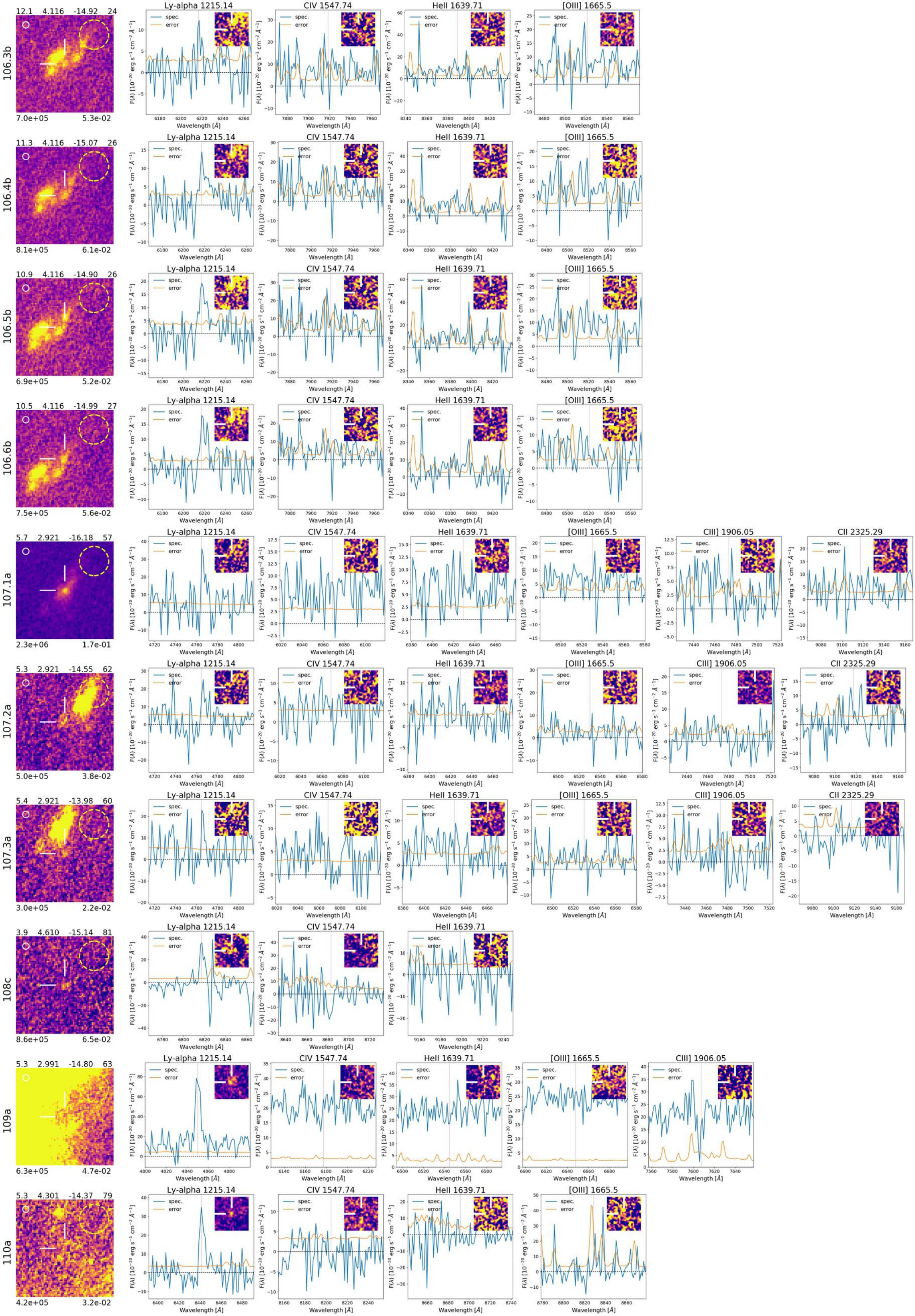}
\caption{Continued.}
\label{MI_3}
\end{figure*}

\begin{figure*}
\centering
\includegraphics[width=4.6cm]{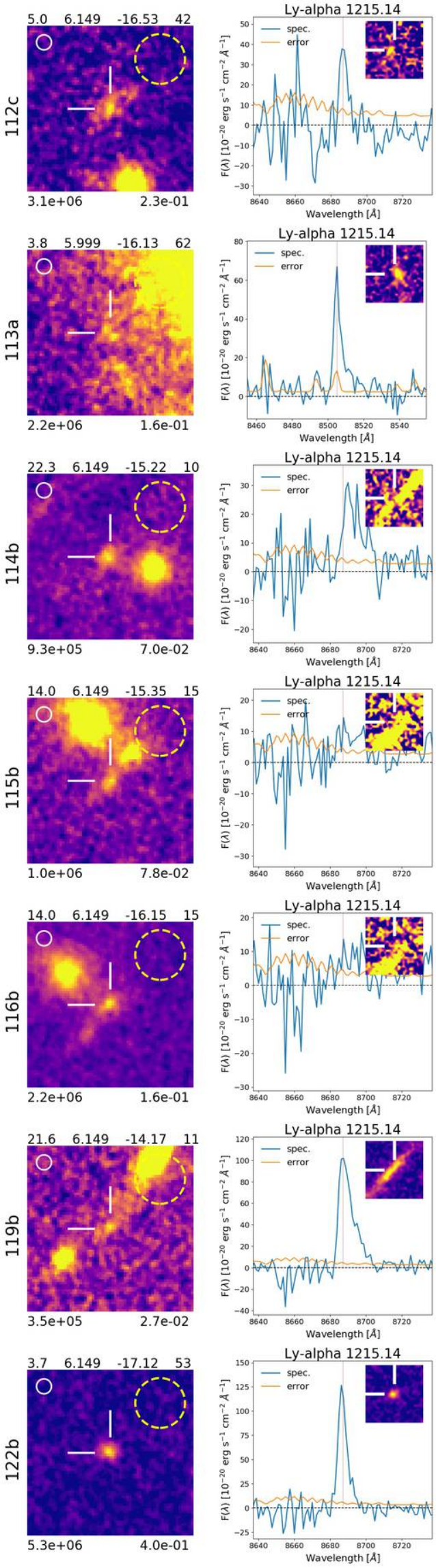}
\caption{Continued.}
\label{MI_3}
\end{figure*}

\section{Revisiting magnification for object 14 at z=3.223}
\label{individual14}

\begin{figure*}
        \centering
        \includegraphics[width=\linewidth]{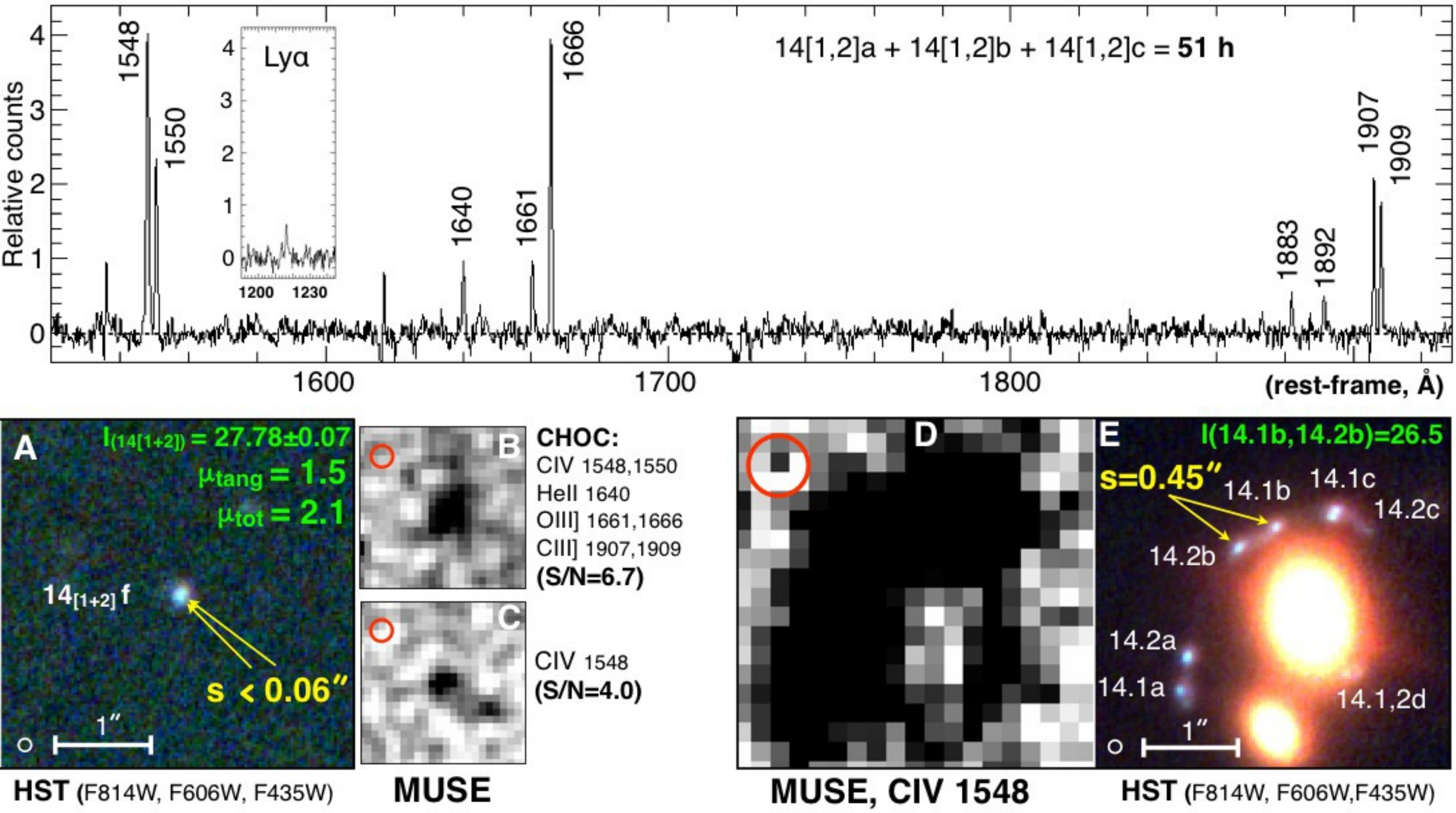}
        \caption{Revisited analysis of the double lensed source 14, made of a pair of knots, at the depth of MDLF. The top panel shows the continuum-subtracted one-dimensional spectrum of the pair obtained from the sum of the three multiple images 14a + 14b + 14c, equivalent to $\simeq 51$ hours of integration. High-ionization lines are detected at S/N~$\sim 10-50$, while the inset shows the deficient \lya\ emission (the same scale on Y-axis is adopted), $\sim 8$ times fainter than \civalone\ $\lambda 1548$, or $\sim 4$ times fainter than \ciiiblue. The bottom-left panel A shows the HST color image (red = F105W, green=F814W, blue=F606W) for the least magnified image (image 14$_{[1+2]}$f), where the upper limit on the separation between knots 1 and 2 is quoted ($s < 0.06''$). The panels C and B show the narrow-band (NB) continuum-subtracted MUSE images of the same 14$_{[1+2]}$f, centered at the \civalone\ $\lambda 1548$ line and at the weight average of seven lines (the CHOC complex, see text for details). The NB images have been smoothed with a Gaussian kernel ($\sigma=1$~pix). The red circles indicate the MUSE PSF of $0\farcs6$. The same HST color image for the most magnified images 14.2b and 14.1b is shown in the bottom-right E panel, while the corresponding \civalone\ $\lambda 1548$ MUSE narrow-band image is in panel D. The small open white circle shown in the HST cutouts marks the F814W PSF.
        }
        \label{fig_id14}
\end{figure*}

Source 14 is split into six multiple images that with the MDLF have now all been confirmed spectroscopically. The new lens model of \citet{bergamini20} nicely reproduces the positions of all of them. Here we revisit the magnification initially reported by \citet{vanz_id14} and based on the observed flux ratios among the images involved. Indeed, the observed flux ratio among the least and the most magnified images provide a guess of the relative magnification among them (see Appendix~\ref{muerr}). Under the condition the measured flux is not affected by systematics, it is possible to rescale the more stable predicted magnification of the least magnified image by the observed flux ratio to obtain the magnification of the most magnified one. \citet{vanz_id14} did this calculation starting from the least magnified image of source 14 that, however, at that time was not spectroscopically confirmed and was mainly based on the photometric redshift identification (ASTRODEEP, ID=1141 with magnitude F814=29.1). With the MDLF we now confirmed the sixth image f as ID=1127, with F814W=$27.78\pm0.07$. The previous wrong identification of image 14f lead to significant overestimations of the magnification value ($\mu \simeq 40$). The inclusion of the 
correct image 14f (ID=1127) and after properly computing the rescaling and propagation errors (flux ratio $7.29 \pm 0.83$ and $\mu(14f)=2.08\pm0.02$), the new value turns out to be $\mu_{tot}=15.2\pm1.7$ for image 14.1b (and similarly for 14.2b). Such a magnification is in line with the estimate provided by the lens model, $\mu_{tot}=19.4^{+11}_{-5.9}$ (see Table~\ref{ratios} and Figure~\ref{fig_id14}). This fact highlights the importance of having spectroscopically confirmed multiple images.

It is worth mentioning that image ID14f has been confirmed through the detection of \civalone\ $\lambda 1548$ at S/N~$\approx4$, while \lya\ is deficient in this source (as shown in Figure~\ref{fig_id14} and discussed by \citealt{vanz_id14}). Figure~\ref{fig_id14} shows such \civalone\ detection for image 14f (the least magnified) and image 14.(1,2)[a,b,c] (the most magnified). For the faintest one (14f), we also calculate the continuum-subtracted weight average of seven narrow-band MUSE images (with $dv = 200$ \kms) centered at the position of the ultraviolet transitions \civ, \heii, \oiiiuv,\ and \ciiidoub, in which each doublet is resolved and $-$ following the wavelength order $-$ arise from the Carbon, Helium, Oxygen, and Carbon complex (CHOC, 1548, 1550, 1640, 1661, 1666, 1907, 1909), respectively. The weights (that follow the relative line ratios) have been extracted from the mean stacked spectrum reported in Sect.~\ref{stack}. The CHOC ultraviolet signature is detected at S/N=6.7 for image 14f an reaches its peak emission at the systemic redshift (see Figure~\ref{individual14}). As discussed in \citet{vanz_id14} and shown in Figure~\ref{fig_id14}, there is a deficiency of \lya\ emission in source 14 (at variance from the typical 
positive correlation among \lya\ equivalent width and ultraviolet CHOC nebular lines, e.g., \citealt{feltre20}).
Therefore, without the availability of the rest-frame optical lines (e.g., \oiidoub, \oiiidoub, \hb, \ha) the redshift confirmation for this kind of \lya-deficient sources 
is left to high-ionization lines, which, if present (as in the case of source 14 discussed above), can be properly combined to gain in depth through the UV CHOC complex indicator. 

From the comparison between the new image 14f and the pair 14.1b-14.2b, it is also possible to set a rough lower limit on the tangential stretch the most magnified images are subjected, 14.1b,14.2b (or 14.1a, 14.2a). From the lens model of \citet{bergamini20}, the tangential magnification on image 14f is $\mu_{T}=1.5\pm0.05$ with a small error, being far from the critical lines. Such a value is still too low to make the two knots spatially resolvable with HST imaging (Figure~\ref{fig_id14}), implying that the upper limit on the separation among the two on image 14f is not larger than the HWHM (i.e., with a separation of $s_{f}<0.06''$). On the other hand, the pair 14b (14.1 and 14.2) are well separated by $0.45''\pm0.03$, suggesting that the relative tangential magnification is $>7.5$, and that the tangential one for images 14b is $\mu_{T}(14b) \simeq \mu_{T}(14f) \times 7.5 \gtrsim 11$. This value is in line with the value provided by the lens model ($7-20$), that, however, is affected by large uncertainties due to the proximity of the objects to the critical lines. Adopting the above estimate of $\mu_{T} > 11$ and the effective radius of $0.045''$ ($1.5\pm0.5$ pix) as estimated in \citet{vanz_id14}, the sizes of each knot of the pair is plausibly smaller than 30 pc, while the two are separated by $\sim 390$ pc in the source plane \citep[][]{bergamini20}).

\section{Spectral stacking}
\label{stack_appendix}

\begin{figure*}
        \centering
        \includegraphics[width=\linewidth]{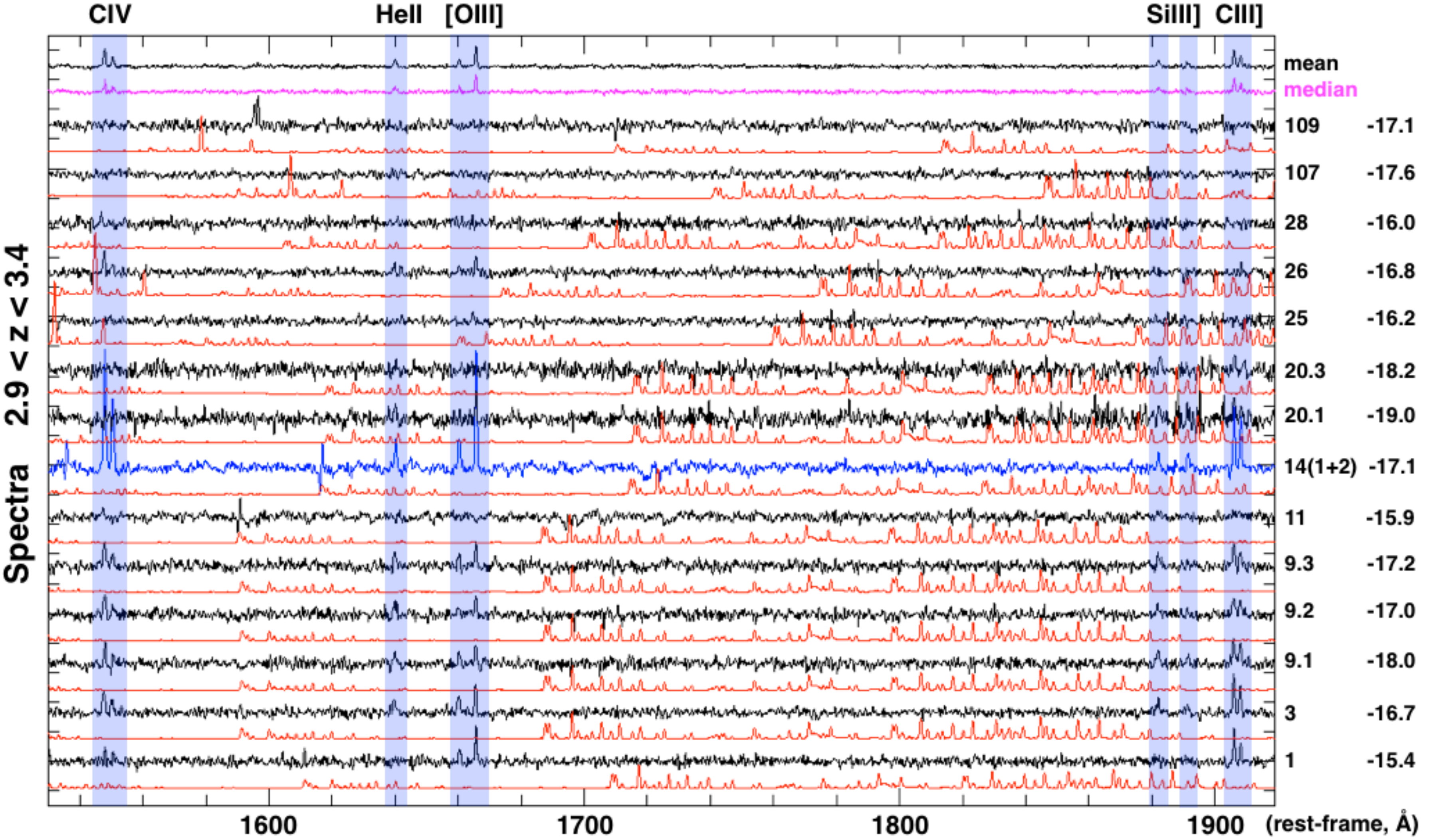}
        \caption{One-dimensional spectra (black lines) of a subset of sources with $2.9<z<3.4$ and average absolute magnitude M$_{\rm UV}=-17,$  shown versus the the rest-frame wavelength. Each spectrum is the weighted average of multiple images eventually producing net integration time of $17.1-51.3$ hours each. The ID and absolute magnitudes are reported on the rightmost two columns. The red lines show the error spectra associated to each spectrum, properly scaled and shifted for clarity below each black line. The red spectra show the pattern of the skylines. The blue spectrum indicates source 14, in which the high-ionization lines are prominently detected (see Figure~\ref{comparison_id14} and relative caption for details). The two spectra on the top panel are the mean (black) and median (magenta) of the sample included in this figure. The vertical transparent stripes mark the location of the typical high-ionization lines (labeled on the top axis).}
        \label{spectra_faint}
\end{figure*}

Section~\ref{stack.section} presents the stacking of continuum-subtracted spectra and the detection maps (continuum-subtracted S/N spectra). 
We select here a sub-sample of sources in the redshift range of $2.9<z<3.4$ (14 entries, $<z>=3.2$) such that the complex of lines \civ, \heii, \oiiiuv,\ and \ciiidoub\ lies in the deeper wavelength interval probed by MUSE ($6000-8300$~\AA, see Figure~\ref{rms}), also avoiding the crowded region of intense sky lines ($\lambda > 8300$~\AA). The spectra of the selected sample have integration time ranging from 17.1 to 51 hours depending on the presence of (usable) multiple images. Figure~\ref{spectra_faint} shows individual and stacked spectra. Among the 14 sources with absolute magnitude ranging between $-15.4$ and $-19$ (with a median of $-17.0$), more than 50\% show high-ionization lines with S/N ratios $>3$. Source 14 is the emitter with the most prominent lines detected with S/N ratio exceeding 50 (Figure~\ref{spectra_faint} and~\ref{fig_id14}). The mean and median stacks show evident nebular emission lines, all of them well detected with S/N $>10$.

\begin{figure*}
        \centering
        \includegraphics[width=\linewidth]{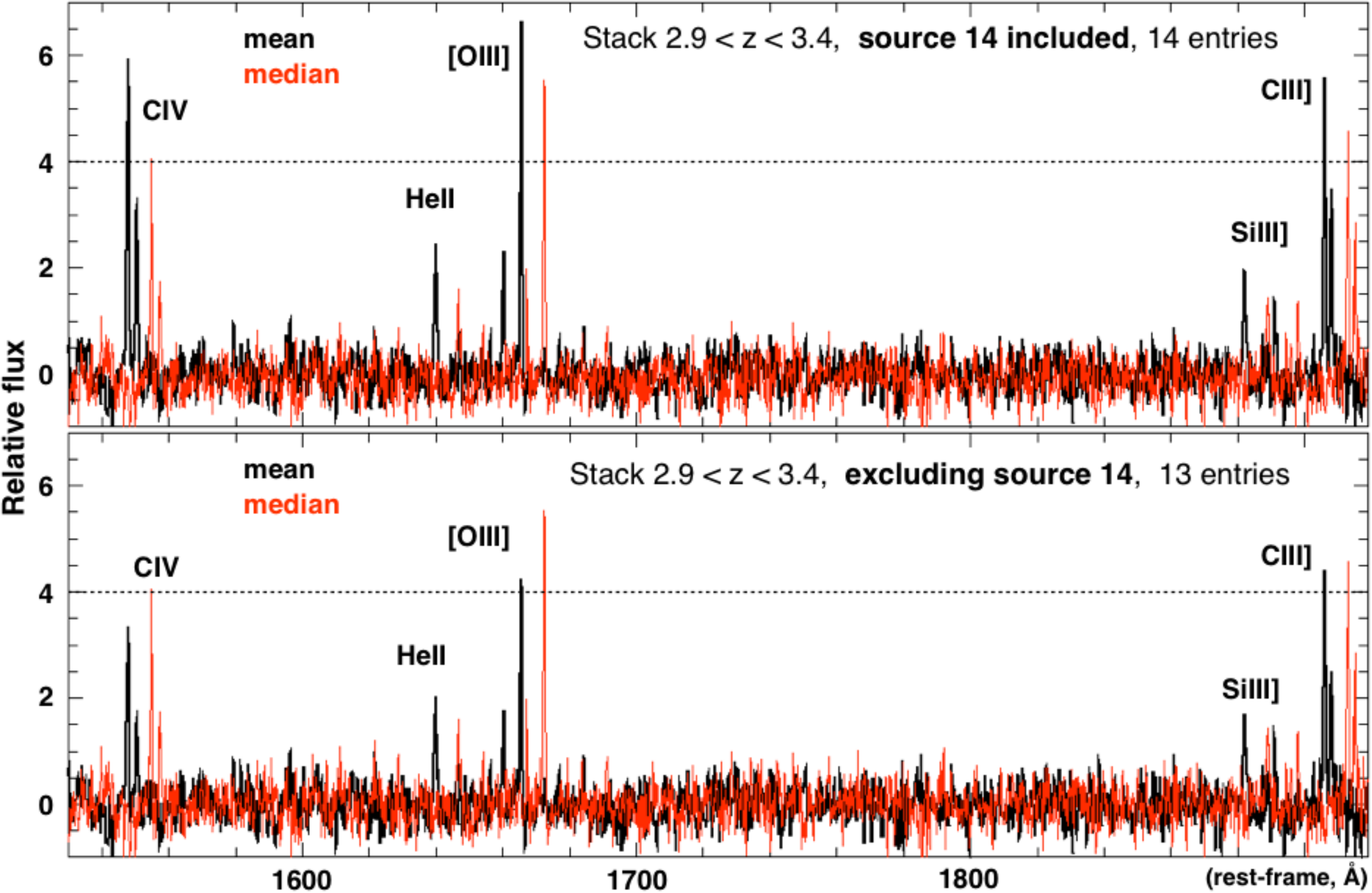}
        \caption{Mean (black line) and median (red line) stacked detection maps calculated from the set of spectra reported in Figure~\ref{spectra_faint}. The zoomed region including \civ, \heii, \oiiiuv,\ and \ciiidoub\ (from left to right) is shown. In the top panel, all sources are included (14 entries), while in the bottom panel source 14 is excluded, that is the source showing prominent nebular high-ionization lines (see Figure~\ref{fig_id14}, spectrum in blue; see also Figure~\ref{spectra_faint}). The median stacks (red lines) are rigidly redshifted by a fixed quantity for illustrative purposes only.}
        \label{comparison_id14}
\end{figure*}

The emission lines of the mean stack show values at the peak systematically higher than the median stack. The presence of source 14 with the highest S/N significantly affects the resulting average. This is shown in Figure~\ref{comparison_id14} where the mean and median detection maps (i.e., stacked continuum-subtracted and inversely weighted by their errors) are shown for all sources ($2.9<z<3.4$, 14 entries) and after excluding only source 14. In the case where source 14 is excluded, the mean and median results are fully compatible, while the inclusion of source 14 boosts the mean. Overall, the presence of high-ionization lines detected on individual spectra and in the median stack (with or without source 14) show that at faint luminosity regimes ($-15 < $~M$_{\rm UV}<-18$), the occurrence of nebular high-ionization lines appears to be common.

\section{Magnification from relative flux ratios or angular separations: individual cases}
\label{muerr}

We focus in this section on the magnification uncertainty of a subset of sources, specifically those discussed in the main text. 
Such analysis is not complete, however it is included here to describe 
a key method to overcome the large magnification uncertainty in the most magnified cases, where systematic uncertainties in the lens models may dominate. While statistical errors on magnifications have been discussed in \citet{bergamini20}, a complementary and more robust method 
exploits flux ratios (or relative angular separations between clumps) among multiple images of the same family to estimate relative magnifications \citep[][]{vanz16_id11,vanz_id14}.
This is based on the following assumptions: (1) the multiple images do not cross or intercept the caustics on the source plane so that individual lensed images are produced, and (2) the images are well-detected and the inferred magnitudes free from significant contamination.

Such a magnification ratio can be rescaled to the magnification of the least magnified image, which typically has $\mu < 5$ and is far from the critical lines and, therefore, subject to small uncertainties from pure model prediction ($<20\%$, \citealt{Meneghetti_2017}). Under these conditions, the magnification of a lensed object subject to a complex geometry (e.g., close to the critical line) can be recovered with a relatively low uncertainty by propagating the error on $\mu$ of the least magnified image and the uncertainty associated to the photometry (flux ratios). 

A challenging object discussed in this work is source 14 (see Sect.~\ref{14} and Appendix~\ref{individual14}), with magnification larger than 10 which arises from a complex lens geometry. In this case, the MDLF allowed us to confirm the redshift of the least magnified of the six multiple images, with $\mu=2.08\pm0.02$, and to infer the magnification of the knots belonging to source 14 from the measured flux ratios based on HST photometry. Therefore, the total magnification can be estimated by rescaling the flux, while  the tangential magnification can be estimated by rescaling the relative angular separation (see Appendix~\ref{individual14}). Table~\ref{ratios} reports the results obtained from the application of this method, by comparing the magnification estimated directly from the lens model with the one derived using this method. The agreement in the case of source 14 is within 30\% and in general within 50\% ($\mu_{tot-model}/\mu_{tot-rescale}=0.6\div 1.5$).
Among the sources presented in Table~\ref{ratios}, source 1 is a challenging case, for which this method does not work properly, since all three multiple images are highly magnified, including the least magnified (1a) with $\mu = 56.6 \pm 9.6$, due to its vicinity to the critical line. 
From pure flux rescaling the inferred $\mu(1c)$ is $\sim 500$, while the lens model predicts $\mu(1c) = 78.1^{+19.1}_{13.4}$. 

\begin{figure}
        \centering
        \includegraphics[width=\linewidth]{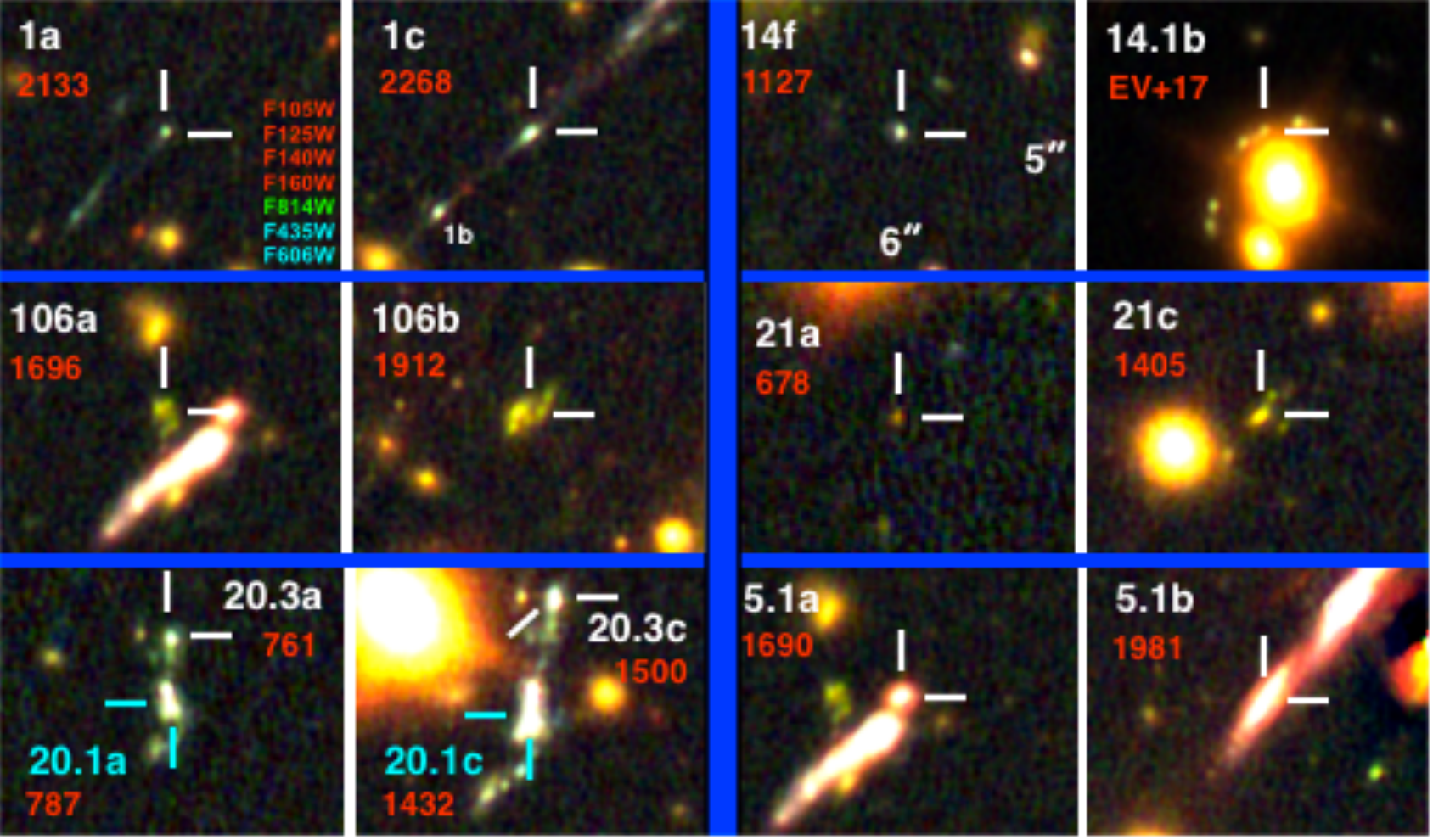}
        \caption{RGB HST images reported in Table~\ref{ratios} are shown on two columns (divided by the vertical blue bar). The least magnified image of source 1 is shown on the top-left (1a), and the most magnified (1c) in the adjacent cutout on the right. The same applies for the other images. The size of the cutouts is $6'' \times 5''$. The ASTRODEEP identifier is quoted in red, while our internal ID is in white. For source 20 (bottom-left), two clumps are reported: 20.3a,  20.3c (marked with white segments), and 20.1a, 20.1c (cyan segments).}
        \label{examples_muerr}
\end{figure}

Apart from source 1, all the magnifications of the other sources reported in the table agree well with the model prediction, on average. However, in some cases such prediction are not within the formal 1-$\sigma$ statistical error derived from the model (68\% interval), suggesting that systematic errors can have a dominant role \citep[e.g.,][]{Meneghetti_2017}. Despite of that, this preliminary test demonstrates the good predicting power of the lens model in the moderate-to-high magnification regime.
Figure~\ref{examples_muerr} shows the HST RGB cutouts of the objects reported in Table~\ref{ratios}. 

\begin{table*}
\caption{\label{ratios} Comparison of magnification values from the method based on flux ratios (or clump angular separations) and from the lens model for a subset of sources discussed in the main text.}
\centering
\begin{tabular}{lclcclc}
\hline\hline
 ID & z & Least $\mu$ ($\pm 1\sigma$)\tablefootmark{($\#$)} & Ratio\tablefootmark{($\star$)}        & Rescaled $\mu$ ($\pm 1\sigma$) & Model $\mu^{+}_{-} 68$ (95,99)\% & Sect/Fig\\
\hline
1c\tablefootmark{I} & 3.2370 & $\mu_{tot}(1a)=56.6\pm9.6$ & $8.89\pm 1.86$ (F)       & $\mu_{tot} = 502 \pm 136$ & $\mu_{tot} = 78.1^{+19.1~(34.9,59.9)}_{-13.4~(20.4,26.1)}$ & {\small \ref{general}/\ref{dyn},\ref{examples}}\\ \\

5.1b  & 1.8961 &$\mu_{tot}(5.1a)=3.63\pm0.09$ & $4.33\pm 0.13$ (F)       & $\mu_{tot} = 15.7 \pm 0.6$ & $\mu_{tot} = 9.1^{+0.6~(1.2,2.0)}_{-0.6~(0.9,1.1)}$ & {\small \ref{multiple}/\ref{sys5}}\\ \\

5.[4,2)]c\tablefootmark{II}  & 1.8961 &$\mu_{tang}(5.(4,2)a) = 2.4\pm0.1$ & $3.07\pm0.36$ (A)       & $\mu_{tang} = 7.5 \pm 0.9$ & $\mu_{tang} = 5 - 9$ & {\small \ref{multiple}/\ref{sys5}} \\ \\
14.1b\tablefootmark{III} & 3.2226 &$\mu_{tot}(14f) = 2.08\pm0.02$ & $7.29\pm 0.83$ (F)  & $\mu_{tot}=15.2\pm1.7$ & $\mu_{tot} = 19.4^{+11.1~(29.6,69.6)}_{-5.9~(8.1,9.4)}$ & {\small \ref{14}/\ref{dyn},\ref{fig_id14}}\\ \\

14.[1,2]b\tablefootmark{IV}  & 3.2226 &$\mu_{tang}(14f) = 1.5\pm0.1$ & $>7.5$ (A)       & $\mu_{T}>11.2 $& $\mu_{T} \simeq 7-20$ & {\small \ref{14}/\ref{dyn},\ref{fig_id14}}\\ \\
20.3c  & 3.2190 &$\mu_{tot}(20.3a) = 2.24\pm0.03$ & $2.97\pm0.22$ (F)       & $\mu_{tot}=6.7 \pm 0.5$& $\mu_{tot}= 5.0^{+0.1~(0.2,0.3)}_{-0.1~(0.2,0.3)}$ & {\small \ref{multiple}/\ref{system20}} \\ \\
20.1c  & 3.2190 &$\mu_{tot}(20.1a) = 2.20\pm0.03$ & $3.47\pm0.14$ (F)       & $\mu_{tot}=7.6 \pm 0.4$& $\mu_{tot}= 6.2^{+0.2~(0.3,0.5)}_{-0.2~(0.3,0.4)}$ & {\small \ref{multiple}/\ref{system20}} \\ \\
21c\tablefootmark{II}  & 5.1093 &$\mu_{tot}(21a) = 2.04\pm0.02$ & $7.49\pm 0.84$ (F)       & $\mu_{tot} = 15.3\pm1.7$ & $\mu_{tot} = 11.6^{+0.3~(0.6,0.8)}_{-0.3~(0.5,0.6)}$ & {\small \ref{multiple}/\ref{system21}} \\ \\
106b  & 4.1162 &$\mu_{tot}(106a) = 4.54\pm0.11$ & $1.77\pm 0.11$ (F)       & $\mu_{tot} = 8.1\pm0.5$ & $\mu_{tot} = 11.8^{+0.6~(1.2,1.8)}_{-0.5~(0.8,1.3)}$ & {\small \ref{multiple}/\ref{sys106}}\\
\hline
\hline
\end{tabular}
\tablefoot{\\
\tablefootmark{$(\#$)}{The magnification value of the least magnified multiple image.}\\
\tablefootmark{($\star$)}{Flux or angular separation ratios are indicated with `F' or `A', respectively, and are calculated on images with IDs reported in columns 1 and 3.}\\
\tablefoottext{I}{The flux ratio calculated among images 1a (the least magnified) and 1c (the most magnified), based on the F814W band. We note that $\mu_{tot}$(1a) is possibly subjected to large uncertainty, we rely on the lens model for the magnification of this object (see Appendix~\ref{muerr}).}\\
\tablefoottext{II}{Ratio of the angular separations between knots 5.4 and 5.2 for group images `a' and `c'.}\\
\tablefoottext{III}{Based on flux ratio between images 14.2b (or 14.1b), and the least magnified image 14f, taking into account that 14f is the sum of the two knots (14.1+14.2, see Sect.~\ref{14}).}\\
\tablefoottext{IV}{Based on the relative angular separation among knots 14.2b $-$ 14.1b and the upper limit on 14.1f (see Appendix~\ref{individual14}).}\\
}
\end{table*}

\end{appendix}

\end{document}